\documentclass[10pt,final,journal,twocolumn,singlespaced]{IEEEtran}
\usepackage[linesnumbered,ruled,vlined]{algorithm2e}
\usepackage{algpseudocode}
\usepackage{amsmath}
\usepackage{amssymb}
\usepackage{array}
\usepackage{bm}
\usepackage{booktabs}
\usepackage{color}
\usepackage{cite}
\usepackage{epsfig}
\usepackage{epstopdf}
\usepackage{flafter}
\usepackage{float}
\usepackage{ifthen}
\usepackage{multirow}
\usepackage{makecell} 
\usepackage{pifont}
\usepackage{subfigure}
\usepackage{times}
\usepackage{threeparttable}
\usepackage{subfigure,graphicx}
\usepackage{float}
\usepackage{color,multirow}
\usepackage{makecell}
\usepackage{cite} 
\usepackage{url}  
\usepackage{ifthen}  
\usepackage[T1]{fontenc}
\usepackage{graphicx,mdframed}
\usepackage{comment,bbold}

\definecolor{orange}{RGB}{255,107,0}

\def\red{\color{red}}

\newcommand{\W}{\boldsymbol{W}}

\newcommand{\Y}{\boldsymbol{Y}}

\newcommand{\X}{\boldsymbol{X}}
\newcommand{\C}{\boldsymbol{C}}

\newcommand{\A}{\boldsymbol{A}}

\newcommand{\B}{\boldsymbol{B}}
\renewcommand{\S}{\boldsymbol{S}}

\newcommand{\s}{\boldsymbol{s}}
\renewcommand{\c}{\boldsymbol{c}}

\newcommand{\y}{\boldsymbol{y}}

\newcommand{\T}{{\!\top\!}}

\newcommand{\tX}{\underline{\bm X}}

\newcommand{\tY}{\underline{\bm Y}}
\newcommand{\tG}{\underline{\bm G}}

\newtheorem{theorem}{Theorem}
\newtheorem{lemma}{Lemma}
\newtheorem{proposition}{Proposition}

\newtheorem{remark}{Remark}

\begin{document}
\title{Hyperspectral Super-Resolution via Interpretable Block-Term Tensor Modeling}

\author{Meng Ding, Xiao Fu, Ting-Zhu Huang, Jun Wang, and Xi-Le Zhao

\thanks{This work of M. Ding, T. Huang, and X. Zhao are supported in part by the National Natural Science Foundation of China (NSFC) under Grants 61876203 and 61772003. X. Fu is supported in part by National Science Foundation under Projects ECCS 1608961 and ECCS 1808159, and the Army Research Office under Project ARO W911NF-19-1-0247. 
J. Wang is supported in part by the NSFC under Grant U19B2014, in part by the National Research Program of China under Grant 9020302, in part by the Foundation of National Key Laboratory of Science and Technology on Communications, and in part by the Innovation Fund of NCL (IFN) under IFN2019102.

M. Ding, T. Huang, and X. Zhao are  with the School of Mathematical Sciences at University of Electronic Science and Technology of China, Chengdu, China. e-mail: dingmeng56@163.com, tingzhuhuang@126.com, and xlzhao122003@163.com. The work is developed during M. Ding's visit to Oregon State University.

X. Fu is with the School of Electrical Engineering and Computer Science, Oregon State University (OSU), Corvallis, OR 97331, United States. email: xiao.fu@oregonstate.edu.

J. Wang is with the National Key Laboratory of Science and Technology on Communications at University of Electronic Science and Technology of China, Chengdu, China. email: junwang@uestc.edu.cn.}
}
\maketitle

\begin{abstract}
This work revisits coupled tensor decomposition (CTD)-based hyperspectral super-resolution (HSR). HSR aims at fusing a pair of hyperspectral and multispectral images to recover a super-resolution image (SRI). 
The vast majority of the HSR approaches take a low-rank matrix recovery perspective. The challenge is that theoretical guarantees for recovering the SRI using low-rank matrix models are either elusive or derived under stringent conditions.
A couple of recent CTD-based methods ensure recoverability for the SRI under relatively mild conditions, leveraging on algebraic properties of the {\it canonical polyadic decomposition} (CPD) and the {\it Tucker} decomposition models, respectively.
However, the latent factors of both the CPD and Tucker models have no physical interpretations in the context of spectral image analysis, which makes incorporating prior information challenging---but using priors is often essential for enhancing performance in noisy environments. This work employs an idea that models spectral images as tensors following the block-term decomposition model with multilinear rank-$(L_r,L_r,1)$ terms (i.e., the {\sf LL1} model) and formulates the HSR problem as a coupled {\sf LL1} tensor decomposition problem. 
Similar to the existing CTD approaches, recoverability of the SRI is shown under mild conditions.
More importantly, the latent factors of the {\sf LL1} model can be interpreted as the key constituents of spectral images, i.e., the endmembers' spectral signatures and abundance maps. This connection allows us to easily incorporate prior information for performance enhancement.
A flexible algorithmic framework that can work with a series of structural information is proposed to take advantages of the model interpretability. The effectiveness is showcased using simulated and real data.
\end{abstract}

\section{Introduction}
The spatial and spectral resolution tradeoff in spectral image sensing has been a well-known effect \cite{Yokoya2017HSRoverview}. To be specific, hyperspectral sensors acquire data with high spectral resolution but low spatial resolution. However, multispectral sensors produce images that exhibit high spatial resolution---but the spectral resolution is often coarse. Since both types of information are of great interest, many hyperspectral super-resolution (HSR) approaches have been proposed \cite{Yokoya2017HSRoverview}. The goal of HSR is to fuse a pair of co-registered hyperspectral image (HSI) and multispectral image (MSI) to produce a super-resolution image (SRI) that has the ``best of the two''---i.e., high resolutions in both the spectral and spatial domains.

Many classic HSR techniques (e.g., \cite{Yokoya2012HSR,Simoes2015HSR,Wei2015HSRSylvester}) view the SRI as a low-rank matrix, which is reminiscent of the {\it linear mixture model} (LMM) for spectral images \cite{Bioucas2012HUOverview}. The HSI and MSI are modeled as ``marginalized'' data of the SRI by degrading from the spatial and spectral modes, respectively.
Consequently, recovering the SRI can be understood as low-rank matrix estimation from downsampled data---which is an ill-posed inverse problem. 
Recoverability for the SRI using low-rank estimation techniques had been elusive until a couple of recent works \cite{Li2018HSRmatrix,liu2019there}. 
Nevertheless, as pointed out in \cite{Li2018HSRmatrix,liu2019there}, matrix estimation-based approaches have recoverability guarantees for the SRI under somewhat stringent conditions, e.g., when every pixel only contains a small number of endmembers.

In 2018, Kanatsoulis {\it et al.} proposed a {\it coupled tensor decomposition} (CTD) framework for HSR \cite{Kanatsoulis2018HSR}. There, the spectral images are modeled as tensors with low-rank canonical polyadic decomposition (CPD) representations. Utilizing the algebraic properties of CPD, recoverability of the SRI is shown to hold under mild conditions. Later on, the work in \cite{Prevost2020HSR} adopted a similar idea but used the Tucker tensor model, which facilitates a fast algorithm leveraging on the higher-order singular value decomposition (SVD).
Note that a number of works from the computer vision and remote sensing communities also take tensor perspectives for HSR; see, e.g., \cite{Dian2017HSR,Li2018HSR}. Nonetheless, these works put more emphasis on the computation side, but no recoverability support was offered.

Compared to the low-rank matrix-based methods, e.g., those in \cite{Yokoya2012HSR,Simoes2015HSR,Wei2015HSRSylvester}, the tensor methods in \cite{Kanatsoulis2018HSR,Prevost2020HSR} are advantageous in terms of recoverability guarantees. Nevertheless, the performance gain obtained by the tensor methods is not always obvious. One possible reason is that the model parameters (or, the latent factors) of the tensor models used in \cite{Kanatsoulis2018HSR,Prevost2020HSR} do not have physical interpretations in the context of HSR. Consequently, it is hard to incorporate prior information into the CTD frameworks for performance enhancement. On the other hand, since the matrix-based approaches have clear connections to the LMM, the latent factors there can be interpreted as
two key constituents of spectral images, i.e., the spectral signatures of endmembers (i.e., materials captured in the image) and the corresponding abundance maps---which makes using prior information (e.g., nonnegativity and spatial smoothness of the abundance maps) fairly easy through adding constraints and regularization terms to their optimization criteria---see examples in \cite{Wycoff2013HSR,Lanaras2015HSR,Wei2015HSRsparse}. Note that for real-world data, since noise and modeling errors always exist, taking advantage of prior information oftentimes plays an essential role in producing high-quality image fusion results.

In this work, our objective is an alternative CTD approach. Our aim at a framework that offers similar recoverability guarantees as in the CPD and Tucker-based methods \cite{Kanatsoulis2018HSR,Prevost2020HSR}---and at the same time has physical interpretations for its model parameters. This way, prior information can be swiftly incorporated in the framework to enhance the HSR performance under challenging scenarios. To this end, our idea is to employ the tensor decomposition model with multilinear rank-$(L_r,L_r,1)$ block-terms (or, the {\sf LL1} model for short) \cite{Lathauwer2008BTD2} for modeling the spectral images. The {\sf LL1} model was recently connected to HSIs in \cite{Qian2017Unmixing}, where the latent factors of the {\sf LL1} model were linked to the LMM model. This connection was exploited for hyperspectral unmixing in \cite{Qian2017Unmixing}. Nonetheless, using the {\sf LL1} model for HSR has not been considered, to our best knowledge.

\noindent
{\bf Contributions:} Our detailed contributions are as follows:

\noindent
$\bullet$ {\bf Recoverability Guarantees.} We formulate the HSR problem as a coupled {\sf LL1} tensor decomposition problem, and show that the formulated criteria can provably recover the SRI under mild conditions. Similar properties have been observed in existing tensor approaches that are based on the CPD or Tucker models. Nonetheless, our analysis and recoverability conditions are new and specialized for the {\sf LL1} modeling. As in \cite{Kanatsoulis2018HSR,Prevost2020HSR}, we also show that the {\sf LL1} model guarantees recoverability under a realistic yet challenging scenario, i.e., the case where the spatial degradation process is unknown. The recoverability analysis ensures that one does not lose theoretical guarantees when using the advocated {\sf LL1} model.

\noindent
$\bullet$ {\bf Flexible Algorithmic Framework.}
We recast the formulated criteria to optimization forms that are convenient to incorporate regularization and constraints. In particular, we propose a change of variable scheme together with a Shatten-$p$ function-based low-rank promotion regularization, for approximating the {\sf LL1} model. This way, the working optimization problems admit continuously differentiable objectives, and thus are relatively easy to tackle. More importantly, this reformulation strategy makes the spectral signatures and abundance maps explicitly present in the optimization objectives, and thus constraints and regularization can be naturally imposed.
We showcase the flexibility of this framework via incorporating nonnegativity of the endmembers and the spatial smoothness of the abundance maps. We propose to handle the formulated problems using a unified inexact and accelerated {\it block coordinate descent} (BCD) algorithmic framework.

\noindent
$\bullet$ {\bf Extensive Experiments.} We test the proposed algorithms over four different semi-real datasets (namely, semi-real data derived from the Salinas, Pavia University, Indian Pines, and Jasper Ridge datasets\footnote{https://rslab.ut.ac.ir/data}) and different performance metrics.
In addition, we test the algorithms on a pair of co-registered real HSI and MSI that were provided in the recent paper \cite{Yang2018HSR}.

\begin{figure}[t!]
\centering
\includegraphics[width=0.5\linewidth]{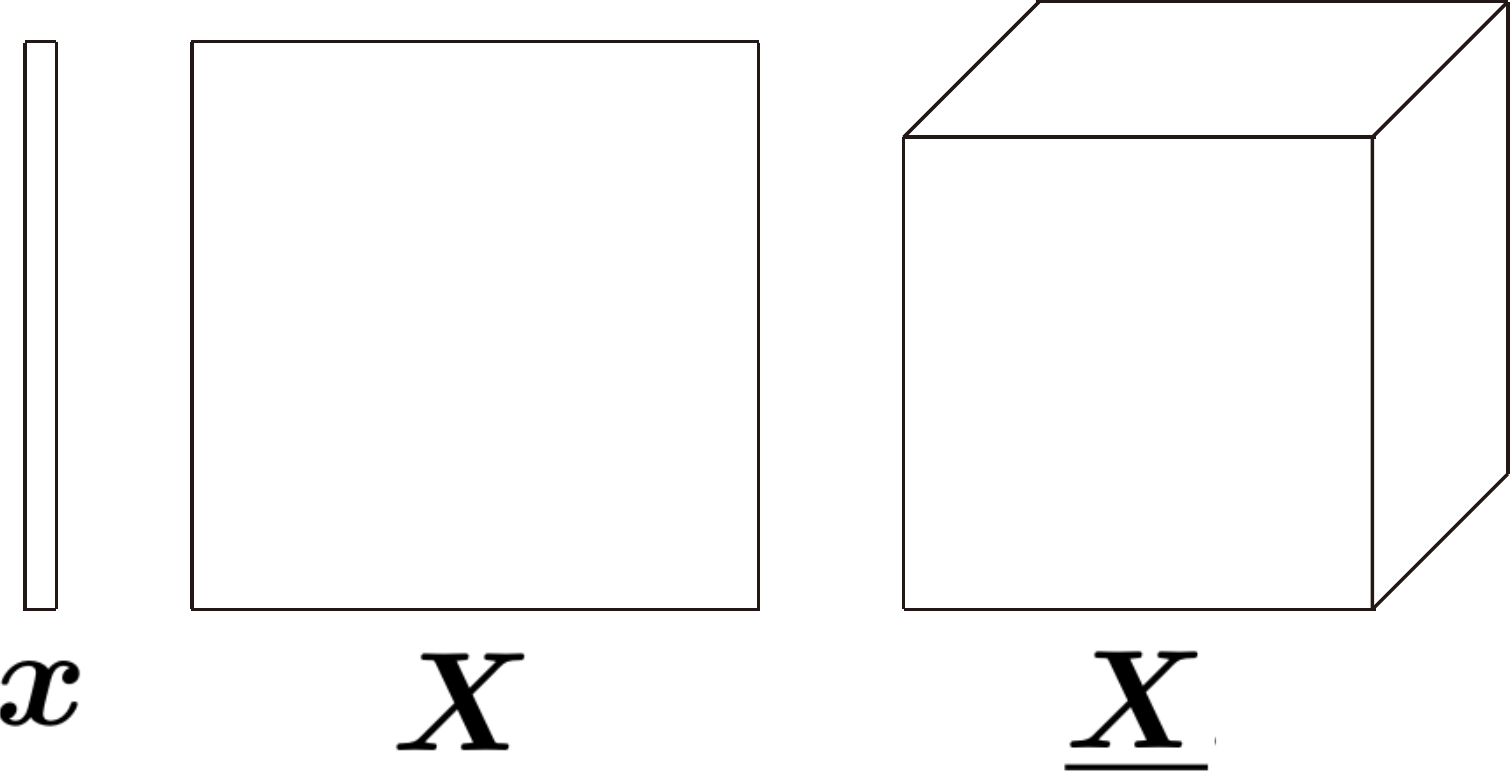}\hspace{1cm}
~\includegraphics[width=0.32\linewidth]{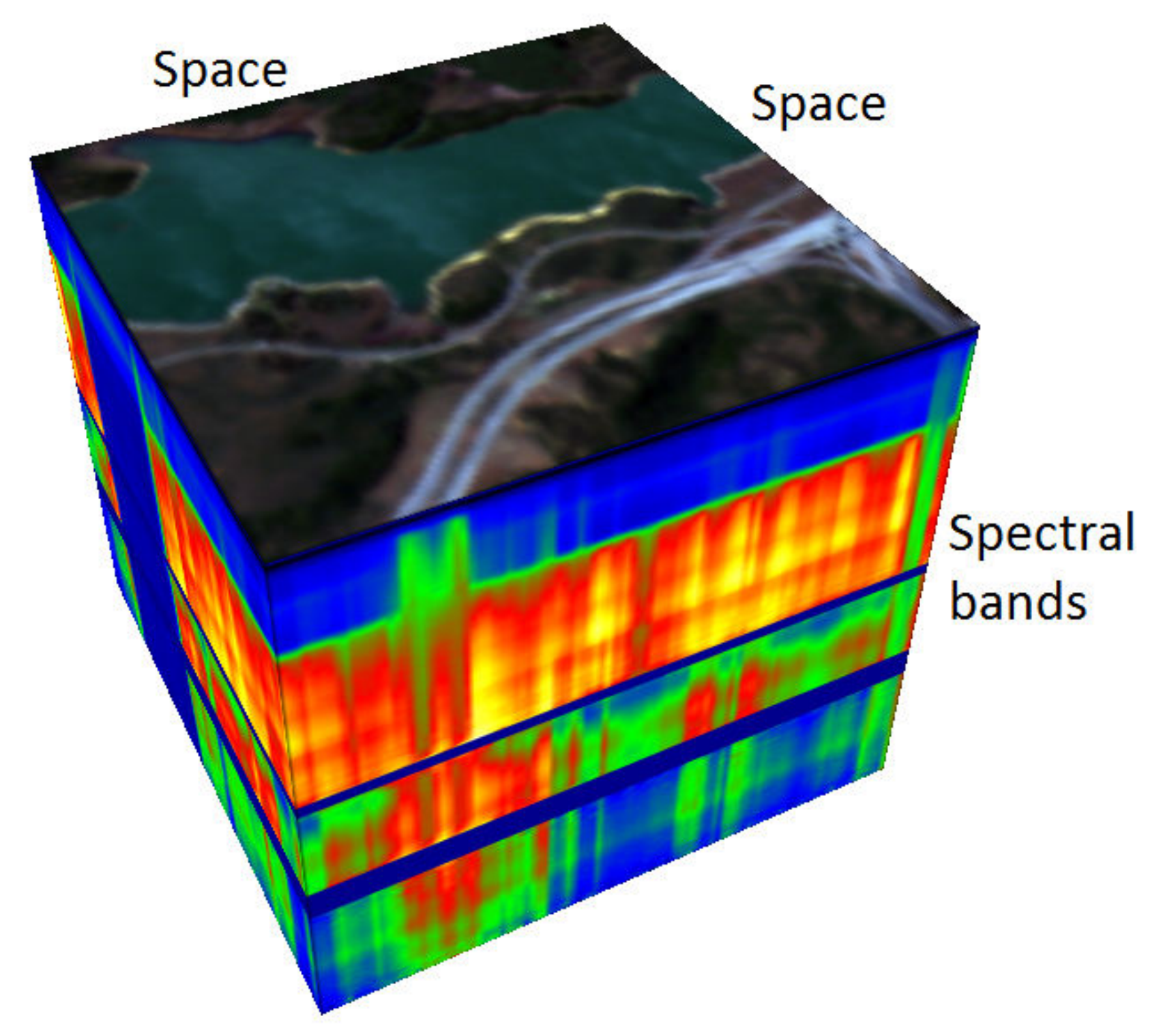}
\caption{Left: A vector, matrix, and a tensor. Right: A spectral image can be understood as a third-order tensor. Every spectral pixel is measured over a large number of spectral bands.}
  \label{fig:cube}
\end{figure}

\smallskip

A conference version that contains part of the work was published in Proc. IEEE CAMSAP 2019 \cite{Zhang2019HSR}. In the journal version, we additionally included more recoverability analysis results, the flexible algorithmic framework, extensive semi-real experiments, and the real experiment.

\color{black}
\smallskip

\noindent
{\bf Notation.}
A scalar, a vector, a matrix, and a tensor are denoted as $x$, $\bm{x}$, $\bm{X}$, and $\underline{\bm{X}}$, respectively. $[\bm{x}]_i$, $[\bm{X}]_{i,j}$, and $[\underline{\bm{X}}]_{i,j,k}$ denote the $i$-th, $(i,j)$-th, and $(i,j,k)$-th element of $\bm{x}\in \mathbb{R}^{I}$, $\bm{X}\in \mathbb{R}^{I\times J}$, and $\underline{\bm{X}}\in \mathbb{R}^{I\times J\times K}$, respectively.
The Matlab notation $\bm{X}(:,j)$ or $\bm{x}_j$ represents the $j$-th column of a matrix $\bm{X}\in \mathbb{R}^{I\times J}$ in an exchangeable manner. $\underline{\bm{X}}(i,j,:)$ and $\underline{\bm{X}}(:,:,k)$ denote the $(i,j)$-th tube and the $k$-th slab of $\underline{\bm{X}}$, respectively. The Frobenius norms of $\bm{X}$ and $\underline{\bm{X}}$ are denoted as $\|\bm{X}\|_{F}=\sqrt{\sum_{i,j}[\bm{X}]_{i,j}^{2}}$ and $\|\underline{\bm{X}}\|_{F}=\sqrt{\sum_{i,j,k}[\underline{\bm{X}}]_{i,j,k}^{2}}$, respectively. Given two vectors $\bm{x}\in \mathbb{R}^{M}$, $\bm{y}\in \mathbb{R}^{N}$ and two matrices $\bm{X}\in \mathbb{R}^{I\times J}$, $\bm{Y}\in \mathbb{R}^{P\times Q}$, the outer product $\bm{x}\circ \bm{y}$, $\bm{X}\circ \bm{y}$, and the Kronecker product $\bm{X}\otimes \bm{Y}$ are a $M\times N$ matrix, a $I\times J\times N$ tensor, and a $IP\times JQ$ matrix, respectively. The Khatri-Rao product $\odot_c$ is defined as $\bm{X}\odot_c \bm{Y}=[\bm{x}_1\otimes \bm{y}_1,\cdots,\bm{x}_J\otimes \bm{y}_J]\in \mathbb{R}^{IP\times J}$ when $J=Q$. Let $\bm{X}=[\bm{X}_1,\cdots,\bm{X}_R]\in \mathbb{R}^{I\times \sum_{r}J_r}$ and $\bm{Y}=[\bm{Y}_1,\cdots,\bm{Y}_R]\in \mathbb{R}^{P\times \sum_{r}Q_r}$ be two partitioned matrices, the partitioned Khatri-Rao product $\odot$ is defined as $\bm{X}\odot \bm{Y}=[\bm{X}_1\otimes \bm{Y}_1,\cdots,\bm{X}_R\otimes \bm{Y}_R]\in \mathbb{R}^{IP\times \sum_{r}J_rQ_r}$ \cite{Lathauwer2008BTD2}. The $i$-th singular value of $\bm{X}$ is denoted as $\sigma_{i}(\bm{X})$ and the largest one as $\sigma_{\textrm{max}}(\bm{X})$. $\widehat{\bm{X}}$ is used as the estimator of $\bm{X}$. 

\section{Background}
In this section, we first briefly introduce some preliminaries pertaining to our framework. 
\subsection{Tensor Modeling for Spectral Images}
Third-order tensors $\tX\in\mathbb{R}^{I\times J\times K}$ can be understood as generalization for vectors and matrices (see Fig.~\ref{fig:cube}).
Apparently, spectral images are tensors, which admit two spatial dimensions and one spectral dimension---also see Fig.~\ref{fig:cube}.

One of the key differences between tensors and matrices is that the definition for tensor rank is nonsingular---which consequently makes rank decomposition for tensors have various forms.
The arguably most popular tensor rank decomposition model is the so-called {\it canonical polyadic decomposition} (CPD) model \cite{Hitchcock1927CPD}, which is also known as the parallel factor analysis (PARAFAC) model.
Under the CPD model, a tensor with CP rank $R$ can be written as
\[ \tX =\sum_{r=1}^R \A(:,r)\circ\B(:,r) \circ \C(:,r).      \]

\begin{figure}
\centering
\includegraphics[width=0.85\linewidth]{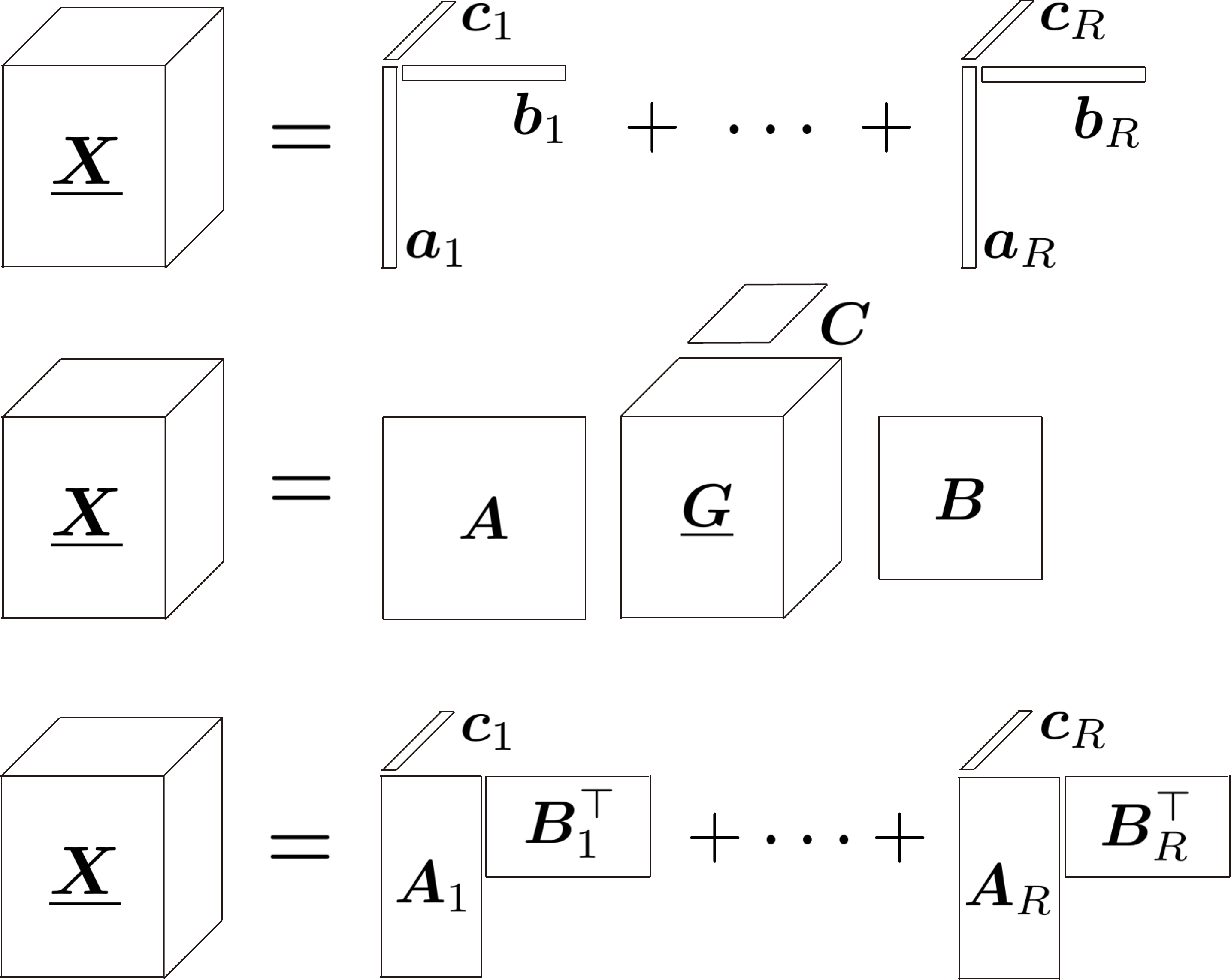}
\caption{Illustration of three tensor decompositions. Top: CPD. Middle: Tucker. Bottom: {\sf LL1}.}
  \label{fig:tensor_comparison}
\end{figure}
Besides, the Tucker model \cite{Tucker1966Tucker} has also been widely used in the imaging society. Under the Tucker model, a tensor can be decomposed as
\[\tX =\sum_{i=1}^I\sum_{j=1}^J\sum_{k=1}^K  \tG(i,j,k) \A(:,i)\circ\B(:,j) \circ \C(:,k).\]
Both the CPD and Tucker models are able to ``ecode'' dependence across different dimensions of the data, and thus are meaningful low-rank models (see illustrations in Fig.~\ref{fig:tensor_comparison}). In fact, both the CPD and Tucker models have been utilized for modeling hyperspectral and multispectral images in the literature. In particular, \cite{Kanatsoulis2018HSR} and \cite{Prevost2020HSR} employed coupled CPD and coupled Tucker decomposition models, respectively, for hyperspectral super-resolution.
In both cases, leveraging the low-rank structure of the tensor models, the recoverability of SRI can be proved---under mild conditions.
Note that most low-rank matrix estimation-based approaches, e.g., those in \cite{Yokoya2012HSR,Simoes2015HSR,Wei2015HSRsparse,Wycoff2013HSR}, do not have recoverability guarantees (except for some recent ones under relatively stringent conditions \cite{Li2018HSRmatrix,liu2019there}).

\subsection{Linear Mixture Model (LMM) and Matrix-based HSR}
The CPD and Tucker based HSR frameworks are appealing due to the SRI recoveriability guarantees under mild conditions. However, the performance in practice is not always better than the heuristic-driven low-rank matrix estimation-based approaches.
One reason may be that the matrix models often exploit the physical interpretations of their latent factors. Imposing structural information of the latent factors helps regularize the HSR criteria with prior knowledge. This is often essential for performance enhancement when noise and modeling errors are present.
On the other hand, both the CPD and the Tucker models do not have physical interpretations for their latent factors. Therefore, in spite of the recoverability appeal, the challenge for incorporating prior information makes it hard to further improve the performance of the CPD/Tucker-based HSR approaches.

\begin{figure}[t]
\centering
\includegraphics[width=0.98\linewidth]{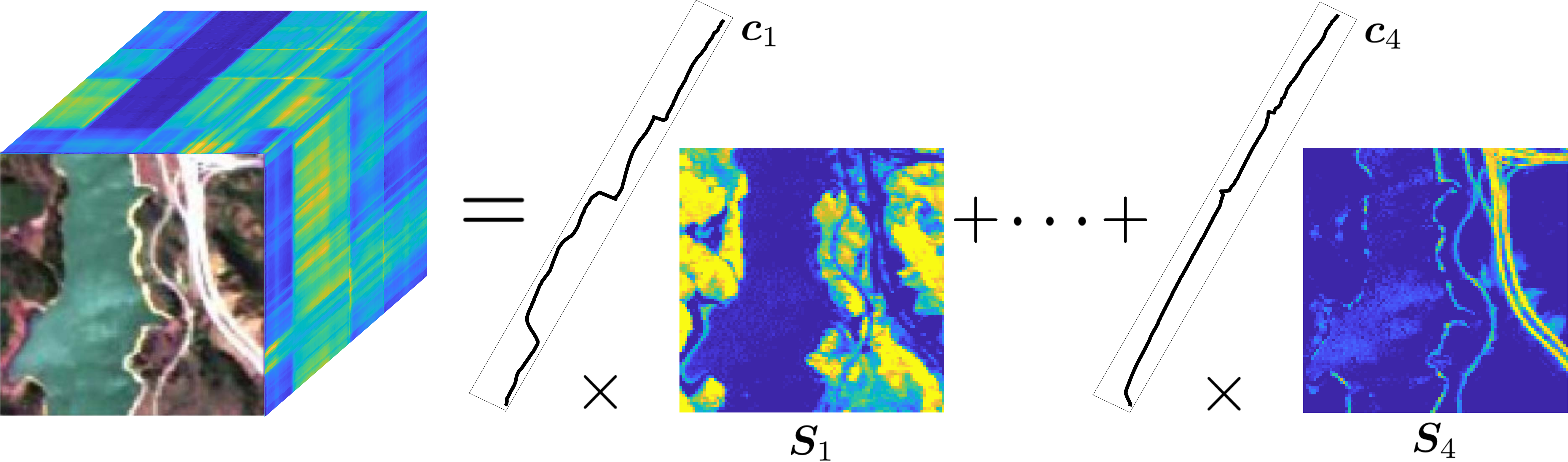}
\caption{The LMM of spectral images. The vector $\bm{c}_{r}$ denotes the spectral signature of material $r$; the matrix $\bm{S}_{r}$ denotes the corresponding abundance map.}
  \label{fig:Ridge_physical}
\end{figure}

To see how the matrix factorization-based approaches exploit the physical interpretations, consider the LMM \cite{Bioucas2012HUOverview} of spectral pixels. 
In the noise-free case, a spectral pixel of an $I\times J\times K$ image can be written as follows:
\[   \y_\ell = \C \bm s_\ell\in\mathbb{R}^K,~\ell=1,2,\ldots,N,      \]
where $K$ denotes the number of spectral bands, $\C=[\c_1,\ldots,\c_R]\in\mathbb{R}^{K \times R}$ collects $R$ different spectral signatures of materials (endmembers) contained in the pixel, $N$ is the total number of pixels, and $\s_\ell(r)$ is the proportion of endmember $r$ in pixel $\ell$ (i.e., the abundance of endmember $r$). Putting all the pixels together one has
\begin{equation}\label{eq:LMM}
    \Y =\C\S^{\top},
\end{equation}
where 
\begin{equation}\label{eq:S}
\S=\left[\s_1,\ldots,\s_N\right]^{\top}
\end{equation}
and $\Y=[\y_1,\ldots,\y_N]$. Note that $\S(:,r)$ can be reshaped to an $I\times J$ matrix, which is often referred to as the abundance map of endmember $r$ (see Fig. \ref{fig:Ridge_physical}).
Both the HSI and MSI can be expressed following the above LMM. If the HSI and MSI are co-registered, the two LMMs' latent factors are coupled together.
Then, the HSR problem can be formulated as a coupled matrix factorization problem; see details in \cite{Wycoff2013HSR,Wei2015HSRsparse,Wei2015HSRSylvester}. 
In addition, the latent factors in the LMM (i.e., $\C$ and $\S$) both have strong physical interpretations, and thus a variety of prior knowledge can be readily utilized. For example, both $\C$ and $\S$ should be nonnegative \cite{Wycoff2013HSR,Wei2015HSRSylvester}, and the abundance map ${\rm mat}(\S(:,r))$ should exhibit low total variation (TV) across both the row and the column dimensions \cite{Simoes2015HSR}. 
As a consequence, these priors can be incorporated into the computational frameworks to enhance the HSR performance.

\subsection{Interpretable Tensor Modeling for Spectral Images}
Recently, a relatively new tensor model, namely, the block-term decomposition into multilinear rank-$(L_r,L_r,1)$ terms model (or simply the {\sf LL1} model) \cite{Lathauwer2008BTD2}, has been connected to spectral images \cite{Qian2017Unmixing}. The model is very similar to the LMM, but has an extra assumption---i.e., the abundance maps are low-rank matrices. To be specific, instead of looking at the LMM model from a matrix view as in \eqref{eq:LMM}, one can re-write the LMM as follows:
\begin{equation}\label{eq:LMM_tensor}
    \tY = \sum_{r=1}^R \bm S_r  \circ \C(:,r),
\end{equation}
where 
\begin{equation}\label{eq:S_r}
\S_r ={\rm mat}(\S(:,r))\in\mathbb{R}^{I\times J}
\end{equation}
represents the abundance map in its (more natural) matrix form, and $\C(:,r)=\bm c_r\in\mathbb{R}^{K}$ the $r$-th endmember; see Fig.~\ref{fig:Ridge_physical} for illustration.
Let ${\rm rank}(\S_r)=L_r \leq \min\{I,J\}$, the above can be re-written as
\begin{equation}\label{eq:LMM_tensor_LL1}
\tY = \sum_{r=1}^R (\bm A_r\B_r^\T)  \circ \C(:,r),
\end{equation}
where $\A_r\in\mathbb{R}^{I\times L_r}$ and $\B_r\in\mathbb{R}^{J\times L_r}$ are full column-rank matrices and $\S_r=\A_r\B_r^\T$. The tensor model in \eqref{eq:LMM_tensor_LL1} is exactly the {\sf LL1} model.

The {\sf LL1} model was used in hyperspectral imaging for unmixing the endmembers \cite{Qian2017Unmixing,Xiong2019HUBTD}. This is well motivated. First, since the abundance maps are correlated across the two spatial dimensions, assuming $\S_r$ to be low-rank is plausible. Second, since the {\sf LL1} model is unique under mild conditions, the abundance maps and the endmembers can be identified up to certain trivial ambiguities with provable guarantees.

Using the {\sf LL1} model for HSR seems to be appealing, since physical interpretations of the key constituents in spectral images are reflected in the model. 
Nevertheless, although the {\sf LL1} model exhibits promising advantages in hyperspectral unmixing (HU), the HU problem has very different settings and objectives relative to HSR.
Particularly, there are a number of challenges of utilizing the {\sf LL1} model for HSR. First, it is unclear if certain SRI recoverability guarantees, e.g., those were shown under the CPD and Tucker models, still exist under the {\sf LL1} model. Second, numerical optimization involving the {\sf LL1} model while considering structural constraints on the abundance maps and spectral signatures is a challenging problem.

\section{Coupled {\sf LL1} Tensor Decomposition for HSR}
In this section, we propose an {\sf LL1} model-based HSR approach, and discuss its recoverability properties.
\subsection{Problem Formulation}
We follow the setup in \cite{Kanatsoulis2018HSR} for tensor-based HSR. Specifically, we use $\tY_H\in\mathbb{R}^{I_H\times J_H\times K_H}$, $\tY_M \in \mathbb{R}^{ I_M\times J_M\times K_M }$ and $\tY_S\in\mathbb{R}^{I_M\times J_M\times K_H}$ to denote the HSI, MSI, and SRI, respectively. 
All the spectral images are represented in a ``space$\times$ space$\times$ spectrum'' format.
We assume that the HSI and MSI are downsampled/degraded from a SRI. Note that the SRI has the spatial resolution of the MSI (i.e., $I_M\times J_M$) and the spectral resolution of the HSI (i.e., $K_H$).
\begin{figure}[t]
    \centering
    \includegraphics[width=.99\linewidth]{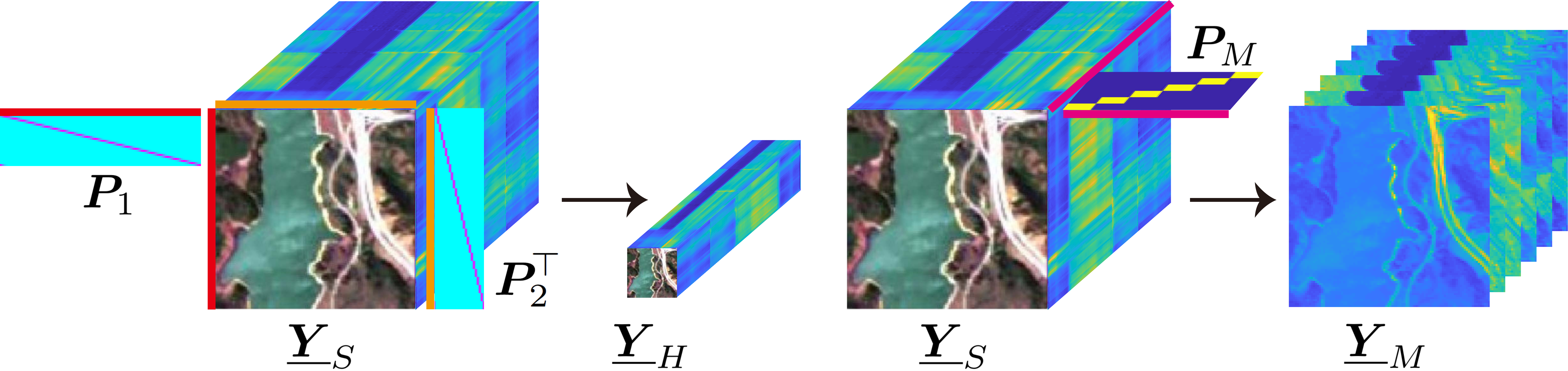}
    \caption{{Illustration of spatial and spectral degradation procedures from SRI to HSI and MSI, respectively.}}
    \label{fig:degradation}
\end{figure}

We assume that the abundance maps are low-rank matrices. Consequently, the SRI follows the {\sf LL1} model:
\begin{equation}\label{BTD_Ys}
\underline{\bm{Y}}_{S} = \sum_{r=1}^{R}(\bm{A}_{r}\bm{B}_{r}^{\top})\circ \bm{C}(:,r),
\end{equation}
where $\bm{A}_{r} \in \mathbb{R}^{I_{M} \times L_{r}}$, $\bm{B}_{r} \in \mathbb{R}^{J_{M} \times L_{r}}$ for $r=1,\ldots,R$, $\bm{C}=[\bm{c}_{1},\ldots,\bm{c}_{R}]\in \mathbb{R}^{K_{H} \times R}$.
The HSI can be understood as a spatially blurred and downsampled version of the SRI \cite{Wei2015HSRsparse,Wycoff2013HSR}. This can be modeled by multiplying two blurring and compressing matrices to the row and column dimensions to every slab $\tY_S(:,:,k)$, i.e.,
\begin{equation}\label{eq:modeproduct12}
  \tY_H(:,:,k) = \bm P_1 \tY_S(:,:,k) \bm P_2^\T, \quad k=1,\ldots,K_H, 
\end{equation}       
where $\bm P_1\in\mathbb{R}^{I_H\times I_M}$ and $\bm P_2\in\mathbb{R}^{J_H\times J_M}$ represent two blurring and downsampling matrices along the two spatial dimensions, respectively. In hyperspectral imaging, the blurring operator can be modeled by some kernel functions, e.g., the widely used Gaussian kernel---see detailed discussions in \cite{Kanatsoulis2018HSR}.

If $\tY_S$ follows the {\sf LL1} model and the spatial degradation model in \eqref{eq:modeproduct12} holds, we have the following:
\begin{align}\label{eq:HSI_degradation}
  \underline{\bm{Y}}_{H}&= \sum_{r=1}^R (\bm P_1 \bm S_r \bm P_2^\T) \circ \C(:,r) \nonumber \\
  &=\sum_{r=1}^{R}(\bm{P}_{1}\bm{A}_{r}(\bm{P}_{2}\bm{B}_{r})^{\top})\circ \bm{C}(:,r).
\end{align}
Similarly, the MSI can be modeled as spectral band-aggregated version of the SRI, i.e.,
\begin{equation}\label{eq:modeproduct}
    \tY_M(i,j,:) =  \bm P_M\tY_S(i,j,:), \quad \forall i,j,
\end{equation}   
where $\bm P_M\in\mathbb{R}^{K_M\times K_H}$ is a band aggregation matrix. The above leads to the following compact representation:
\begin{align}\label{eq:MSI_degradation}
\underline{\bm{Y}}_{M}=\sum_{r=1}^{R}(\bm{A}_{r}\bm{B}_{r}^{\top})\circ \bm{P}_{M}\bm{C}(:,r).
\end{align}
The spatial and spectral degradation procedures are illustrated in Fig. \ref{fig:degradation}.

Under the degradation model, the main task of recovering $\underline{\bm{Y}}_{S}$ boils down to estimating the latent factors $\{\bm{A}_{r}\bm{B}_{r}^{\top}\}_{r=1}^{R}$ (high-resolution abandance maps) and $\bm{C}$ (high-resolution endmembers) from the ``marginalized data'' $\underline{\bm{Y}}_{H}$ and $\tY_M$. Once $\{\bm{A}_{r}\bm{B}_{r}^{\top}\}_{r=1}^{R}$ and $\bm{C}$ are identified, one can readily reconstruct the SRI using $\underline{\bm{Y}}_{S} = \sum_{r=1}^{R}(\bm{A}_{r}\bm{B}_{r}^{\top})\circ \bm{C}(:,r)$.

\subsection{Recoverability Analysis}
A key benefit for tensor modeling is that the recoverability of the SRI can be established via exploiting the algebraic properties of various tensor structures, e.g., the CPD and Tucker decompositions, as shown in \cite{Kanatsoulis2018HSR,Prevost2020HSR}.
It is also of interest to consider the recoverability properties under the {\sf LL1} model. Note that the characteristics of the {\sf LL1} model are different from those of the CPD and Tucker models. 
Hence, custom analysis for the {\sf LL1} model is needed.

To answer the recoverability inquiry, we consider the following formulation:
\begin{mdframed}
 \begin{align}\label{Non_blind_BTD_model}
\textrm{find} \ & \{\bm{A}_{r} \in \mathbb{R}^{I_{M} \times L_{r}},\bm{B}_{r}\in \mathbb{R}^{J_{M} \times L_{r}}\}_{r=1}^{R}, \bm{C} \in \mathbb{R}^{K_{H} \times R} \nonumber\\
\textrm{s.t.} \ &\underline{\bm{Y}}_{H}=\sum_{r=1}^{R}(\bm{P}_{1}\bm{A}_{r}(\bm{P}_{2}\bm{B}_{r})^{\top})\circ \bm{C}(:,r), \nonumber\\
&\underline{\bm{Y}}_{M}=\sum_{r=1}^{R}(\bm{A}_{r}\bm{B}_{r}^{\top})\circ \bm{P}_{M}\bm{C}(:,r).
 \end{align} 
\end{mdframed}
The above can be understood as a coupled {\sf LL1} tensor decomposition criterion: we aim at finding an {\sf LL1} tensor (i.e., the SRI) that fits both of the marginalized tensors (the HSI and MSI). 
The recoverability problem amounts to answering if the solution to \eqref{Non_blind_BTD_model} always reconstructs $\tY_S$. Our analysis shows the following:
\begin{theorem} \label{the:non_blind_BTD}
Assume that the SRI $\tY_S$ follows the {\sf LL1} model with $L_r=L$ for all $r$.
Let $(\{\bm{A}_{r},\bm{B}_{r}\}_{r=1}^{R},\bm{C})$ be the ground-truth latent factors of $\underline{\bm{Y}}_{S}$. Assume that $\underline{\bm{Y}}_{H}$ and $\underline{\bm{Y}}_{M}$ follow the degradation models in \eqref{eq:HSI_degradation} and \eqref{eq:MSI_degradation}, respectively. Suppose that each of $\bm{A}_{r}\in\mathbb{R}^{I_M\times L}$, $\bm{B}_{r}\in\mathbb{R}^{J_M\times L}$, and $\bm{C}\in\mathbb{R}^{K_H\times R}$ is drawn from any absolutely continuous distribution, that $\bm{P}_{1}$, $\bm{P}_{2}$, and $\bm{P}_{M}$ have full row rank, and that $(\{\bm{A}_{r}^{\star},\bm{B}_{r}^{\star}\}_{r=1}^{R},\bm{C}^{\star})$ represent any solution to Problem \eqref{Non_blind_BTD_model}. Then, the ground-truth $\underline{\bm{Y}}_{S}$ is uniquely recovered with probability one by
\[
\underline{\bm{Y}}_{S} = \sum_{r=1}^{R}\left (\bm{A}_{r}^{\star}(\bm{B}_{r}^{\star}\right )^{\top})\circ \bm{C}^{\star}(:,r),
\]
if $I_{M}J_{M}\geq L^{2}R$, $I_{H}J_{H}\geq LR$, and
\[
\min \left(\left\lfloor\frac{I_{M}}{L} \right\rfloor, R \right)+\min \left(\left\lfloor\frac{J_{M}}{L} \right\rfloor, R \right)+\min(K_{M},R)\geq 2R+2.
\]
\end{theorem}
The proof of Theorem \ref{the:non_blind_BTD} is relegated to Appendix \ref{app:identifiability_non_blind}. The criterion and Theorem~\ref{the:non_blind_BTD} consider the case where the spatial and spectral degradation operators are both known. In \cite{Kanatsoulis2018HSR,Prevost2020HSR}, it was argued that in some cases the spatial degradation operators may not be easy to estimate. Indeed, the spatial degradation process may be complex and may vary from case to case, caused by various reasons such as the sensor specifications and the sensing environment---modeling this process {\it per se} may be a challenging task. 
A side benefit of using tensor modeling in \cite{Prevost2020HSR,Kanatsoulis2018HSR} is that the SRI recoverability can still be ensured even if the spatial degradation operators $\bm P_1$ and $\bm P_2$ are unknown. In this work, we show that this benefit remains under the {\sf LL1} model.
To see this, we consider the following criterion:
\begin{mdframed}
\begin{subequations}\label{blind_BTD_model}
 \begin{align}
\textrm{find} \ & \{\widetilde{\bm{A}}_{r}\in \mathbb{R}^{I_{H} \times L_{r}},\widetilde{\bm{B}}_{r}\in \mathbb{R}^{J_{H} \times L_{r}}, \bm{A}_{r},\bm{B}_{r}\}_{r=1}^{R}, \bm{C} \nonumber\\
\textrm{s.t.} \ &\underline{\bm{Y}}_{H}=\sum_{r=1}^{R}(\widetilde{\bm{A}}_{r}\widetilde{\bm{B}}_{r}^{\top})\circ \bm{C}(:,r), \label{eq:hsi_constraint}\\
&\underline{\bm{Y}}_{M}=\sum_{r=1}^{R}(\bm{A}_{r}\bm{B}_{r}^{\top})\circ \bm{P}_{M}\bm{C}(:,r) \label{eq:msi_constraint}
 \end{align} 
\end{subequations}
\end{mdframed}
In the above, $\widetilde{\A}_r$ and $\widetilde{\B}_r$ ``absorb'' the spatial degradation operators (i.e., we use $\widetilde{\A}_r$ and $\widetilde{\B}_r$  to model $\bm P_1\A_r$ and $\bm P_2\B_r$, respectively). For this more challenging case, we also show recoverability as follows:

\begin{theorem}\label{the:blind_BTD}
Under the same settings as in Theorem~\ref{our_non_blind_algorithm}, assume that $K_M\geq 2$ and that $(\{\widetilde{\A}_{r}^{\star},\widetilde{\B}_{r}^{\star}\}_{r=1}^{R},\{\bm{A}_{r}^{\star},\bm{B}_{r}^{\star}\}_{r=1}^{R},\bm{C}^{\star})$ is any solution to Problem \eqref{blind_BTD_model}. Then, 
if $I_{H}J_{H}\geq L^{2}R$ and
$
\min \left(\left\lfloor\frac{I_{H}}{L} \right\rfloor, R \right)+\min \left(\left\lfloor\frac{J_{H}}{L} \right\rfloor, R \right)+\min(K_{M},R)\geq 2R+2,
$ the ground-truth $\underline{\bm{Y}}_{S}$ is uniquely recovered by
$
\underline{\bm{Y}}_{S} = \sum_{r=1}^{R}\left (\bm{A}_{r}^{\star}(\bm{B}_{r}^{\star}\right )^{\top})\circ \bm{C}^{\star}(:,r)
$ with probability one.
\end{theorem}
The proof of Theorem \ref{the:blind_BTD} is given in Appendix \ref{app:identifiability_blind}. A remark is that the conditions in Theorem~\ref{the:blind_BTD} are more restrictive relative to those in Theorem~\ref{the:non_blind_BTD}. 
In particular, smaller $L$ and $R$ are needed for the conditions in Theorem~\ref{the:blind_BTD} to hold---which stands for the price to pay for not knowing $\bm P_1$ and $\bm P_2$.

\begin{remark}
The proofs of Theorems~\ref{the:non_blind_BTD}-\ref{the:blind_BTD} are reminiscent of a number of prior works that consider coupled tensor decomposition \cite{Kanatsoulis2018HSR,Zhang2020CartographyBTD,Sorensen2015CoupledCPDBTD,kanatsoulis2019tensor}. In particular, the work in \cite{Sorensen2015CoupledCPDBTD,Zhang2020CartographyBTD} considered coupled LL1 decomposition under various settings. Nonetheless, the recoverability results in \cite{Sorensen2015CoupledCPDBTD,Zhang2020CartographyBTD}
cannot be directly applied to our case. 
First, the work in \cite{Sorensen2015CoupledCPDBTD} does not consider ``marginalziation'' (e.g., compression by $\bm P_1$, $\bm P_2$ and $\bm P_M$ matrices in our case).
Second, the work in \cite{Zhang2020CartographyBTD} considers special marginalization matrices. There, the marginalization matrices are all slab/fiber selection matrices, which makes the proof relatively simple. In our case, $\bm P_1$, $\bm P_2$ and $\bm P_M$ are general compression matrices, which requires more careful and tailored derivations (e.g., invoking Lemma~\ref{lemma:joint} in Appendix~\ref{app:identifiability_non_blind}) to help establish recoverability. Our proof and theorems therefore cover more cases that could not be covered by the recoverability theorems in \cite{Zhang2020CartographyBTD}.
\end{remark}

\section{Algorithm Design: Exploiting Physical Interpretations}
Besides the recoverability guarantees, another important fact is that the {\sf LL1} tensor model is consistent with the LMM of spectral images. Therefore, the latent factors of the {\sf LL1} model admit physical interpretations. This connection allows us to incorporate prior information about the endmembers and the abundance maps of the materials contained in the HSI and MSI. Note that utilizing prior information is often important for parameter estimation, especially 
in the presence of heavy noise or modeling errors.

\subsection{Reformulation}
To proceed, let us first recast the criterion in \eqref{Non_blind_BTD_model} in a more implementation-friendly form:
\begin{align}\label{eq:our_constrainted_model}
\min_{\{\bm{S}_{r}\}_{r=1}^{R},\bm{C}}\quad
 &\frac{1}{2}\left\|\underline{\bm{Y}}_{H}-\sum_{r=1}^{R}\left(\bm{P}_{1}\bm{S}_{r}\bm{P}_{2}^{\top}\right)\circ \bm{C}(:,r)\right\|_{F}^{2} \nonumber \\
+&\frac{1}{2}\left\|\underline{\bm{Y}}_{M}-\sum_{r=1}^{R}\bm{S}_{r}\circ (\bm{P}_{M}\bm{C}(:,r))\right\|_{F}^{2}\\
+&\sum_{r=1}^{R}\theta_{r}\varphi(\bm{S}_{r})+\frac{\lambda}{2}\|\bm{C}\|_{F}^{2} \nonumber\\
\textrm{s.t.} \ \bm{S}_{r}&\geq 0,\textrm{rank}(\bm{S}_{r})= L_r, r=1,\ldots,R, \bm{C}\geq 0, \nonumber
\end{align}
where $\theta_r,\lambda \geq 0$ are regularization parameters, and $\varphi(\cdot)$ is a regularization function that promotes small 2D total variation.

To explain the above reformulation, first notice that we have lifted the equality constraints in \eqref{Non_blind_BTD_model} to the cost function using two fitting terms,
which is more robust to noise and easier to handle.
Second, we have replaced $\A_r\B_r^\T$ by $\S_r$ under the constraint ${\rm rank}(\S_r)= L_r$. The reason is that we hope to ``bring out'' $\S_r$ in our formulation, since $\S_r$ is the abundance map of endmember $r$, and in many cases one hopes to add structural constraints/regularization on $\S_r$. Third, we have imposed nonnegativity constraints on $\S_r$ and $\C$, per their physical interpretations.
In addition, we have added a spatial 2D TV regularizer $\varphi(\cdot)$ on $\S_r$ (which will be detailed soon). This is motivated by the fact that an endmember oftentimes exhibits very similar abundances in neighboring pixels \cite{Li2012CS,Iordache2012TVHU}. 
The regularization on $\C$ is to combat the scaling/counter-scaling effect in factorization models; see discussions in \cite{Fu2015Joint}.

The reformulation is still very hard to handle. 
One reason is that \eqref{eq:our_constrainted_model} has a nonconvex rank constraint on $\S_r$.
In addition, the commonly used 2D TV regularization for $\S_r$ is $\ell_1$-norm based---thereby being nonsmooth; see, e.g., \cite{Zymnis2007HU,Aggarwal2016HU,Iordache2012TVHU}. 
Hence, the problem is a nonconvex set-constrained nonsmooth optimization criterion, making designing effective and convergence-guaranteed algorithms challenging.
To circumvent these difficulties, we propose to approximate the formulation in \eqref{eq:our_constrainted_model} using the following optimization surrogate:
\begin{align}\label{eq:our_non_blind_model}
\min_{\{\bm{S}_{r}\}_{r=1}^{R},\bm{C}}\quad
 &\frac{1}{2}\left \|\underline{\bm{Y}}_{H}-\sum_{r=1}^{R}(\bm{P}_{1}\bm{S}_{r}\bm{P}_{2}^{\top})\circ \bm{C}(:,r) \right\|_{F}^{2} \nonumber \\
+&\frac{1}{2}\left \|\underline{\bm{Y}}_{M}-\sum_{r=1}^{R}\bm{S}_{r}\circ (\bm{P}_{M}\bm{C}(:,r)) \right\|_{F}^{2} \\
+&\sum_{r=1}^{R}\theta_{r} \varphi\big(\S_r\big)+\sum_{r=1}^{R}\eta_{r}\phi_{p,\tau}(\bm{S}_{r}) +\frac{\lambda}{2} \left\|\bm{C} \right\|_{F}^{2} \nonumber \\
\textrm{s.t.} \ \bm{S}_{r}&\geq {\bf 0}, r=1,\ldots,R,\quad \bm{C}\geq {\bf 0}.\nonumber
\end{align}
Here, $\varphi(\bm{S}_{r})$ is a smoothed 2D TV regularizer as before, and $\phi_{p,\tau}(\bm{S}_{r})$ is low-rank promoting regularizer---which is introduced to serve as a surrogate for the hard constraint ${\rm rank}(\S_r)= L_r$.
As we will see, if $\phi_{p,\tau}(\bm{S}_{r})$ is properly designed, an efficient optimization algorithm for handling \eqref{eq:our_non_blind_model} can be devised.
In the following, we provide design details of the reqularization terms $\varphi(\bm{S}_{r})$ and $\phi_{p,\tau}(\bm{S}_{r})$, respectively.
\subsubsection{$\ell_q$ Function-Based Total Variation Regularization}
To explain, we use the following TV surrogate:
\begin{equation}\label{eq:Lp_TV}
\varphi(\bm{S}_{r})=\varphi_{q,\varepsilon}(\bm{H}_{x}{\bm q}_r)+\varphi_{q,\varepsilon}(\bm{H}_{y}{\bm q}_r),
\end{equation}
where ${\bm q}_r=\textrm{vec}(\bm{S}_{r})$ and $\varphi_{q,\varepsilon}(\bm{x})=\sum(x_{i}^2+\varepsilon)^{\frac{q}{2}}$ with $0<q\leq 1$ and $\varepsilon>0$.
The matrices $\bm{H}_{x}$ and $\bm{H}_{y}$ are the two ``gradient matrices'', which are defined as
\[
\bm{H}_{x}=\bm{H}\otimes \bm{I},\bm{H}_{y}=\bm{I}\otimes \bm{H},
\]
where $\bm{I}\in \mathbb{R}^{J_{M} \times J_{M}}$ is an identity matrix and
\[
\bm{H} =
\left[
\begin{array}{cccccc}
1 & -1 & 0  & \cdots & 0 & 0\\
0 & 1  & -1 & \cdots & 0 & 0\\
\vdots & \vdots  &  \vdots   & \vdots & \vdots & \vdots\\
0 & 0 & \cdots & 0 & 1 & -1\\
-1 & 0 & \cdots & 0 & 0 & 1\\
\end{array}
\right]
\in \mathbb{R}^{I_{M} \times I_{M}}.
\]
Note that when $q<2$, the $\ell_q$ function is known to be effective in promoting sparsity \cite{Fu2016Unmixing}.
The parameter $\varepsilon>0$ is employed for making the function smooth when $0<q\leq 1$.
As $q \rightarrow 0$ and $\varepsilon \rightarrow 0$, $\varphi(\bm{S}_{r}) \rightarrow \|\bm{H}_{x}{\bm q}_r\|_{0}+\|\bm{H}_{y}{\bm q}_r\|_{0}$. 
Hence, $\varphi(\S_r)$ promotes small 2D TV of $\S_r$.

\subsubsection{Schatten-$p$ Function Based Low-rank Rank Regularization} The reguarlization $\phi_{p,\tau}(\bm{S}_{r})$ is introduced for promoting a low-rank $\S_r$---i.e., as a surrogate for the constraint ${\rm rank}(\S_r)=L_r$.
Here, we propose to employ the smoothed Schatten-$p$ function \cite{Bouldin1980Schatten,Mohan2012Iterative} for this purpose. Specifically, given a matrix $\bm{X}$, the smoothed Schatten-$p$ function is defined as follows:
\[
\begin{split}
\phi_{p,\tau}(\bm{X})
&=\sum_{i=1}^{M}(\sigma_{i}(\bm{X})^{2}+\tau)^{p/2}\\
&=\textrm{tr}((\bm{X}\bm{X}^{\top}+\tau\bm{I})^{p/2}),
\end{split}
\]
where $p>0$.
In the second line we have used the convention $\A^p = \bm U\bm \Lambda^p\bm U^\T$ for real symmetric matrix $\bm A$, in which $\A=\bm U\bm \Lambda\bm U^\T$ denotes the eigendecomposition of $\A$.
The Shattern-$p$ function can be understood as an $\ell_p$ function applied to the singular values of the matrix $\X$.
The parameter $\tau>0$ is again for smoothing the function when $0<p\leq 1$. Also, as pointed out in \cite{Wu2019Hyperspectral,Lai2013ImprovedIRLS}, $\phi_{p,\tau}(\bm{X})$ can be understood as a nonconvex approximation for the nuclear norm. 

The reformulation in \eqref{eq:our_non_blind_model} admits a continuously differentiable objective function.
The reformulation has also avoided directly handling the hard rank constraint via using the Schatten $p$-function. This allows us to design efficient first-order optimization algorithms for tackling the problem.
We should mention that a similar reformulation for \eqref{blind_BTD_model} can also be readily obtained as follows:
\begin{equation}\label{eq:our_blind_model}
\begin{split}
\min_{\{\bm{S}_{r},\widetilde{\bm{S}}_{r}\}_{r=1}^{R},\bm{C}}\quad
 &\frac{1}{2}\left\|\underline{\bm{Y}}_{H}-\sum_{r=1}^{R}\widetilde{\bm{S}}_{r}\circ \bm{C}(:,r) \right\|_{F}^{2}\\
+&\frac{1}{2}\left\|\underline{\bm{Y}}_{M}-\sum_{r=1}^{R}\bm{S}_{r}\circ (\bm{P}_{M}\bm{C}(:,r)) \right\|_{F}^{2}\\
+&\sum_{r=1}^{R}\eta_{r} \big(\phi_{p,\tau}\big(\S_r\big)+\phi_{p,\tau}\big(\widetilde{\bm{S}}_{r}\big)\big)\\
+&\sum_{r=1}^{R}\theta_{r}\varphi(\bm{S}_{r}) +\frac{\lambda}{2}\|\bm{C}\|_{F}^{2} \\
\textrm{s.t.} \ \bm{S}_{r}&\geq {\bf 0}, r=1,\ldots,R, \bm{C}\geq {\bf 0},
\end{split}
\end{equation}
in which we use a low-rank $\widetilde{\bm{S}}_{r}$ (where the low-rank property is promoted by $\phi_{p,\tau}\big(\widetilde{\bm{S}}_{r}\big)$) to replace $\widetilde{\bm{A}}_{r}\widetilde{\bm{B}}_{r}^{\top}\in \mathbb{R}^{I_{H}\times J_{H}}$ in the formulation in \eqref{blind_BTD_model}.

\subsection{Inexact Alternating Accelerated Projected Gradient}
Let us denote the optimization problem in \eqref{eq:our_non_blind_model} as
\begin{align}\label{eq:shrothand_J}
    \min_{ \S,\C }&~{{\cal J}_{1}}(\S, \C) \nonumber\\
    {\rm s.t.}&~\S\geq \bm 0,~\C\geq \bm 0.
\end{align}
Note that by our reformulation, the function ${{\cal J}_{1}}(\S, \C)$ is continuously differentiable.

\subsubsection{Basic Algorithmic Structure}
To tackle the optimization problem, we propose to employ the alternating projected gradient (APG) strategy.
To be specific, the updates can be summarized as follows
\begin{align}\label{eq:alternating_non_blind}
\C^{(t+1)}&\leftarrow \max\left\{\C^{(t)} - \alpha^{(t)} \nabla_{\C} {\cal J}_{1}(\S^{(t)},\C^{(t)}),\bm 0\right\} \\
\S^{(t+1)} &\leftarrow \max\left\{\S^{(t)} - \beta^{(t)} \nabla_{\S} {\cal J}_{1}(\S^{(t)},\C^{(t+1)}),\bm 0\right\}, \nonumber 
\end{align}
where the thresholding operators (i.e., $\max\{\cdot,\bm 0\}$) are orthogonal projectors onto the nonnegativity orthant.
Note that both partial gradients exist, under our design of the regularization terms. The expressions of $\nabla_{\C} {\cal J}_{1}(\S^{(t)},\C^{(t)})$ and $\nabla_{\S} {\cal J}_{1}(\S^{(t)},\C^{(t+1)})$ can be found in Appendix \ref{app:gradient_non_blind}. 

The convergence properties of APG algorithm can be shown under the framework in \cite{Xu2013BCD}. To be specific, under proper choices of $\alpha^{(t)}$ and $\beta^{(t)}$, one can show that the gradient projection step decreases ${\cal J}_{1}$ by a sufficiently large quantity---which is the key for establishing convergence of the solution sequence. To be specific, we have the following:
\begin{proposition}\label{pro:non_blind_convergence}
Assume that $\alpha^{(t)}\leq 1/L_{\bm C}^{(t)}$ and $\beta^{(t)}\leq 1/L_{\bm S}^{(t)}$ in all iterations, where
\begin{align}\label{eq:L_non_blind}
L_{\bm C}^{(t)}&=\sigma_{\textrm{max}}\big((\bm{S}^{(t)})^{\top}(\bm{P}_{2}\otimes \bm{P}_{1})^{\top}(\bm{P}_{2}\otimes \bm{P}_{1})\bm{S}^{(t)}\big)\\
&+\sigma_{\textrm{max}}(\bm{P}_{M}^{\top}\bm{P}_{M})\sigma_{\textrm{max}}\big((\bm{S}^{(t)})^{\top}\bm{S}^{(t)}\big)+\lambda, \nonumber \\
L_{\bm S}^{(t)}&=\sigma_{\textrm{max}}\big((\bm{C}^{(t+1)})^{\top}\bm{C}^{(t+1)}\big)\sigma_{\textrm{max}}\big((\bm{P}_{2}\otimes \bm{P}_{1})^{\top}\bm{P}_{2}\otimes \bm{P}_{1} \big) \nonumber \\
&+\sigma_{\textrm{max}}((\bm{C}^{(t+1)})^{\top}\bm{P}_{M}^{\top}\bm{P}_{M}\bm{C}^{(t+1)})+p\max_{r}\eta_{r}\sigma_{\textrm{max}}(\bm{W}^{(t)}_{r})  \nonumber\\
&+q\max_{r}\theta_{r}\big(\sigma_{\textrm{max}}(\bm{H}_{x}^{\top}\bm{U}_{r}^{(t)}\bm{H}_{x})+\sigma_{\textrm{max}}(\bm{H}_{y}^{\top}\bm{V}_{r}^{(t)}\bm{H}_{y})\big), \nonumber 
\end{align}
where $\bm{S}$ is defined as in \eqref{eq:S}, $\bm{W}^{(t)}_{r}=(\bm{S}^{(t)}_{r}(\bm{S}^{(t)}_{r})^{\top}+\tau\bm{I})^{\frac{p-2}{2}}$, $\bm{U}_{r}^{(t)}$ and $\bm{V}_{r}^{(t)}$ are diagonal matrices with 
$[\bm{U}_{r}^{(t)}]_{i,i}=([\bm{H}_{x}{\bm q}_r^{(t)}]_{i}^{2}+\varepsilon)^{\frac{q-2}{2}}$, and $[\bm{V}_{r}^{(t)}]_{i,i}=([\bm{H}_{y}{\bm q}_r^{(t)}]_{i}^{2}+\varepsilon)^{\frac{q-2}{2}}$, $r=1,\ldots, R$.
In addition, assume that $L_{\bm C}^{(t)}<\infty$ and $L_{\bm S}^{(t)}<\infty$ for all $t$.
Then, every limit point of the solution sequence produced by the algorithm in \eqref{eq:alternating_non_blind} is a stationary point of Problem~\eqref{eq:our_non_blind_model}.
\end{proposition}

The proof of Proposition \ref{pro:non_blind_convergence} is shown in Appendix \ref{proof:non_blind_convergence}.

\subsubsection{Per-iteration Complexity}
To implement the algorithm, one needs to compute the gradients w.r.t. $\bm C$ and $\bm S$, respectively.
The major computation burden lies in constructing $\bm W_r^{(t)}$ since it involves full SVD of $\S_r^{(t)}$.
Consequently, the per-iteration complexity is dominated by this step, which costs ${\cal O}(RI_M^2\textrm{max}(I_M,J_M,K_H))$ flops (which is approximately ${\cal O}(I_M^3)$ if $I_M\approx J_M\geq K_H$); see the detailed complexity analysis in Appendix~\ref{app:gradient_non_blind}.

Another part that may incur many computational flops is to compute $\alpha^{(t)}$ and $\beta^{(t)}$ in each iteration.
To reduce the computational burden in each iteration, we first pre-compute a number of terms, i.e., $\sigma_{\textrm{max}}(\bm{P}_{M}^{\top}\bm{P}_{M})$ and $\sigma_{\textrm{max}}\big((\bm{P}_{2}\otimes \bm{P}_{1})^{\top}\bm{P}_{2}\otimes \bm{P}_{1} \big)$, since they do not change over the iterations.
Second, instead of directly computing $L_{\bm C}^{(t)}$ and $L_{\bm S}^{(t)}$, which need the exact values of the terms $\sigma_{\textrm{max}}\big((\bm{S}^{(t)})^{\top}(\bm{P}_{2}\otimes \bm{P}_{1})^{\top}(\bm{P}_{2}\otimes \bm{P}_{1})\bm{S}^{(t)}\big)$, $\sigma_{\textrm{max}}(\bm{H}_{x}^{\top}\bm{U}_{r}^{(t)}\bm{H}_{x})$ and $\sigma_{\textrm{max}}(\bm{H}_{y}^{\top}\bm{V}_{r}^{(t)}\bm{H}_{y})$, we compute their upper bounds---since we only need the inequalities $\alpha^{(t)}\leq 1/L_{\bm C}^{(t)}$ and $\beta^{(t)}\leq 1/L_{\bm S}^{(t)}$ to hold. Take the first term as an example. Its upper bound can be obtained via the following:
\begin{align*}
    &\sigma_{\textrm{max}}\big((\bm{S}^{(t)})^{\top}(\bm{P}_{2}\otimes \bm{P}_{1})^{\top}(\bm{P}_{2}\otimes \bm{P}_{1})\bm{S}^{(t)}\big)\\
    &\quad\quad\quad\leq \sigma_{\max}^2(\S^{(t)})\sigma_{\textrm{max}}\big((\bm{P}_{2}\otimes \bm{P}_{1})^{\top}\bm{P}_{2}\otimes \bm{P}_{1} \big).  
\end{align*} 
Note that the right hand side only costs ${\cal O}(N R^2 )$ flops.
Similar, we compute the upper bound 
$\sigma_{\textrm{max}}(\bm{H}_{y}^{\top}\bm{V}_{r}^{(t)}\bm{H}_{y})\leq \sigma_{\textrm{max}}(\bm{H}_{y}^{\top})\sigma_{\max}(\bm{V}_{r}^{(t)})\sigma_{\rm max}(\bm{H}_{y}),$
where $\sigma_{\max}(\bm{V}_{r}^{(t)})$ is the largest diagonal entry of $\bm{V}_{r}^{(t)}$ and the other two terms are pre-computed.
Using the above approach, the step size computation's complexity is almost negligible relative to that of the gradient computation.

In summary, the per-iteration complexity is in the order of ${\cal O}(I_M^3)$ (assuming $I_M\approx J_M\geq K_H$). For HSR tasks, $I_M$ often lies in the range of several hundreds, which is affordable in most cases.
One way to reduce complexity is to consider patch-by-patch execution of the coupled decomposition task.

\subsubsection{Extrapolation} 
The APG algorithm is conceptually simple and easy to implement. However, as a first-order optimization algorithm, the iteration complexity is normally not low---i.e., empirically, it often takes many iterations for the algorithm to converge to a reasonably ``good'' solution.
One way to improve this situation without increasing per-iteration complexity is to employ the so-called extrapolation strategy. This strategy was first proposed by Nesterov \cite{Nesterov1983Extrapolation} for convex optimization, and then was extended to handle multiblock nonconvex optimization in \cite{Xu2013BCD}. To be specific, instead of computing the partial gradients w.r.t. $\C$ and $\S$ at the current iterates, one can compute them at some extrapolated points $\check{\C}^{(t+1)}$ and $\check{\S}^{(t+1)}$. 
In a nutshell, the extrapolated point, e.g., $\check{\C}^{(t+1)}$, is a linear combination of the current $\C^{(t+1)}$ and the previous $\C^{(t)}$ using specially designed combination coefficients---and we use the combination coefficients designed by Nesterov \cite{Nesterov1983Extrapolation} (cf. line 9 in Algorithm \ref{our_non_blind_algorithm}). 
The variables are updated via gradient projection at those extrapolated points (cf. Algorithm~\ref{our_non_blind_algorithm}). 

\begin{figure}[!ht]
\scriptsize\setlength{\tabcolsep}{0.1pt}
\begin{center}
\begin{tabular}{ccc} 
\includegraphics[width=0.7\linewidth]{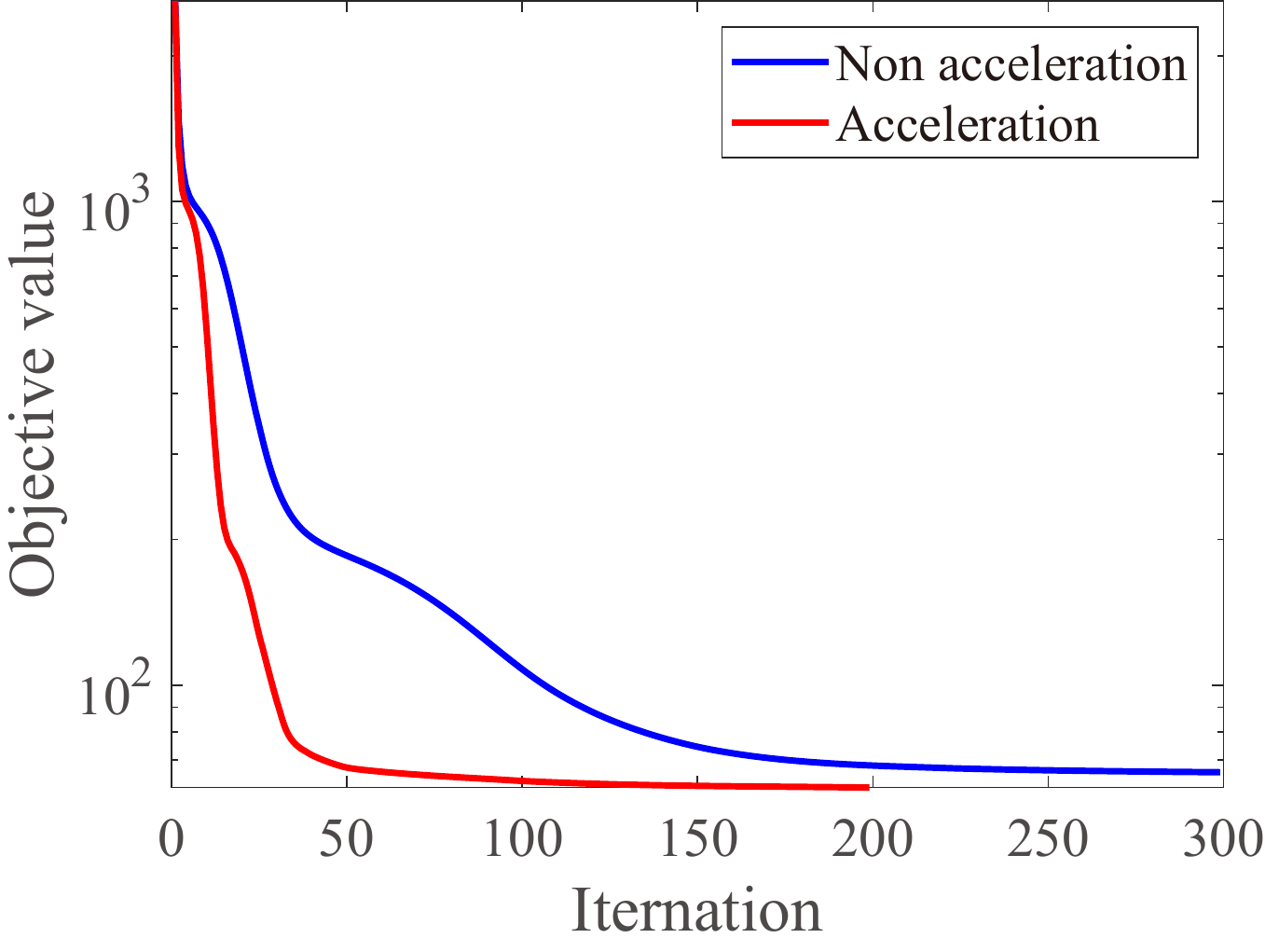}
\end{tabular}
\caption{Comparison of the objective values under SC-\textsf{LL1} with acceleration or not. (The setting of this figure is same as the experiment shown in Table \ref{table:Pavia} in Section \ref{sec:semi_real_non_blind_Pavia}.)}
  \label{fig:acceleration}
  \end{center}\vspace{-0.3cm}
\end{figure}

\begin{algorithm}
\caption{\textsf{SC-LL1} for solving \eqref{eq:our_non_blind_model}.}\label{our_non_blind_algorithm}
\footnotesize
      \textbf{Input:} HSI $\underline{\bm{Y}}_{H}$, MSI $\underline{\bm{Y}}_{M}$, starting points $\bm{C}^{(0)}$, $\bm{S}^{(0)}$, degraded matrices $\bm{P}_{1}$, $\bm{P}_{2}$, $\bm{P}_{M}$.\\ \vspace{0.05cm}
      \quad \textbf{Parameters:} $\lambda$, $\{\theta_{r}\}_{r=1}^{R}$, $\{\eta_{r}\}_{r=1}^{R}$, $\gamma_{1}^{(0)}=\gamma_{2}^{(0)}=1$, $p$, $q$, $\varepsilon$, $\tau$.\\ \vspace{0.1cm}
      \quad $\check{\bm{C}}^{(0)}=\bm{C}^{(0)}$, $\check{\bm{S}}^{(0)}=\bm{S}^{(0)}$.\\ \vspace{0.05cm}
      \quad $t=0$.\\ \vspace{0.05cm}
      \quad \textbf{repeat}\\ \vspace{0.05cm}
      \quad $\%\%$ \textbf{update} $\bm{C}$ $\%\%$ \\ \vspace{0.05cm}
      \quad $\bm{C}^{(t+1)} \leftarrow \max\left\{\check{\bm{C}}^{(t)} - \alpha^{(t)} \nabla_{\check{\bm{C}}} {\cal J}_{1}\left(\bm{S}^{(t)},\check{\bm{C}}^{(t)}\right),\bm 0\right\}$;\\  \vspace{0.1cm}
      \quad $\gamma_{1}^{(t+1)}=\frac{1+\sqrt{1+4\left(\gamma_{1}^{(t)}\right)^{2}}}{2}$;\\ \vspace{0.05cm}
      \quad $\check{\bm{C}}^{(t+1)}=\bm{C}^{(t+1)}+\left(\frac{\gamma_{1}^{(t)}-1}{\gamma_{1}^{(t+1)}}\right)(\bm{C}^{(t+1)}-\bm{C}^{(t)})$;\\ \vspace{0.05cm}
      \quad $\%\%$ \textbf{update} $\bm{S}$ $\%\%$ \\ \vspace{0.05cm}
      \quad $\bm{S}^{(t+1)} \leftarrow \max\left\{\check{\bm{S}}^{(t)} -  \beta^{(t)} \nabla_{\check{\bm{S}}} {\cal J}_{1}\left(\check{\bm{S}}^{(t)},\bm{C}^{(t+1)}\right),\bm 0 \right\}$;\\ \vspace{0.1cm}
      \quad $\gamma_{2}^{(t+1)}=\frac{1+\sqrt{1+4\left(\gamma_{2}^{(t)}\right)^{2}}}{2}$;\\ \vspace{0.05cm}
      \quad $\check{\bm{S}}^{(t+1)}=\bm{S}^{(t+1)}+\left(\frac{\gamma_{2}^{(t)}-1}{\gamma_{2}^{(t+1)}}\right)(\bm{S}^{(t+1)}
      -\bm{S}^{(t)})$;\\ \vspace{0.05cm}
      \quad $t=t+1$;\\ \vspace{0.05cm}
	  \quad \textbf{until} some stopping criterion are satisfied.\\ \vspace{0.05cm}
	  \textbf{Output:} $\widehat{\bm{C}}=\bm{C}^{(t)}$ and $\widehat{\bm{S}}=\bm{S}^{(t)}$. Reconstruct $\underline{\bm{Y}}_{S}$ using $\widehat{\underline{\bm{Y}}}_{S}(i,j,k)=\sum_{i=1}^{R}\widehat{\bm{S}}_{r}(i,j) \widehat{\bm{C}}(k,r)$.
\end{algorithm}

Fig. \ref{fig:acceleration} shows the convergence curves of the original algorithm in \eqref{eq:alternating_non_blind} and the accelerated version. The curves are averaged from 20 trials with Gaussian noise (the signal-to-noise ratio is 30dB). The initialization and noise at each trial are generated randomly. Here, the goal is to fuse an HSI and an MSI whose sizes are $64\times 64\times 103$ and $256\times 256\times 4$, respectively; more details can be found in Section \ref{experiments}. One can see that, under this simulation setting, the accelerated version uses about 50 iterations to reach a fairly low objective value level, while the original algorithm takes more than 200 iterations to reach the same level. Because of such effectiveness, in the experiments, we will always use the accelerated version.

We summarize the proposed accelerated algorithm for solving \eqref{eq:our_non_blind_model} in Algorithm \ref{our_non_blind_algorithm}, which will be referred to as the {\it structured coupled \textsf{LL1} decomposition} (\textsf{SC-LL1}) algorithm in the sequel.

\begin{remark}
In terms of convergence, the {\sf SC-LL1} algorithm has similar properties as those of the original version in \eqref{eq:alternating_non_blind} \cite{Xu2013BCD}. The subtlety is that the accelerated version does not always produce a nonincreasing sequence of the cost function value, which may pose difficulties in convergence analysis. The work in \cite{Xu2013BCD} offered a simple fix via checking the cost value in each iteration. Nevertheless, we observe that such a fix is mostly for the theoretical proof purpose, while does not make practical differences. Checking the objective value in every iteration may increase the computational complexity. Hence, we do not include it in our algorithm.
\end{remark}

\begin{algorithm}
\caption{BSC-\textsf{LL1} for solving \eqref{eq:our_blind_model}.}
\label{our_blind_algorithm}
\footnotesize
      \textbf{Input:} HSI $\underline{\bm{Y}}_{H}$, MSI $\underline{\bm{Y}}_{M}$, starting points $\bm{C}^{(0)}$, $\bm{S}^{(0)}$, $\widetilde{\bm{S}}^{(0)}$, degraded matrix $\bm{P}_{M}$.\\
      \quad \textbf{Parameters:} $\lambda$, $\{\theta_{r}\}_{r=1}^{R}$, $\{\eta_{r}\}_{r=1}^{R}$, $\gamma_{1}^{(0)}=\gamma_{2}^{(0)}=\gamma_{3}^{(0)}=1$, $p$, $q$, $\varepsilon$, $\tau$.\\ \vspace{0.05cm}
      \quad $\check{\bm{C}}^{(0)}=\bm{C}^{(0)}$, $\bar{\bm{S}}^{(0)}=\widetilde{\bm{S}}^{(0)}$, $\check{\bm{S}}^{(0)}=\bm{S}^{(0)}$.\\
      \quad $t=0$.\\ \vspace{0.05cm}
	  \quad \textbf{repeat}\\ \vspace{0.05cm}
	  \quad $\%\%$ \textbf{update} $\bm{C}$ $\%\%$\\ \vspace{0.05cm}
	  \quad $\bm{C}^{(t+1)}\leftarrow \max\left\{\check{\bm{C}}^{(t)} - \alpha^{(t)} \nabla_{\check{\bm{C}}} {\cal J}_{2}\left(\bm{S}^{(t)},\widetilde{\bm{S}}^{(t)},\check{\bm{C}}^{(t)}\right),\bm 0 \right\}$;\\ \vspace{0.05cm}
      \quad $\gamma_{1}^{(t+1)}=\frac{1+\sqrt{1+4\left(\gamma_{1}^{(t)}\right)^{2}}}{2}$;\\ \vspace{0.05cm}
      \quad $\check{\bm{C}}^{(t+1)}=\bm{C}^{(t+1)}+\left(\frac{\gamma_{1}^{(t)}-1}{\gamma_{1}^{(t+1)}}\right)\left(\bm{C}^{(t+1)}-\bm{C}^{(t)}\right)$;\\ \vspace{0.05cm}
      \quad $\%\%$ \textbf{update} $\bm{S}$ $\%\%$\\ \vspace{0.05cm}
      \quad $\bm{S}^{(t+1)}\leftarrow \max\bigg\{\check{\bm{S}}^{(t)} -\beta^{(t)}  \nabla_{\check{\bm{S}}} {\cal J}_{2}\left(\check{\bm{S}}^{(t)},\widetilde{\bm{S}}^{(t)},\bm{C}^{(t+1)}\right),\bm 0 \bigg\}$;\\ \vspace{0.05cm}
      \quad $\gamma_{2}^{(t+1)}=\frac{1+\sqrt{1+4\left(\gamma_{2}^{(t)}\right)^{2}}}{2}$;\\ \vspace{0.05cm}
      \quad $\check{\bm{S}}^{(t+1)}=\bm{S}^{(t+1)}+\left(\frac{\gamma_{2}^{(t)}-1}{\gamma_{2}^{(t+1)}}\right) \left(\bm{S}^{(t+1)}
      -\bm{S}^{(t)} \right)$;\\ \vspace{0.05cm}
      \quad $\%\%$ \textbf{update} $\widetilde{\bm{S}}$ $\%\%$\\  \vspace{0.1cm}
      \quad $\widetilde{\bm{S}}^{(t+1)}\leftarrow \bar{\bm{S}}^{(t)} -\zeta^{(t)} \nabla_{\bar{\bm{S}}} {\cal J}_{2}\left(\bm{S}^{(t+1)},\bar{\bm{S}}^{(t)},\bm{C}^{(t+1)}\right)$;\\ \vspace{0.1cm}
      \quad $\gamma_{3}^{(t+1)}=\frac{1+\sqrt{1+4\left(\gamma_{3}^{(t)}\right)^{2}}}{2}$;\\ \vspace{0.05cm}
      \quad $\bar{\bm{S}}^{(t+1)}=\widetilde{\bm{S}}^{(t+1)}+\left(\frac{\gamma_{3}^{(t)}-1}{\gamma_{3}^{(t+1)}}\right)\left(\widetilde{\bm{S}}^{(t+1)}-\widetilde{\bm{S}}^{(t)}\right)$;\\ \vspace{0.05cm}
      \quad $t=t+1$;\\
	  \quad \textbf{until} some stopping criterion are satisfied.\\ \vspace{0.05cm}
	  \textbf{Output:} $\widehat{\bm{C}}=\bm{C}^{(t)}$ and $\widehat{\bm{S}}=\bm{S}^{(t)}$. Reconstruct $\underline{\bm{Y}}_{S}$ using $\widehat{\underline{\bm{Y}}}_{S}(i,j,k)=\sum_{i=1}^{R}\widehat{\bm{S}}_{r}(i,j) \widehat{\bm{C}}(k,r)$.
\end{algorithm}

\subsection{Algorithm for Semi-blind Cases}
Consider the optimization problem \eqref{eq:our_blind_model} for handling HSR when the spatial degradation operator is unknown. Similarly, we express the problem in
\eqref{eq:our_blind_model} as follows:
\begin{align}\label{eq:shrothand_J_blind}
    \min_{\S,\widetilde{\S},\C}&~{\cal J}_{2}(\S, \widetilde{\S}, \C) \nonumber\\
    {\rm s.t.}&~\S\geq \bm 0, ~\C\geq \bm 0.
\end{align}
We again tackle the problem using APG with the acceleration strategy, and refer the corresponding algorithm as {\it blind structured coupled \textsf{LL1} decomposition} (\textsf{BSC-LL1}) algorithm. The detailed steps are summarized in Algorithm \ref{our_blind_algorithm}. The expressions of 
$\nabla_{\C} {\cal J}_{2}(\S^{(t)},\widetilde{\S}^{(t)},\C^{(t)})$,
$\nabla_{\S} {\cal J}_{2}(\S^{(t)},\widetilde{\S}^{(t)},\C^{(t+1)})$, and $\nabla_{\widetilde{\S}} {\cal J}_{2}(\S^{(t+1)},\widetilde{\S}^{(t)},\C^{(t+1)})$, and the step sizes $\alpha^{(t)}$, $\beta^{(t)}$ and $\zeta^{(t)}$ (and the corresponding Lipchitz constants $L_{\bm C}^{(t)}$, $L_{\bm S}^{(t)}$, and $L_{\widetilde{\S}}^{(t)}$) can be found in Appendix \ref{pro:our_blind_algorithm}.

\section{Experiments}
\label{experiments}
In this section, we present various experiments on semi-real data and real data to demonstrate the effectiveness of the proposed HSR framework. 

\subsection{Experiment Setup}
We benchmark our algorithm using CNMF \cite{Yokoya2012HSR}, HySure \cite{Simoes2015HSR}, FUSE \cite{Wei2015HSRSylvester}, SCOTT \cite{Prevost2020HSR}, STEREO \cite{Kanatsoulis2018HSR}.  
In particular, SCOTT and STEREO are the Tucker and CPD model based HSR approaches, respectively.
All simulations are coded using MATLAB 2019b and the experiments are run on a desktop with 3.4 GHz i7 CPU and 16 GB RAM.

\begin{figure*}[!ht]
\scriptsize\setlength{\tabcolsep}{0.3pt}
\begin{center}
\begin{tabular}{cccccccc}
\includegraphics[width=0.135\textwidth]{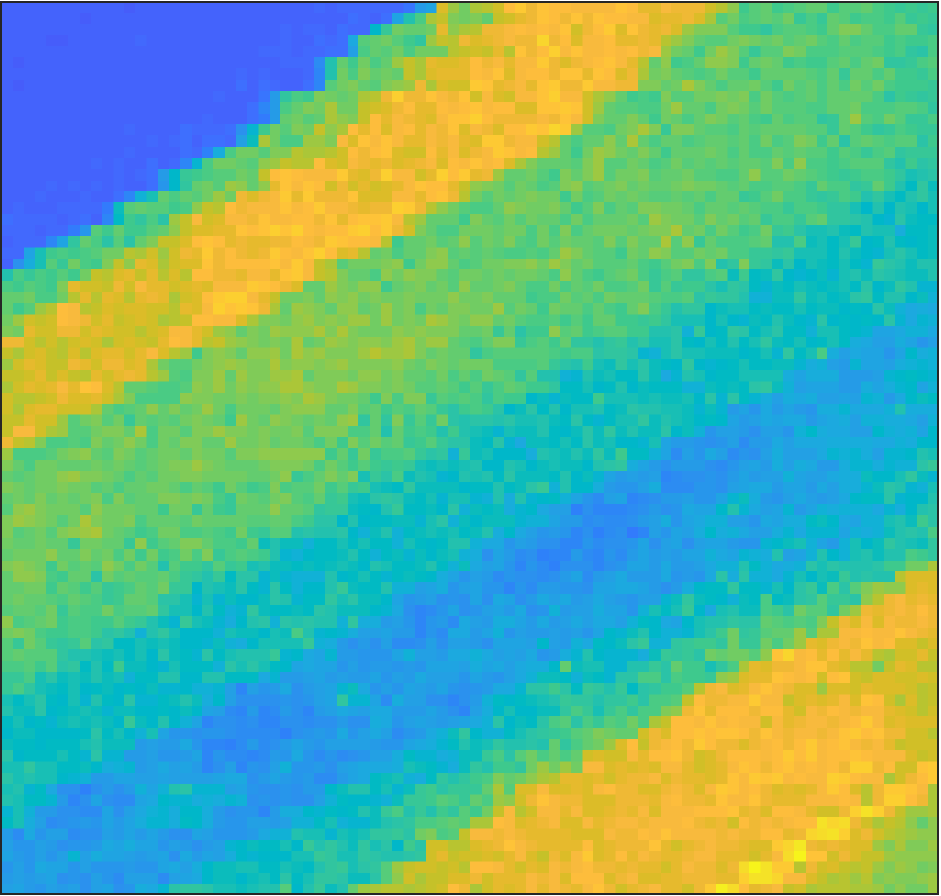}&
\includegraphics[width=0.135\textwidth]{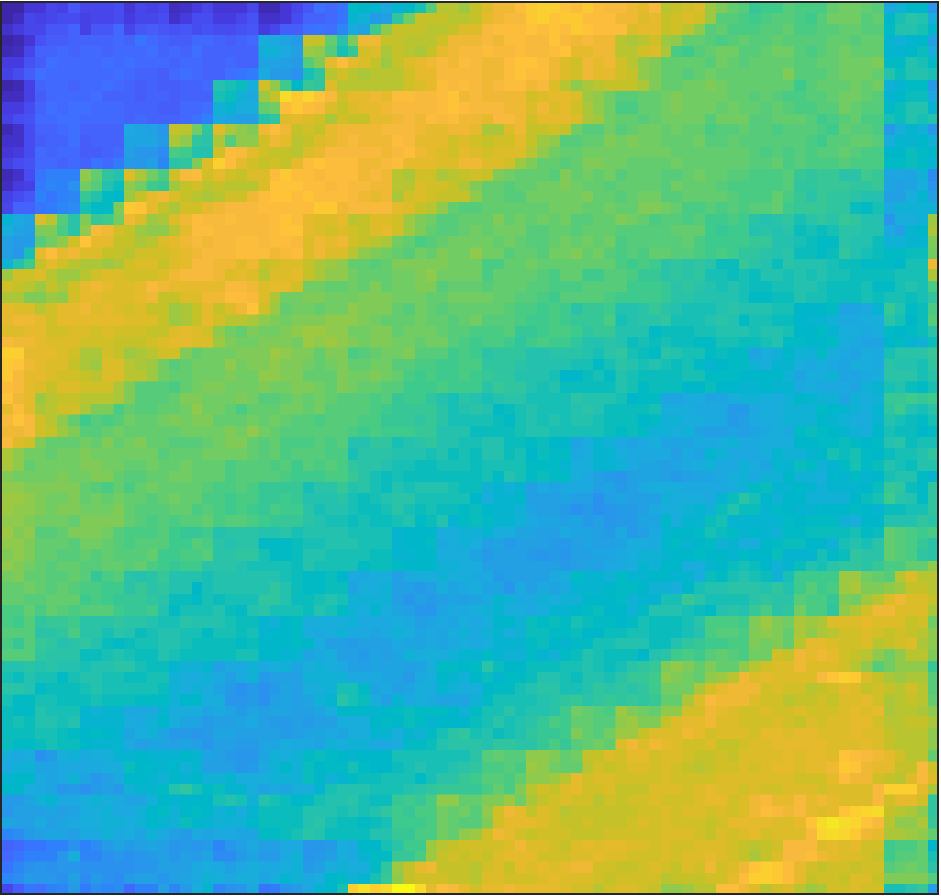}&
\includegraphics[width=0.135\textwidth]{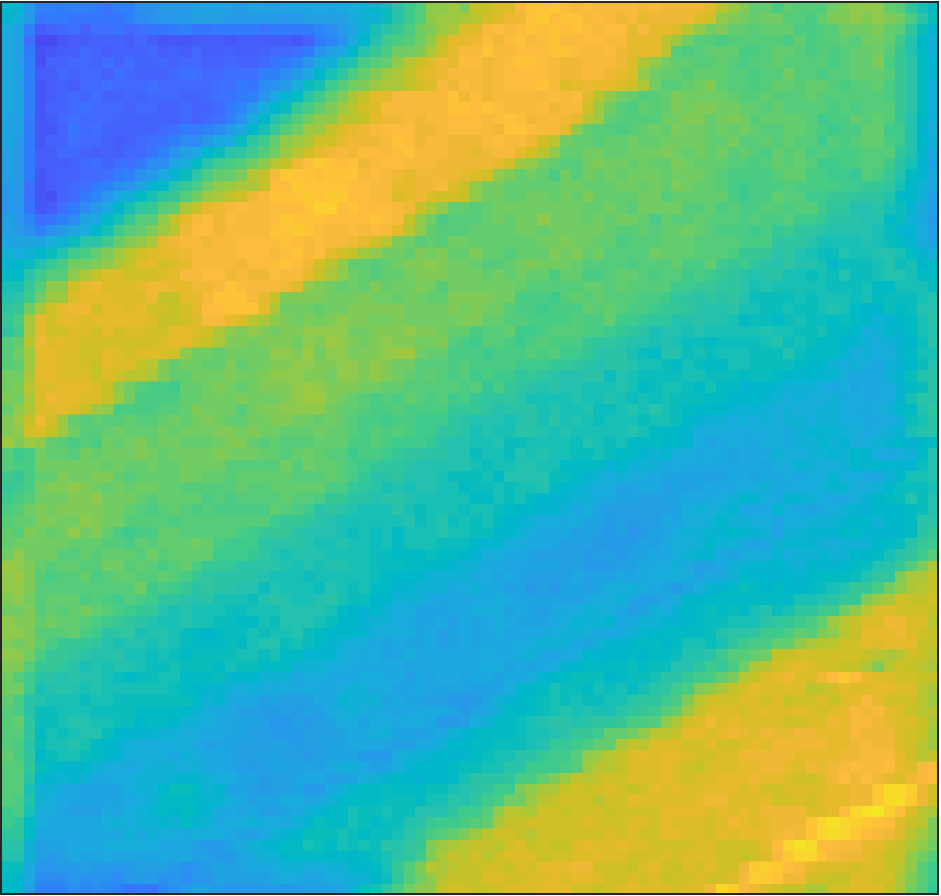}&
\includegraphics[width=0.135\textwidth]{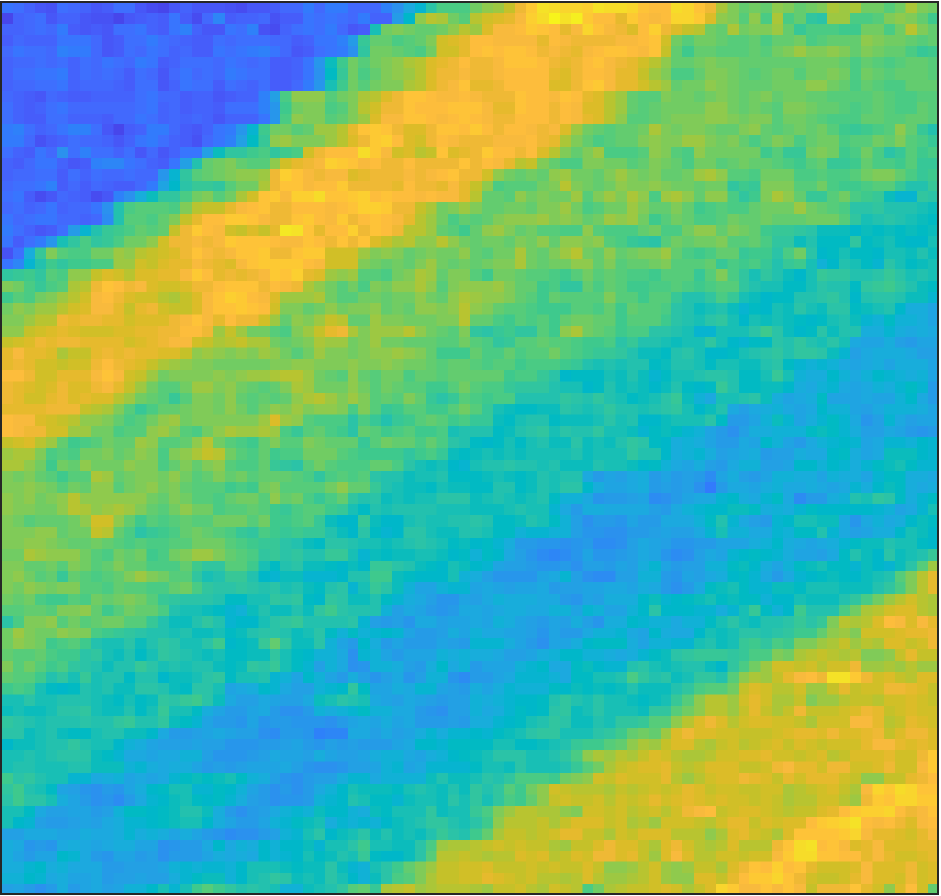}&
\includegraphics[width=0.135\textwidth]{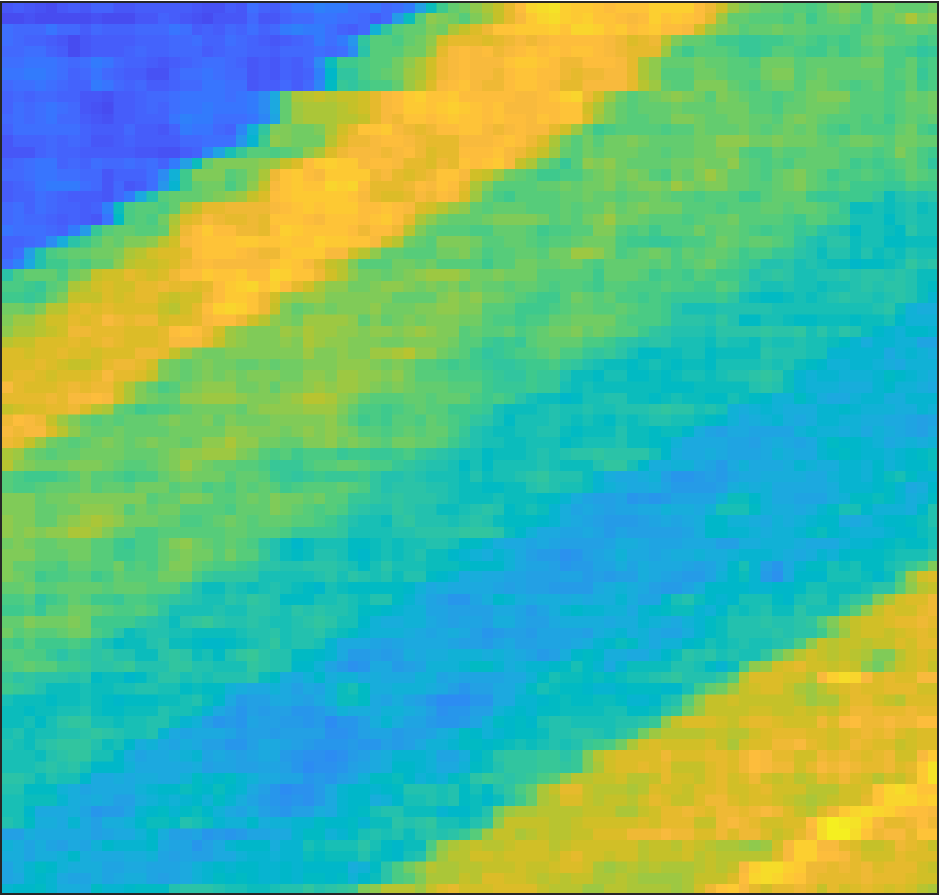}&
\includegraphics[width=0.135\textwidth]{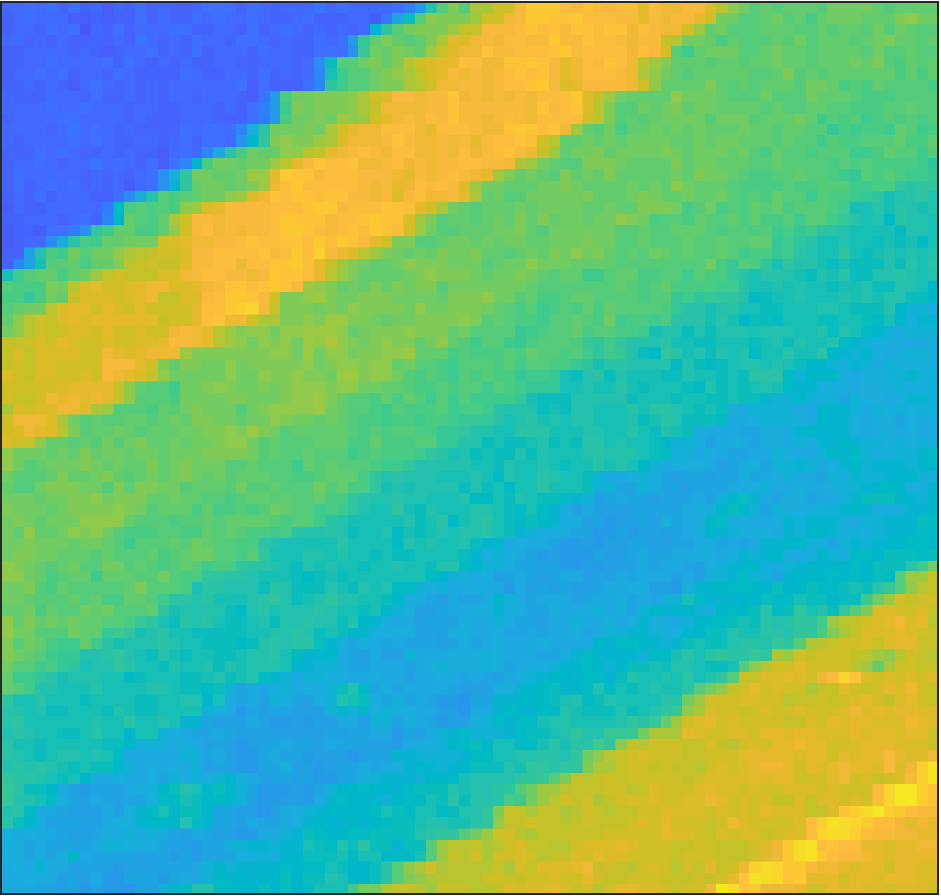}&
\includegraphics[width=0.135\textwidth]{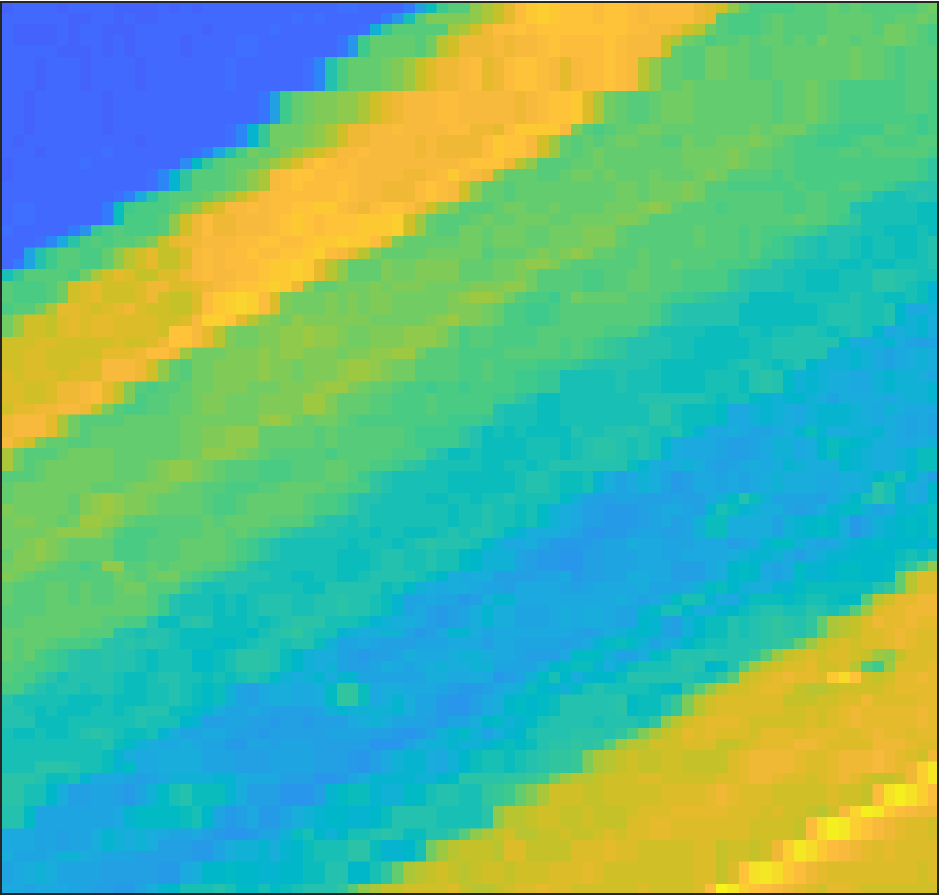}&
\hspace{0.01cm}
\includegraphics[width=0.018\textwidth]{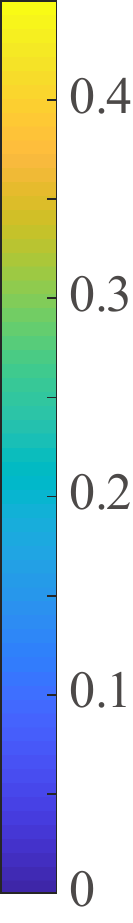}\\
\includegraphics[width=0.135\textwidth]{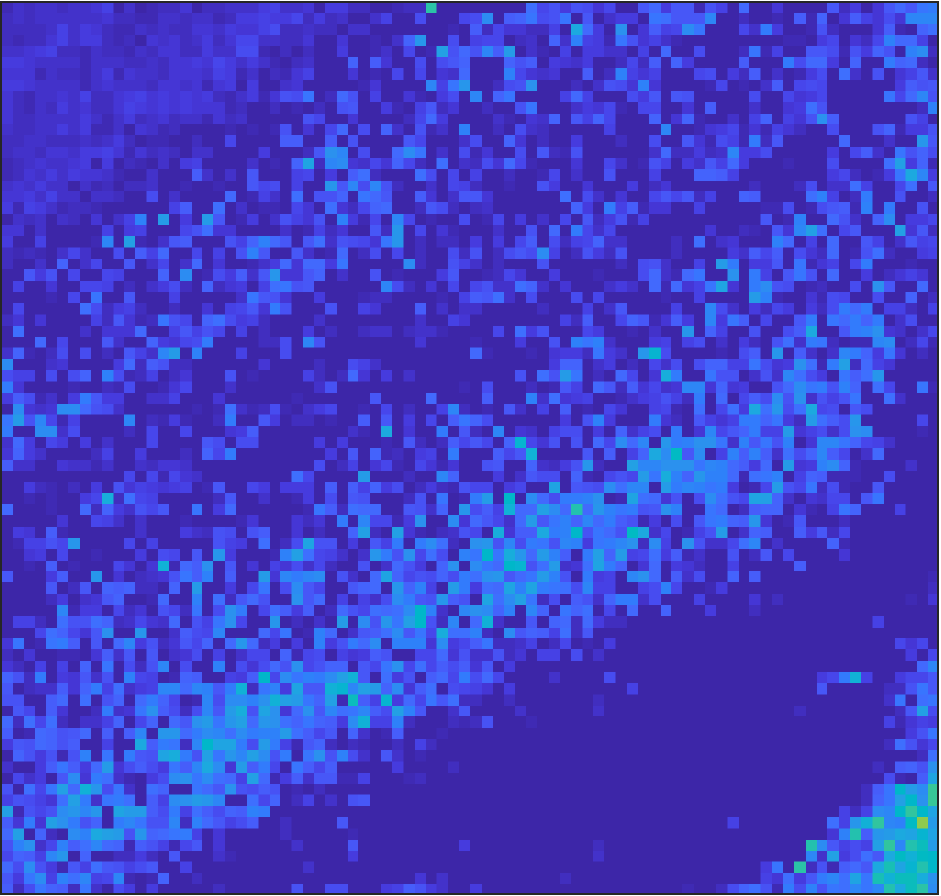}&
\includegraphics[width=0.135\textwidth]{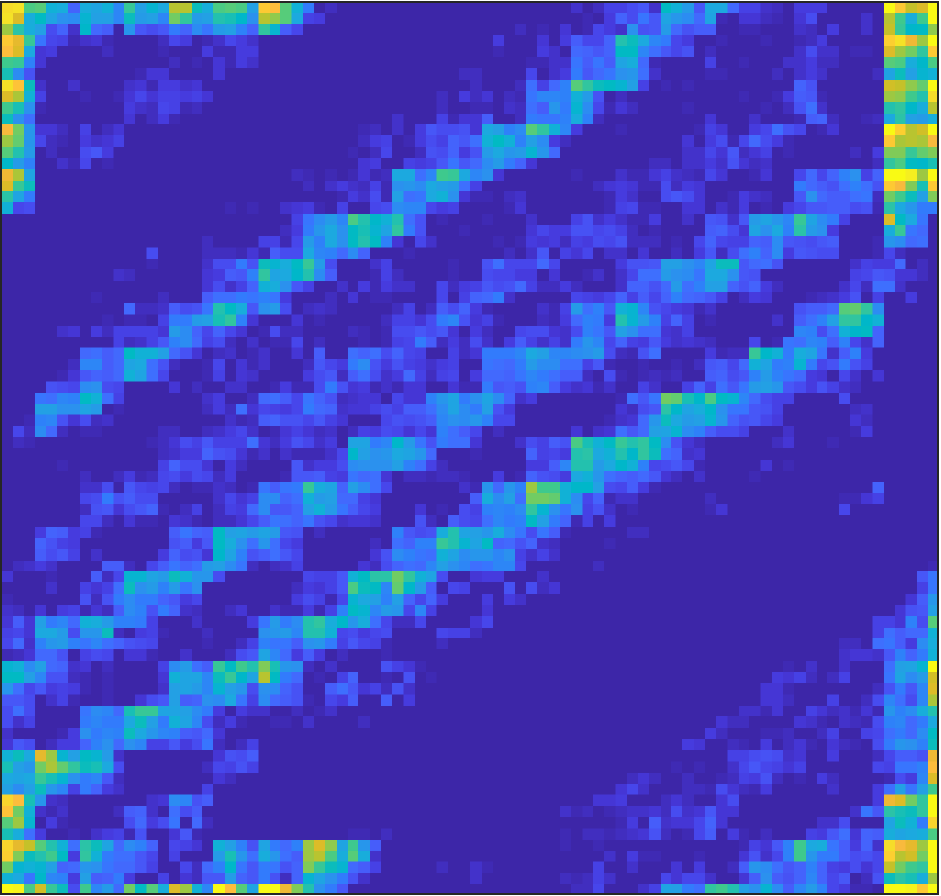}&
\includegraphics[width=0.135\textwidth]{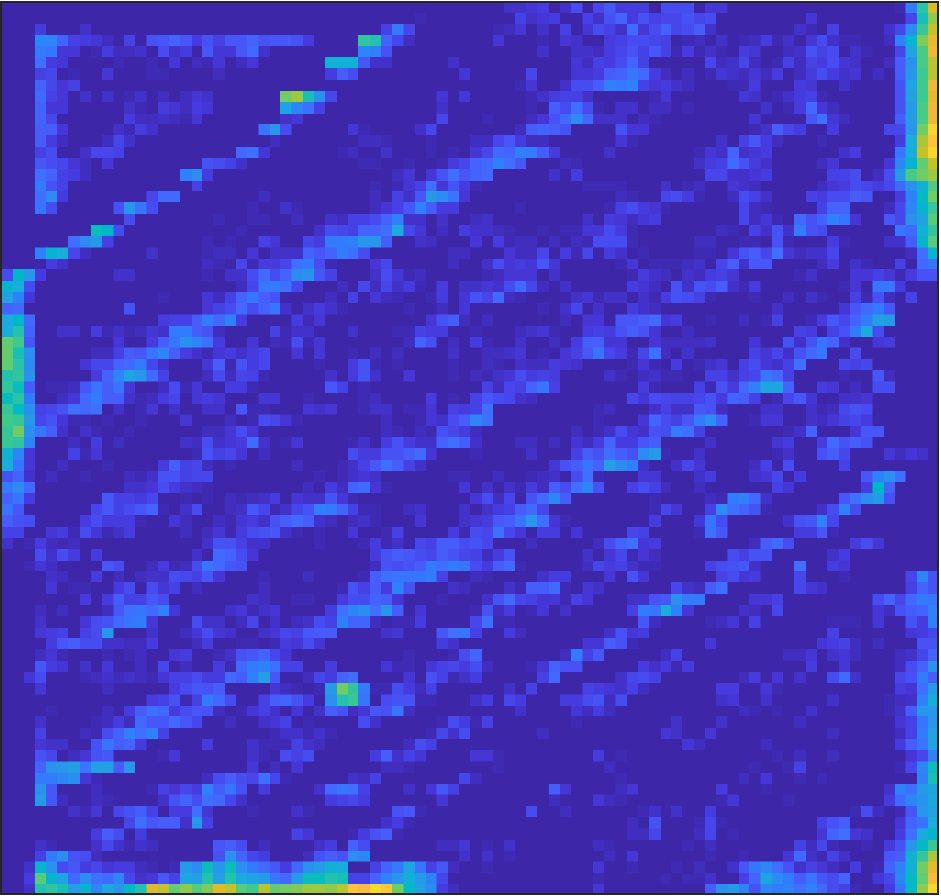}&
\includegraphics[width=0.135\textwidth]{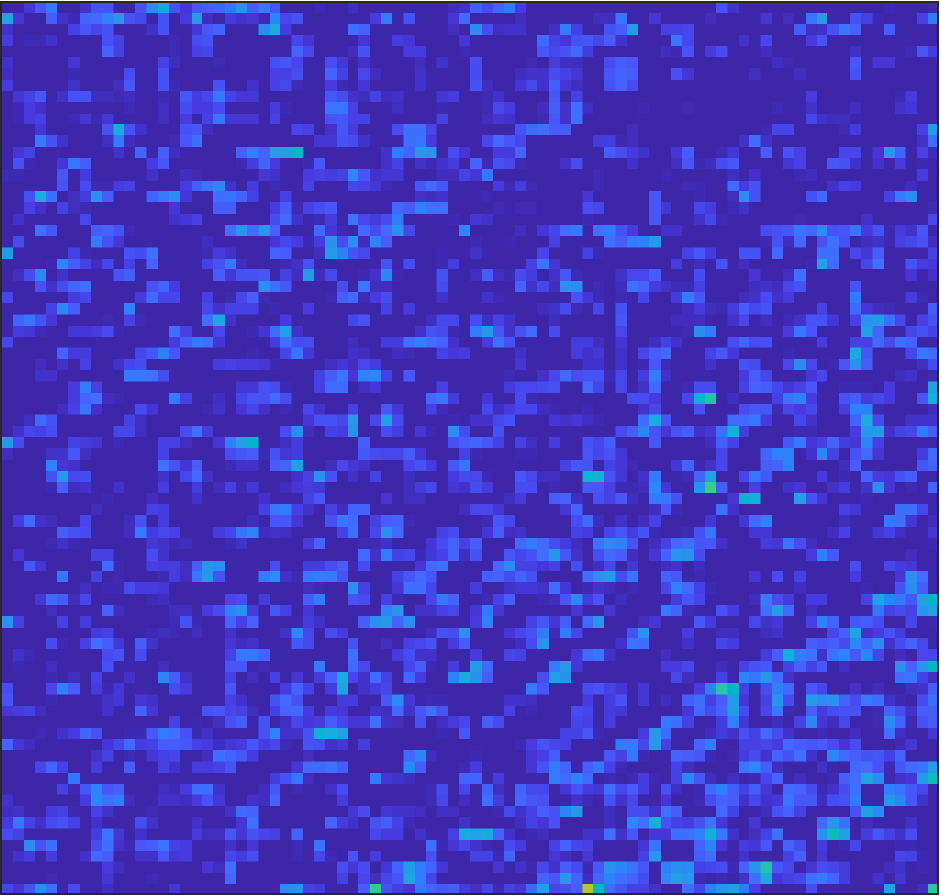}&
\includegraphics[width=0.135\textwidth]{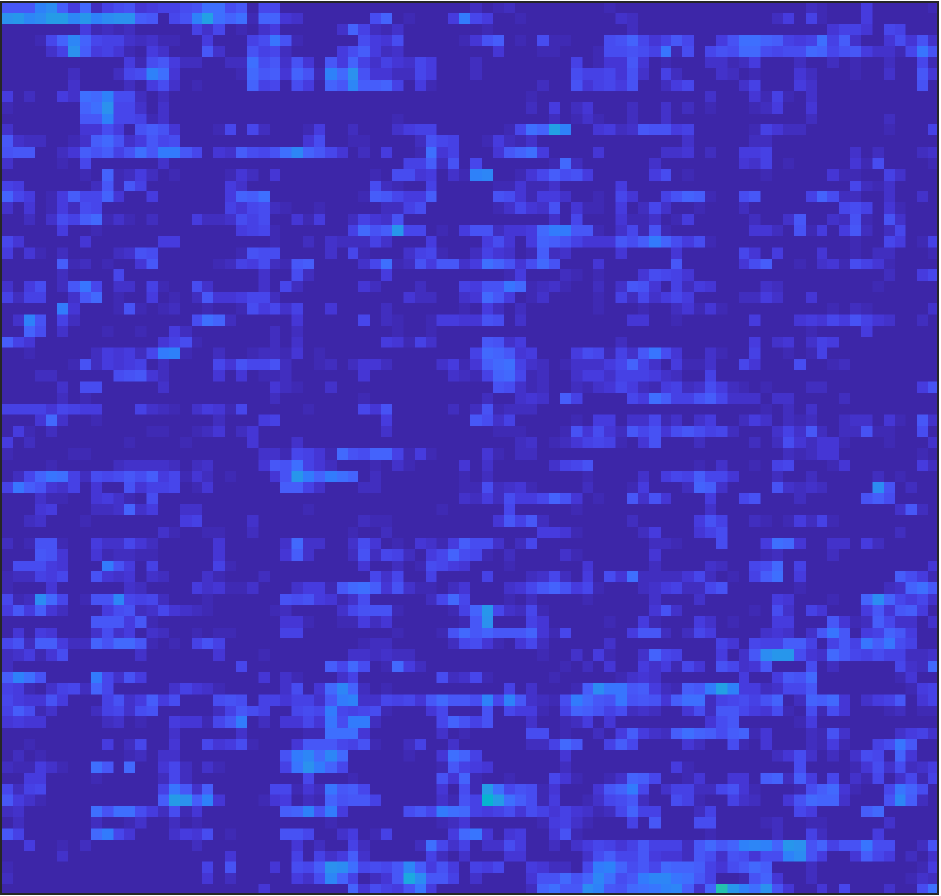}&
\includegraphics[width=0.135\textwidth]{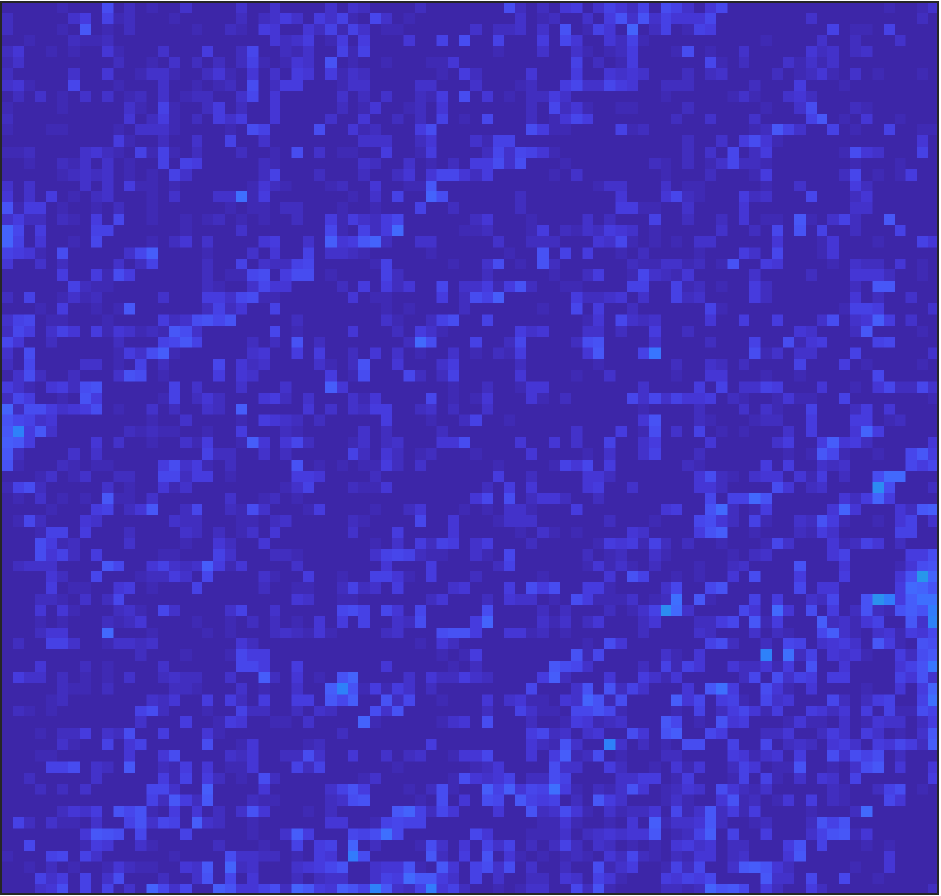}&
\includegraphics[width=0.135\textwidth]{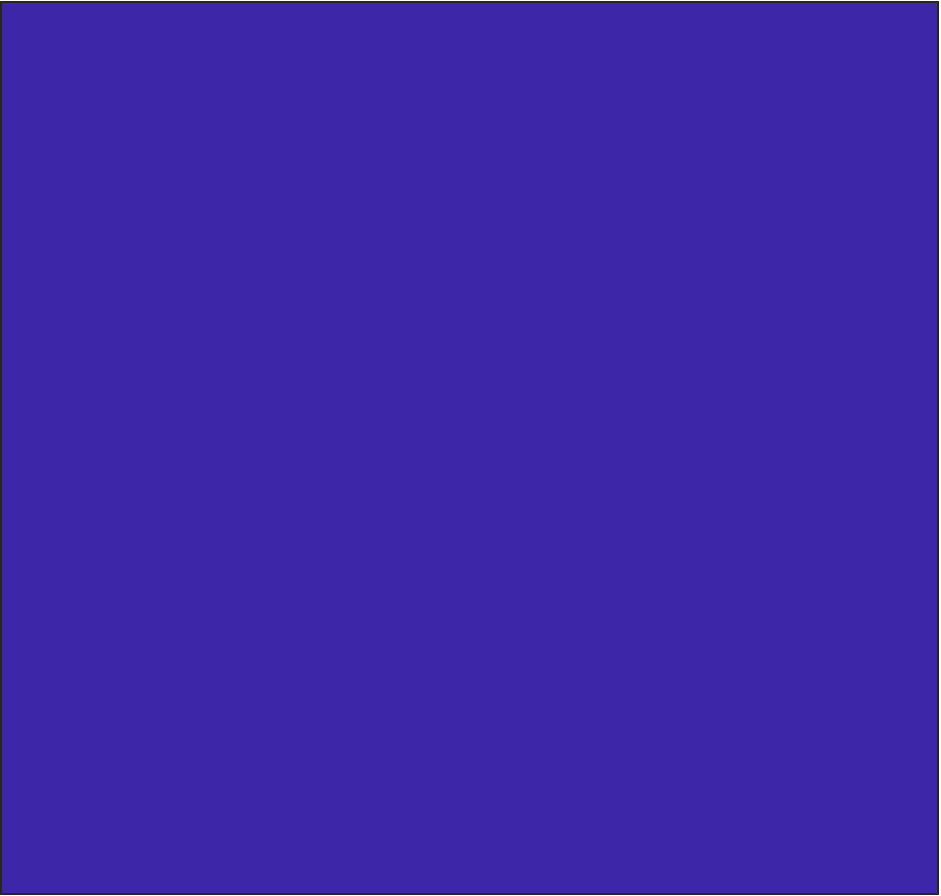}&
\hspace{0.05cm}
\includegraphics[width=0.021\textwidth]{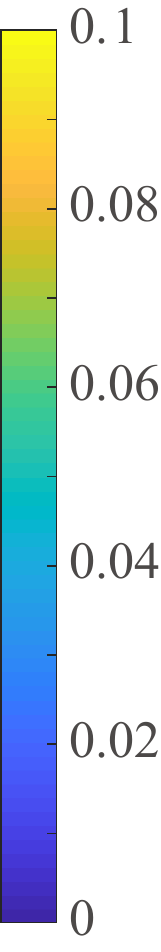}\\
\includegraphics[width=0.135\textwidth]{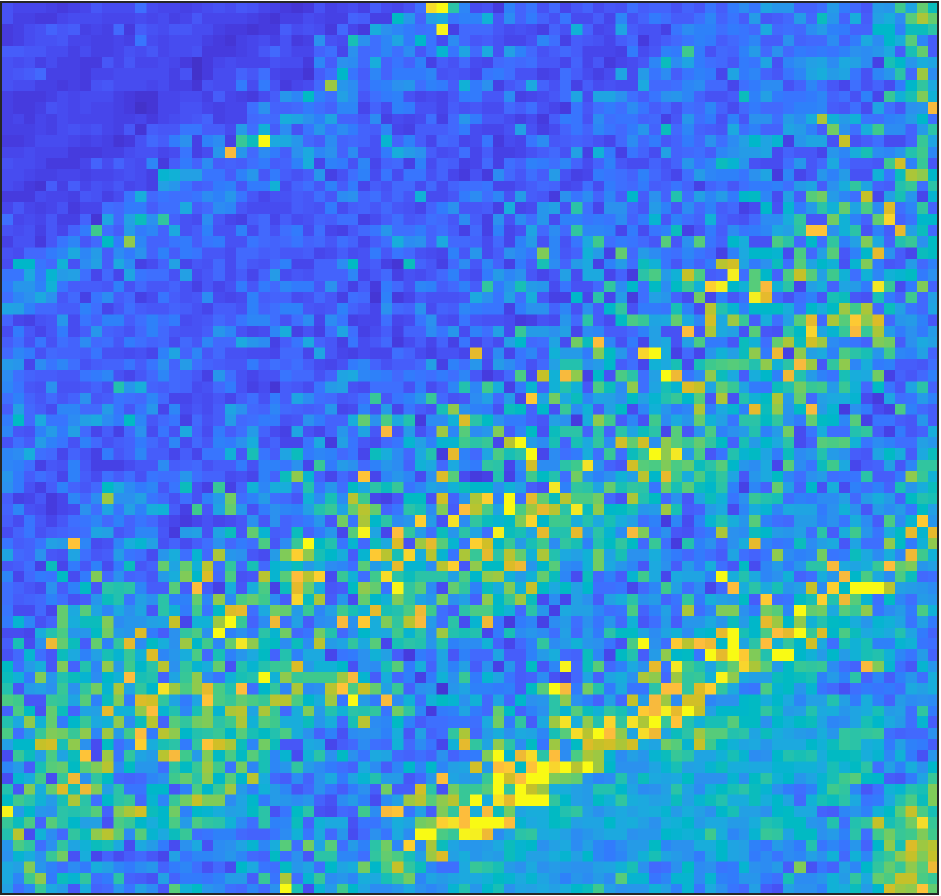}&
\includegraphics[width=0.135\textwidth]{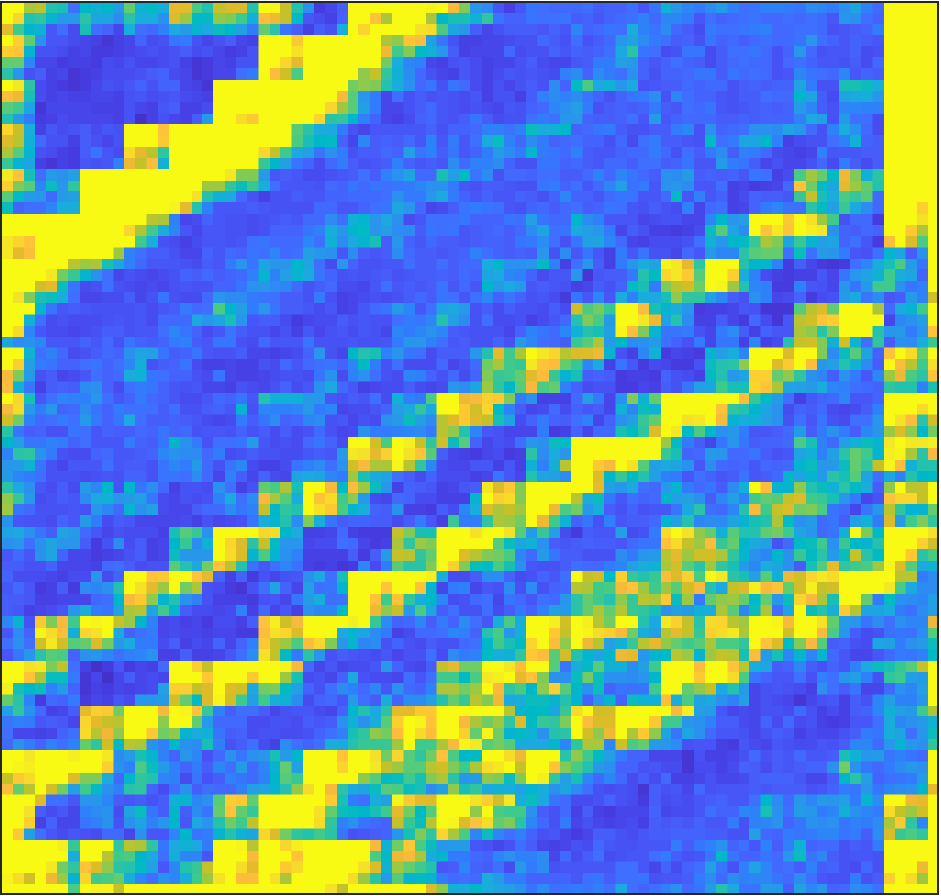}&
\includegraphics[width=0.135\textwidth]{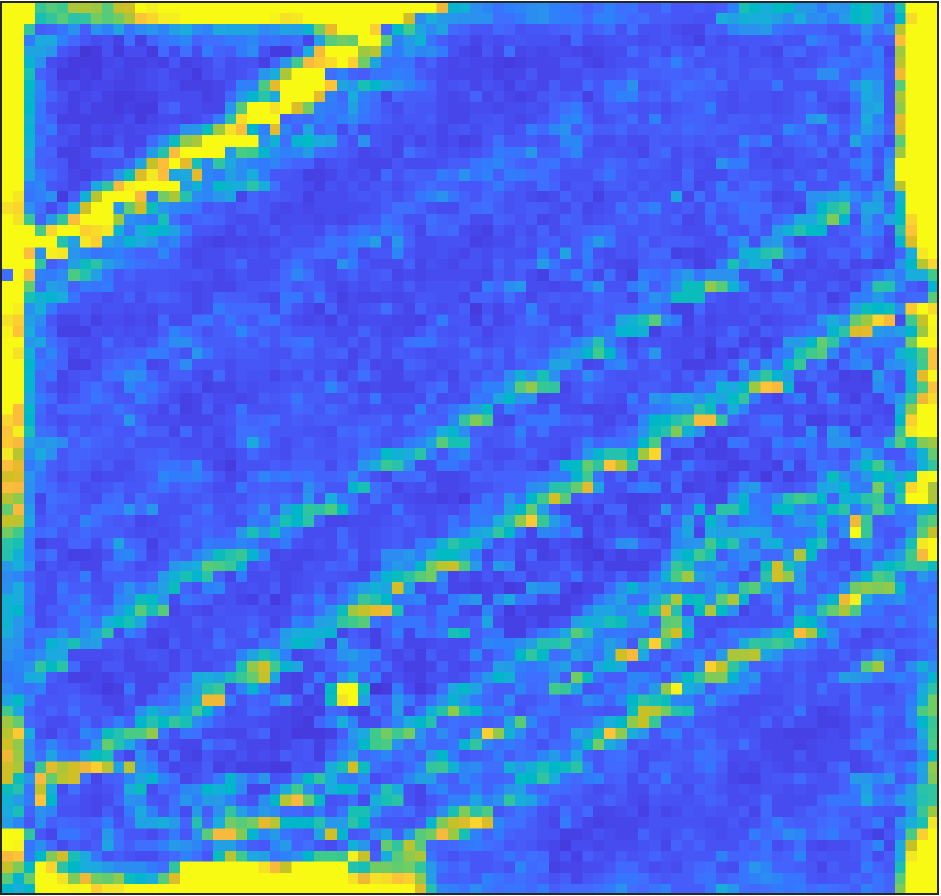}&
\includegraphics[width=0.135\textwidth]{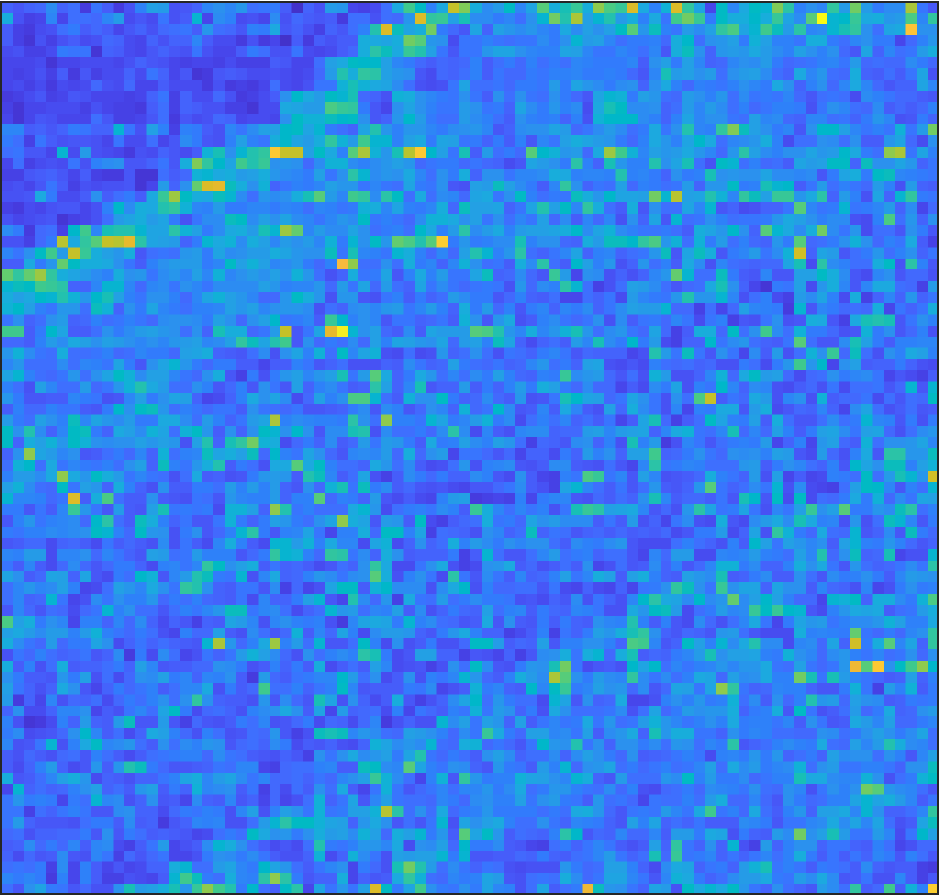}&
\includegraphics[width=0.135\textwidth]{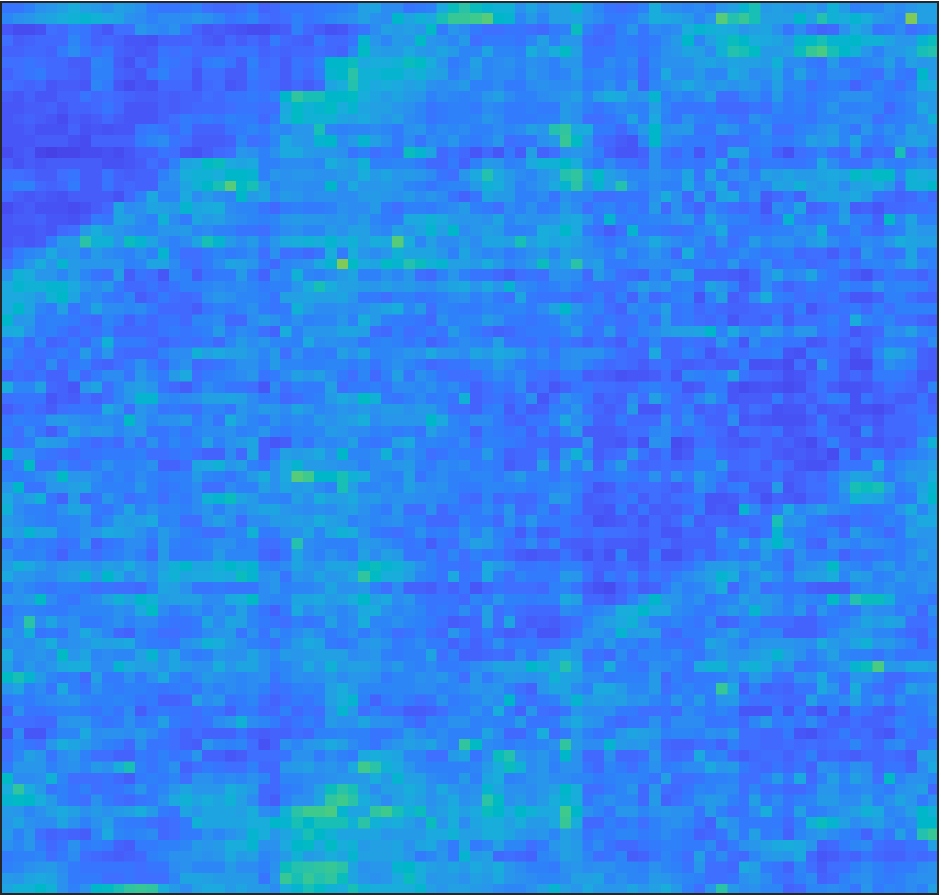}&
\includegraphics[width=0.135\textwidth]{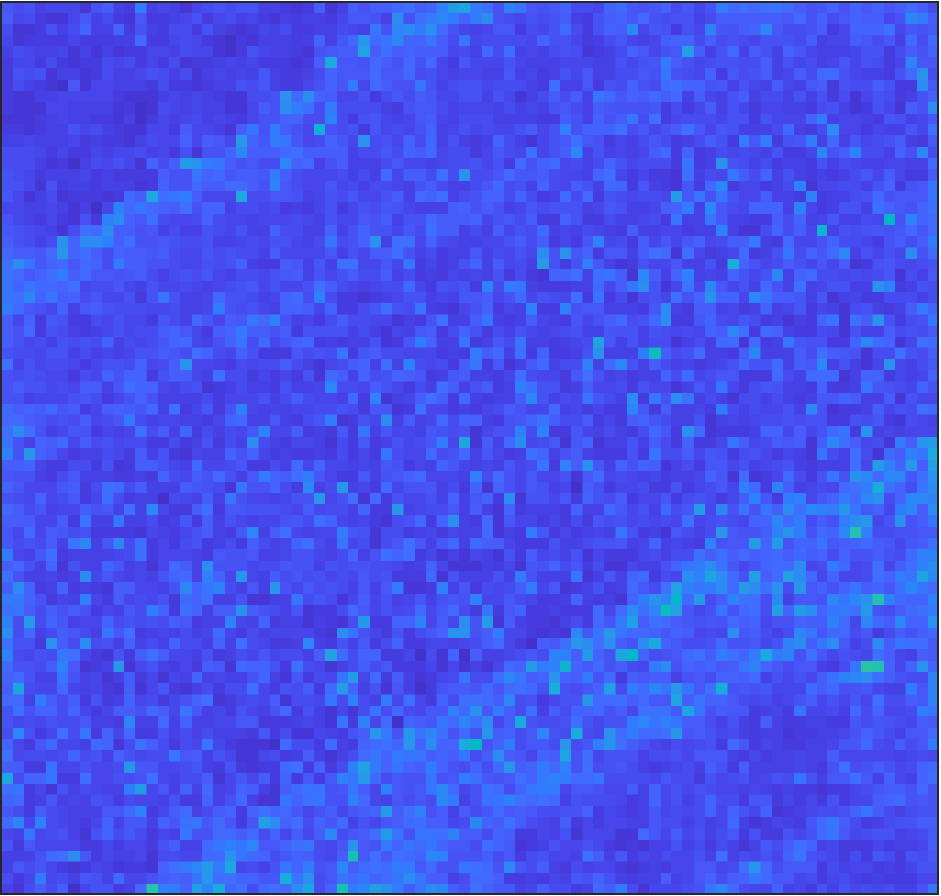}&
\includegraphics[width=0.135\textwidth]{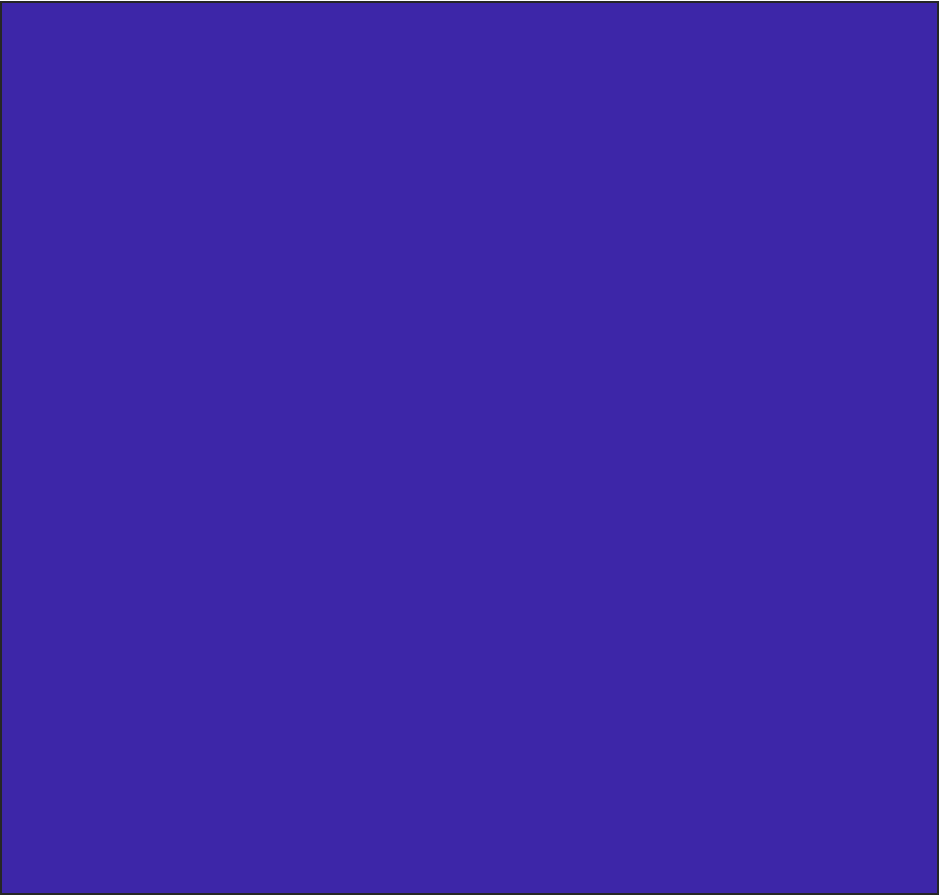}&
\hspace{0.05cm}
\includegraphics[width=0.021\textwidth]{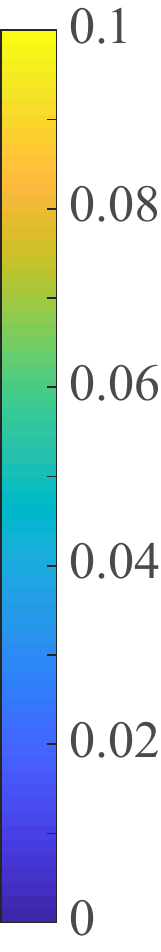}\\
(a) CNMF & (b) HySure & (c) FUSE & (d) SCOTT & (e) STEREO & (f) \textsf{SC-LL1} & (g) SRI\\
\end{tabular}
\caption{The recovered  SRIs, the corresponding residual images (the 34-th band), and the SAM maps of Salinas. First row: the recovered  SRIs of the 34-th band; Second row: the corresponding residual images of the 34-th band; Third row: the SAM maps.}
  \label{fig:Salinas_image}
  \end{center}\vspace{-0.3cm}
\end{figure*}

\subsubsection{Degradation Model}
For the semi-real data experiments, we follow the convention using a real hyperspectral image to act as the ``ground-truth" SRI---so that the recovery performance can be measured \cite{Kanatsoulis2018HSR,Wei2015HSRSylvester,Prevost2020HSR,Simoes2015HSR,Yokoya2012HSR}. The pair of simulated HSI and MSI images are generated following the Wald's protocol \cite{Wald1997Fusion}. The degradation from the SRI to the HSI is as follows: the SRI is first blurred by a $9\times 9$ Gaussian kernel and then downsampled every 4 pixels along each spatial dimension. For the degradation from the SRI to the MSI, we follow the setting in \cite{Kanatsoulis2018HSR}. 
Specifically, we form the degradation matrix $\bm{P}_{M}$ according to the specifications of two multispectral sensors, namely, LANDSAT\footnote{https://landsat.gsfc.nasa.gov/} and QuickBird\footnote{https://www.satimagingcorp.com/satellite-sensors/quickbird/}.  The spectral degradation matrix $\bm{P}_{M}$ is a band selection and aggregating matrix; see details in \cite{Kanatsoulis2018HSR}. In addition, zero-mean i.i.d. Gaussian noise is added to HSI and MSI with signal-to-noise ratio (SNR) being 30 dB, if not otherwise specified.
All the experiments are averaged from 20 random trials with different noise terms.

\subsubsection{Metrics}

In the semi-real data simulations, to evaluate the quality of the recovered SRIs, we employ a number of widely used metrics from the literature \cite{Loncan2015HSRoverview,Yokoya2017HSRoverview}. 
In particular, we employ the reconstruction signal-to-noise ratio (R-SNR), structural similarity index (SSIM), cross correlation (CC), universal image quality index (UIQI),  root mean square error (RMSE), relative dimensional global error (ERGAS), and spectral angle mapper (SAM).
The detailed definitions can be found in \cite{Loncan2015HSRoverview,Wei2016Fusion}.
Higher R-SNR, SSIM, CC, and UIQI values and lower RMSE, ERGAS, and SAM values indicate a more preferred reconstruction performance.

\subsubsection{Parameter Selection}

In the proposed algorithms, we set $\theta_{1}=\cdots=\theta_{R}=\theta$ and $\eta_{1}=\cdots=\eta_{R}=\eta$. To tune the parameters, we use the following idea: we select the parameters that give the lowest RMSE values between the estimated and the observable MSI and HSI entries. In addition, we fix the parameters $p=0.5$, $\tau=1$ in the Schatten-$p$ function and $q=0.5$, $\varepsilon=10^{-3}$ in the TV regularization \eqref{eq:Lp_TV}.
The entries of the initialization terms $\bm{S}^{0}$, $\widetilde{\bm{S}}^{0}$, and $\bm{C}^{0}$ are drawn uniformly at random from 0 to 1. We terminate the proposed algorithm when the relative error of the objective value between two consecutive iterations is below $10^{-4}$ or when the number of iterations exceeds 300 and 600 for \textsf{SC-LL1} and \textsf{BSC-LL1}, respectively.  
We tune the parameters of the baseline algorithms following the respective papers' instructions.

\subsection{Semi-real Experiments with Known $\bm P_1$ and $\bm P_2$}
We first test the proposed method in cases where all $\bm{P}_{1}$, $\bm{P}_{2}$, and $\bm{P}_{M}$ are known.
\subsubsection{Salinas Dataset} The first experiment uses a subscene of the Salinas dataset with a size of $80\times 84\times 204 $ (after removing 20 bands corrupted by water absorption), which is collected by AVIRIS sensor \cite{Green1998AVIRIS} over Salinas Valley \cite{Gualtieri1999SVM}. Applying the described spatial degradation and the LANDSAT spectral degradation, we generate $\underline{\bm{Y}}_{H}\in \mathbb{R}^{20\times 21\times 204}$ and $\underline{\bm{Y}}_{M}\in \mathbb{R}^{80\times 84\times 6}$. We set $R=6$, which is according to the number of materials that was reported in the literature \cite{Gualtieri1999SVM}. Fig. \ref{fig:Salinas_image} presents the 34-th band of the estimated SRIs (a band that is not contained in the MSI), the corresponding residual images (i.e., $\underline{\Y}_S(:,:,k) - \underline{\widehat{\Y}}_S(:,:,k)$ for $k=34$), and the SAM maps. From Fig. \ref{fig:Salinas_image}, one can see that \textsf{SC-LL1} has small residues across all pixels, while other algorithms' residual maps are less smooth. The proposed method also outputs an SAM map that is closer to the ideal one (cf. the rightmost column).
Both results indicate that the proposed method produces an estimated SRI that captures the details of the SRI well.

\begin{figure*}[!ht]
\scriptsize\setlength{\tabcolsep}{0.1pt}
\begin{center}
\begin{tabular}{cccc}
\includegraphics[width=0.99\textwidth]{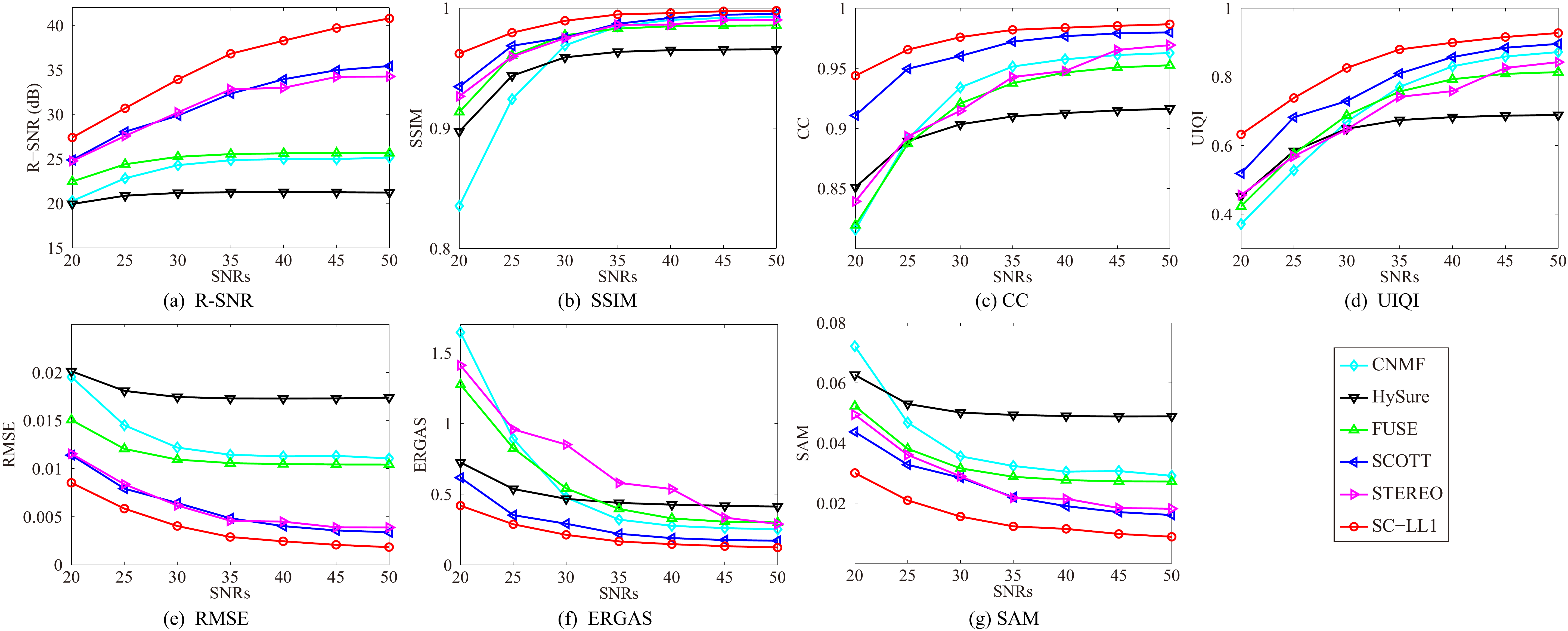}\\
\end{tabular}
\caption{Reconstruction metrics for Salinas under different noises.}
  \label{fig:Salinas_noise}
  \end{center}\vspace{-0.3cm}
\end{figure*}

We also test all the algorithms methods under different SNRs that range from 20 dB to 50 dB.  Fig. \ref{fig:Salinas_noise} shows the averaged evaluation results under different noise levels. One can see that the proposed \textsf{SC-LL1} method consistently outperforms the baselines under all metrics, showing promising performance. More detailed numerical comparison under SNR=30dB can be found in Table~\ref{table:Salinas}. 

\begin{table}[!ht]
  \centering
  \caption{Performance for Salinas with the degradation known.}
  \resizebox{\linewidth}{!}{
  \begin{tabular}{c|c|c|c|c|c|c}\hline

    \hline
    Method (ideal)  & \multicolumn{1}{c|}{CNMF} & \multicolumn{1}{c|}{HySure} & \multicolumn{1}{c|}{FUSE}  & \multicolumn{1}{c|}{SCOTT} & \multicolumn{1}{c|}{STEREO}  & \multicolumn{1}{c}{\textsf{SC-LL1}} \\ \hline
    R-SNR $(\infty)$    & 24.30  & 21.18  & 25.24  & 29.86  & 30.22  & \textbf{33.92} \\
    SSIM $(1)$    & 0.9688 & 0.9589 & 0.9773 & 0.9755 & 0.9749 & \textbf{0.9894} \\
    CC $(1)$      & 0.9341 & 0.9034 & 0.9208 & 0.9603 & 0.9147 & \textbf{0.9760} \\
    UIQI $(1)$    & 0.6693 & 0.6488 & 0.6877 & 0.7294 & 0.6470 & \textbf{0.8258} \\
    RMSE $(0)$    & 0.0122 & 0.0174 & 0.0109 & 0.0064 & 0.0062 & \textbf{0.0040} \\
    ERGAS $(0)$   & 0.4768 & 0.4674 & 0.5414 & 0.2916 & 0.8509 & \textbf{0.2128} \\
    SAM   $(0)$   & 0.0356 & 0.0501 & 0.0316 & 0.0285 & 0.0289 & \textbf{0.0155} \\
    \hline
    \end{tabular}}%
  \label{table:Salinas}%
\end{table}%

\subsubsection{Pavia University Dataset} \label{sec:semi_real_non_blind_Pavia}
For the second experiment, we use a subimage of the Pavia University. This dataset is captured by the ROSIS sensor \cite{Kunkel1988ROSIS}. The sizes of the SRI and the HSI are $256\times 256\times 103$ and $64\times 64\times 103$, respectively. We generate the MSI with a size of $256\times 256\times 4$ through the QuickBird spectral degradation pattern. We set $R=4$ as the number of endmembers in this simulation.

Table \ref{table:Pavia} shows the reconstruction performance of algorithms under SNR=30dB. Similar to the previous experiment, one can see that \textsf{SC-LL1} consistently evaluates the best over different metrics. In particular, the R-SNR output by the proposed algorithm is at least 2dB higher than that of the best baseline, which is considered a notable margin (improvement by 58\%).

\begin{table}[!ht]
  \centering
  \caption{Performance for Pavia University with the degradation known.}
  \resizebox{\linewidth}{!}{
    \begin{tabular}{c|c|c|c|c|c|c}\hline

    \hline
    Method (ideal) & \multicolumn{1}{c|}{CNMF} & \multicolumn{1}{c|}{HySure} & \multicolumn{1}{c|}{FUSE}  & \multicolumn{1}{c|}{SCOTT} & \multicolumn{1}{c|}{STEREO}  & \multicolumn{1}{c}{\textsf{SC-LL1}} \\ \hline
    R-SNR $(\infty)$  & 20.30  & 16.01  & 20.20  & 21.38  & 24.25  & \textbf{26.33} \\
    SSIM $(1)$    & 0.9431 & 0.8931 & 0.9345 & 0.9178 & 0.9459 & \textbf{0.9726} \\
    CC  $(1)$     & 0.9803 & 0.9451 & 0.9770 & 0.9816 & 0.9898 & \textbf{0.9934} \\
    UIQI $(1)$    & 0.9113 & 0.8544 & 0.9010 & 0.8768 & 0.9063 & \textbf{0.9460} \\
    RMSE $(0)$    & 0.0218 & 0.0356 & 0.0220 & 0.0192 & 0.0138 & \textbf{0.0109} \\
    ERGAS $(0)$   & 0.5459 & 0.8570 & 0.5381 & 0.4415 & 0.3288 & \textbf{0.2628} \\
    SAM $(0)$     & 0.0809 & 0.1114 & 0.0827 & 0.0776 & 0.0709 & \textbf{0.0532} \\
    \hline
    \end{tabular}}%
  \label{table:Pavia}%
\end{table}%

\subsubsection{Indian Pines Dataset} 
The third dataset that we use is a subscene of the Indian Pines data, which is again acquired by the AVIRIS sensor. This subscene consists of $R=16$ different prominent materials as reported in \cite{Baumgardner2015Indian}. After removing water-absorption contaminated bands, we have an SRI $\underline{\bm{Y}}_{S}$ that has a size of ${144\times 144\times 200}$. Then, the HSI $\underline{\bm{Y}}_{H}\in \mathbb{R}^{36\times 36\times 200}$ is generated using the aforementioned spatial degradation and the MSI $\underline{\bm{Y}}_{S}\in \mathbb{R}^{144\times 144\times 6}$ is generated using the LANDSAT spectral degradation specification. 

\begin{table}[!ht]
  \centering
  \caption{Performance for Indian Pines with the degradation known.}
  \resizebox{\linewidth}{!}{
    \begin{tabular}{c|c|c|c|c|c|c}\hline

    \hline
    Method (ideal)  & \multicolumn{1}{c|}{CNMF} & \multicolumn{1}{c|}{HySure} & \multicolumn{1}{c|}{FUSE}  & \multicolumn{1}{c|}{SCOTT} & \multicolumn{1}{c|}{STEREO}  & \multicolumn{1}{c}{\textsf{SC-LL1}} \\ \hline
    R-SNR $(\infty)$   & 26.49  & 24.71  & 26.48  & 25.02  & 27.39  & \textbf{28.78} \\
    SSIM $(1)$    & 0.9124 & 0.9151 & 0.9269 & 0.8929 & 0.9339 & \textbf{0.9515} \\
    CC $(1)$      & 0.8468 & 0.8497 & 0.8653 & 0.8363 & 0.8506 & \textbf{0.9162} \\
    UIQI $(1)$    & 0.6229 & 0.6438 & 0.6703 & 0.6117 & 0.6532 & \textbf{0.7696} \\
    RMSE   $(0)$  & 0.0153 & 0.0187 & 0.0153 & 0.0181 & 0.0138 & \textbf{0.0117} \\
    ERGAS  $(0)$  & 0.1989 & 0.2226 & 0.1844 & 0.2124 & 0.1789 & \textbf{0.1369} \\
    SAM   $(0)$   & 0.0432 & 0.0510 & 0.0419 & 0.0517 & 0.0406 & \textbf{0.0339} \\
    \hline
    \end{tabular}}%
  \label{table:Indian}%
\end{table}%

\begin{figure*}[!ht]
\scriptsize\setlength{\tabcolsep}{0.3pt}
\begin{center}
\begin{tabular}{cccc}
\includegraphics[width=0.25\textwidth]{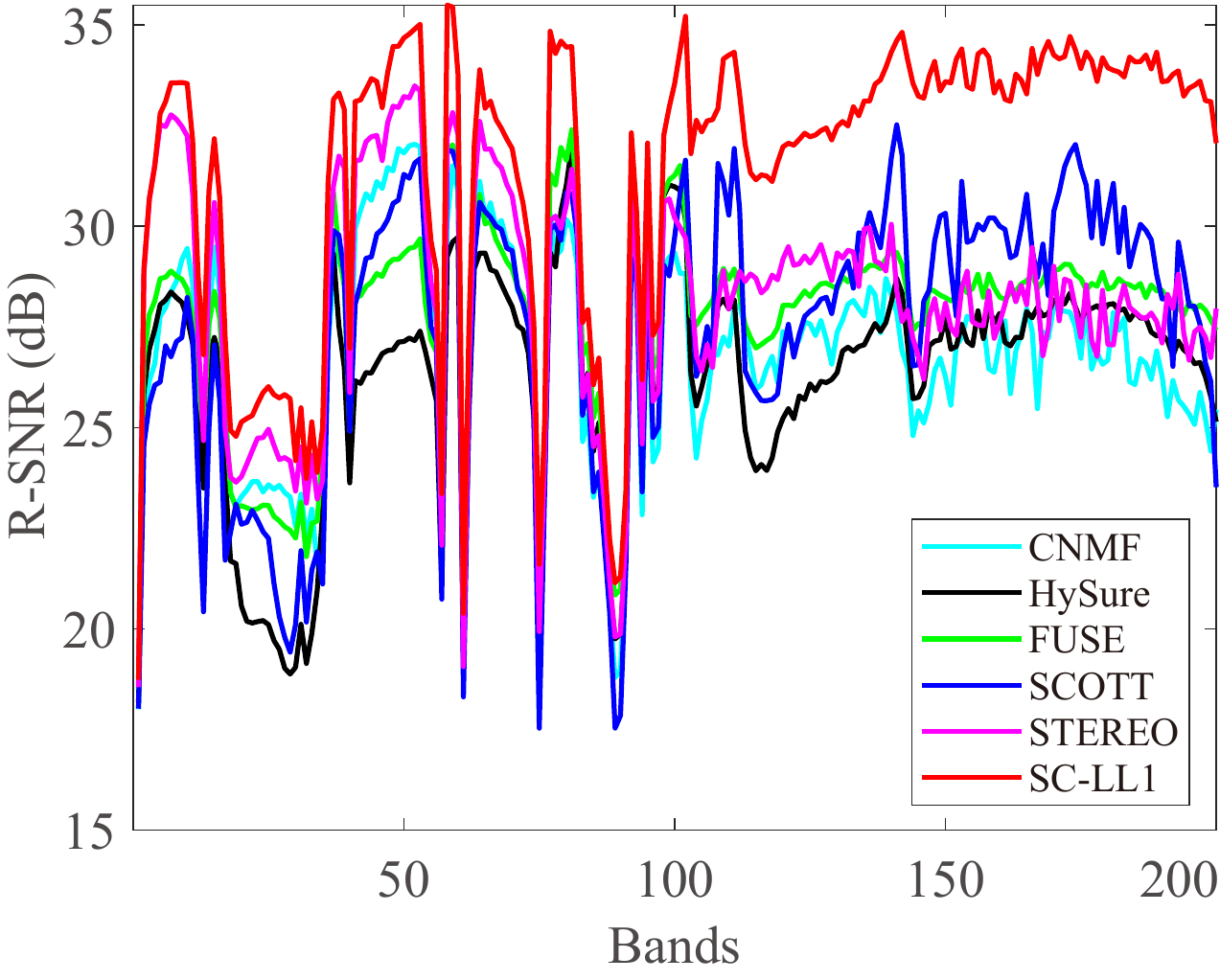}&
\includegraphics[width=0.25\textwidth]{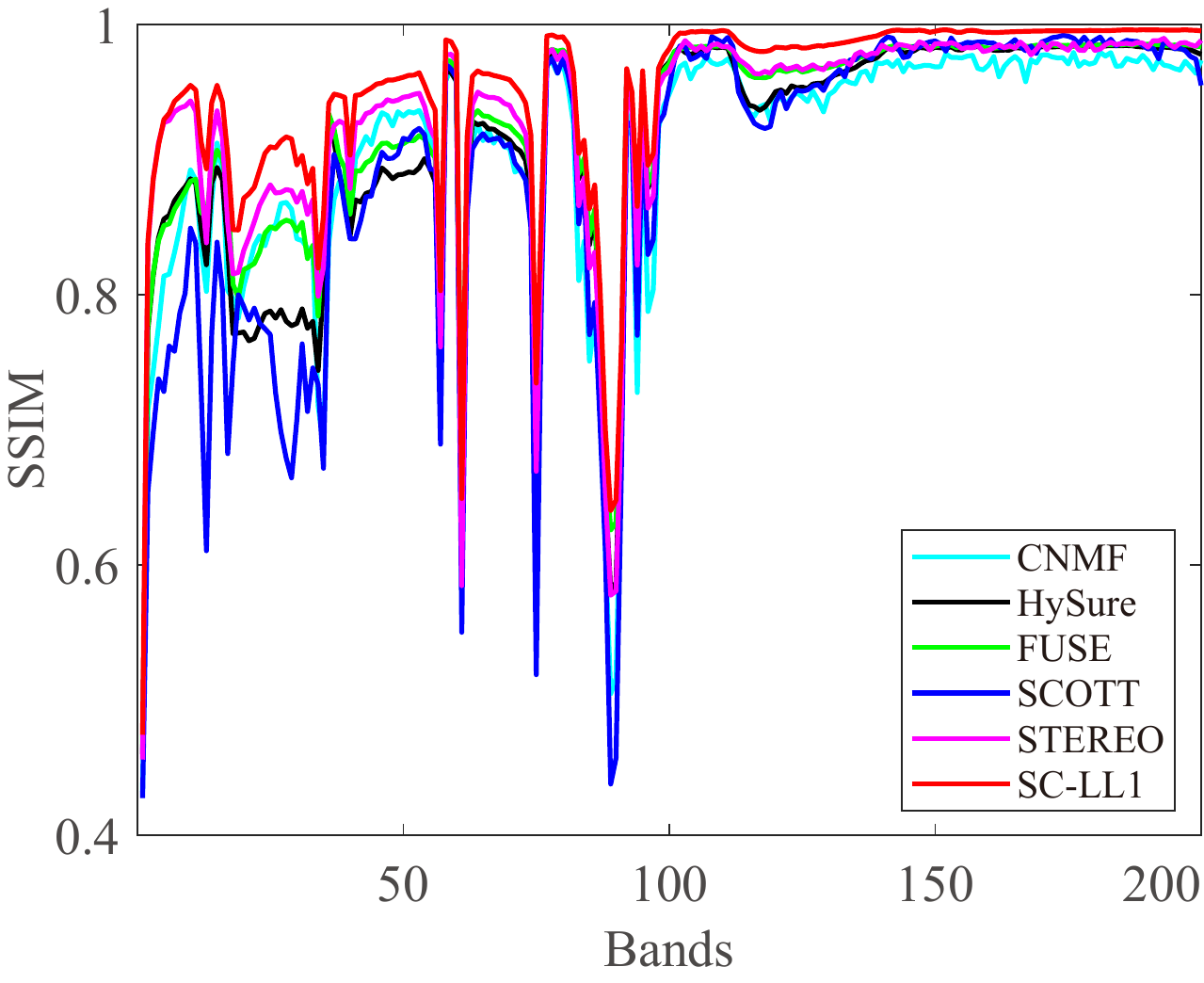}&
\includegraphics[width=0.25\textwidth]{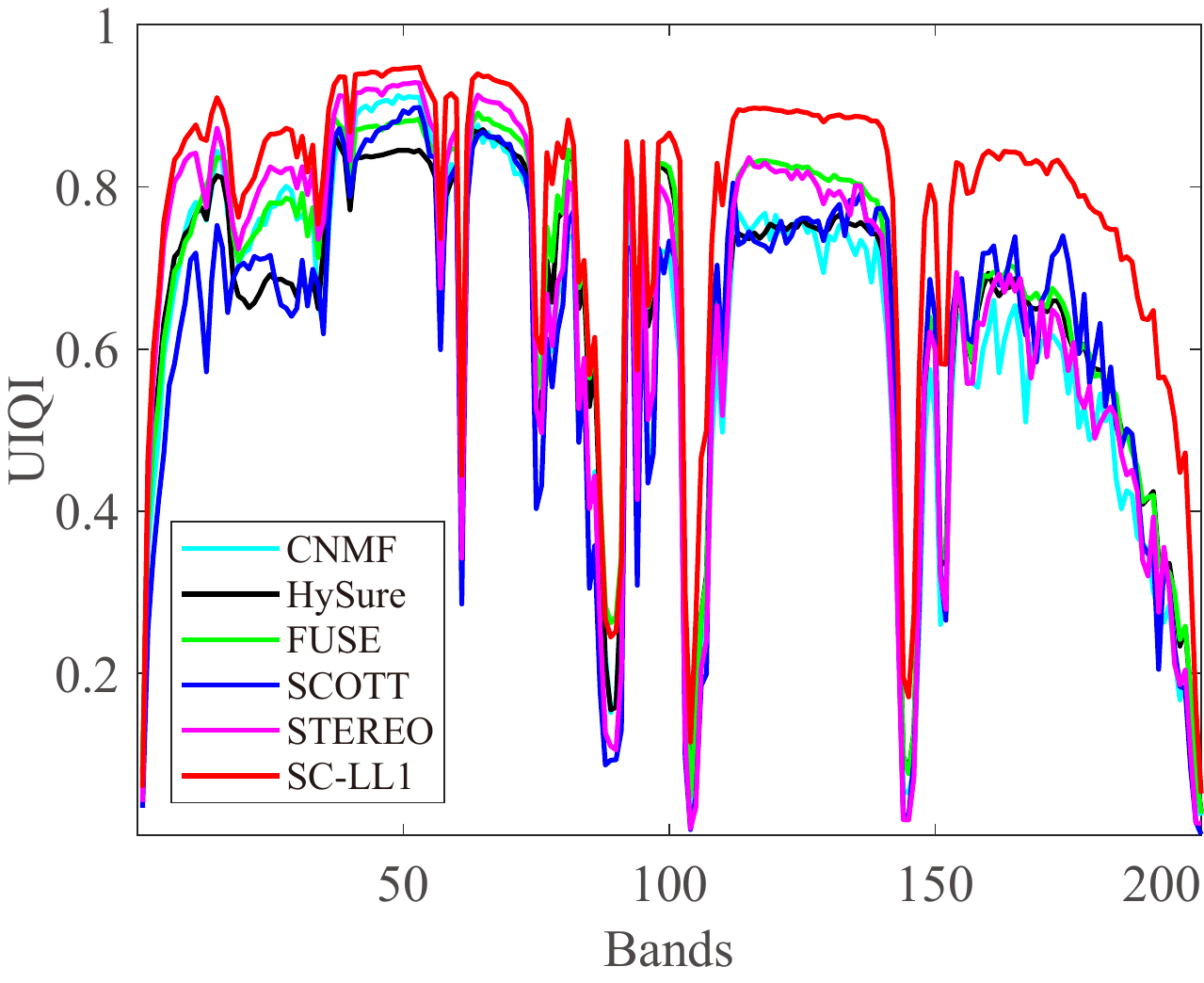}&
\includegraphics[width=0.25\textwidth]{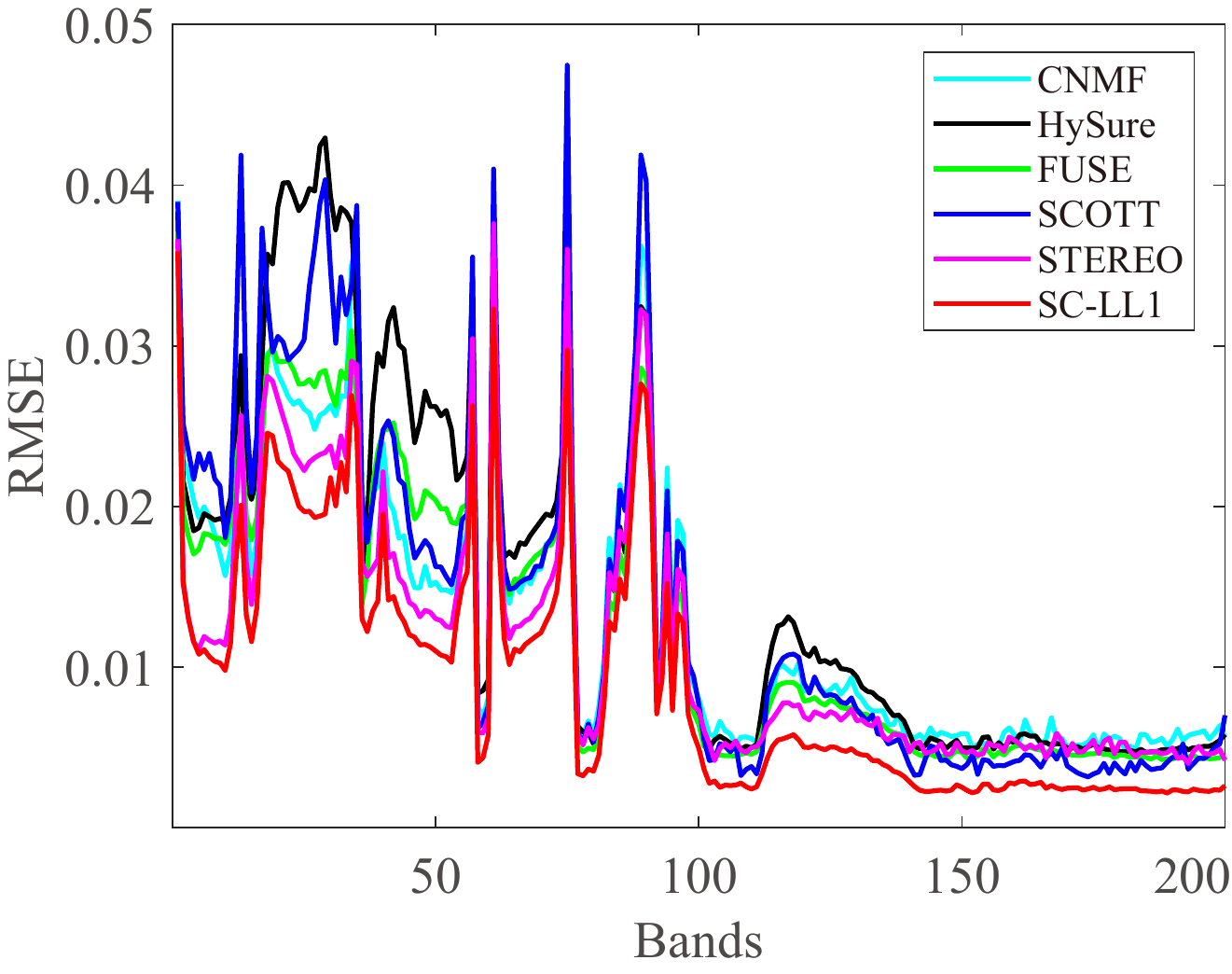}\\
(a) R-SNR & (b) SSIM & (c) UIQI & (d) RMSE\\
\end{tabular}
\caption{R-SNR, SSIM, UIQI, and RMSE values of each band of the Indian Pines image.}
  \label{fig:Indian_rsnr}
  \end{center}\vspace{-0.3cm}
\end{figure*}

From Table \ref{table:Indian}, one can see that the proposed method again exhibits the most promising performance over all evaluation metrics.
In addition, Fig. \ref{fig:Indian_rsnr} shows the R-SNR, SSIM, UIQI, and RMSE curves against the spectral bands. Again, the propose approach has a more favorable performance over different frequency bands.

\subsubsection{Jasper Ridge Dataset} The last dataset that we employ under the settings where $\bm P_1,\bm P_2$ are known is the Jasper Ridge data with a size of $100\times 100\times 198$ (after removing bands corrupted by dense water vapor and atmospheric effects). The HSI and MSI are with sizes of $25\times 25\times 198$ and $100\times 100\times 6$, respectively. We use $R=4$ following \cite{Zhu2014Spectral}. Table \ref{table:Ridge} shows the performance of all methods. Similar as before, the proposed \textsf{SC-LL1} exhibits promising performance over this dataset.

\begin{table}[!ht]
  \centering
  \caption{Performance for Jasper Ridge with the degradation known.}
  \resizebox{\linewidth}{!}{
    \begin{tabular}{c|c|c|c|c|c|c}\hline

    \hline
    Method (ideal)  & \multicolumn{1}{c|}{CNMF} & \multicolumn{1}{c|}{HySure} & \multicolumn{1}{c|}{FUSE}  & \multicolumn{1}{c|}{SCOTT} & \multicolumn{1}{c|}{STEREO}  & \multicolumn{1}{c}{\textsf{SC-LL1}} \\ \hline
    R-SNR $(\infty)$   & 25.37  & 20.07  & 22.63  & 26.46  & 25.42  & \textbf{27.16} \\
    SSIM $(1)$    & 0.9544 & 0.9409 & 0.9319 & 0.9556 & 0.9415 & \textbf{0.9731} \\
    CC $(1)$      & 0.9886 & 0.9700 & 0.9814 & 0.9902 & 0.9840 & \textbf{0.9921} \\
    UIQI $(1)$    & 0.8678 & 0.7960 & 0.8452 & 0.8595 & 0.8401 & \textbf{0.9022} \\
    RMSE   $(0)$  & 0.0156 & 0.0288 & 0.0215 & 0.0138 & 0.0156 & \textbf{0.0127} \\
    ERGAS $(0)$   & 0.3740 & 0.6312 & 0.4877 & 0.3490 & 0.5183 & \textbf{0.3279} \\
    SAM   $(0)$   & 0.0796 & 0.1168 & 0.0938 & 0.0836 & 0.0933 & \textbf{0.0676} \\
    \hline
    \end{tabular}}%
  \label{table:Ridge}%
\end{table}%

\subsection{Semi-real Experiments with Unknown $\bm P_1$ and $\bm P_2$}
\label{exp:semi-blind}

\begin{figure*}[!ht]
\scriptsize\setlength{\tabcolsep}{0.3pt}
\begin{center}
\begin{tabular}{cccccccc}
\includegraphics[width=0.135\textwidth]{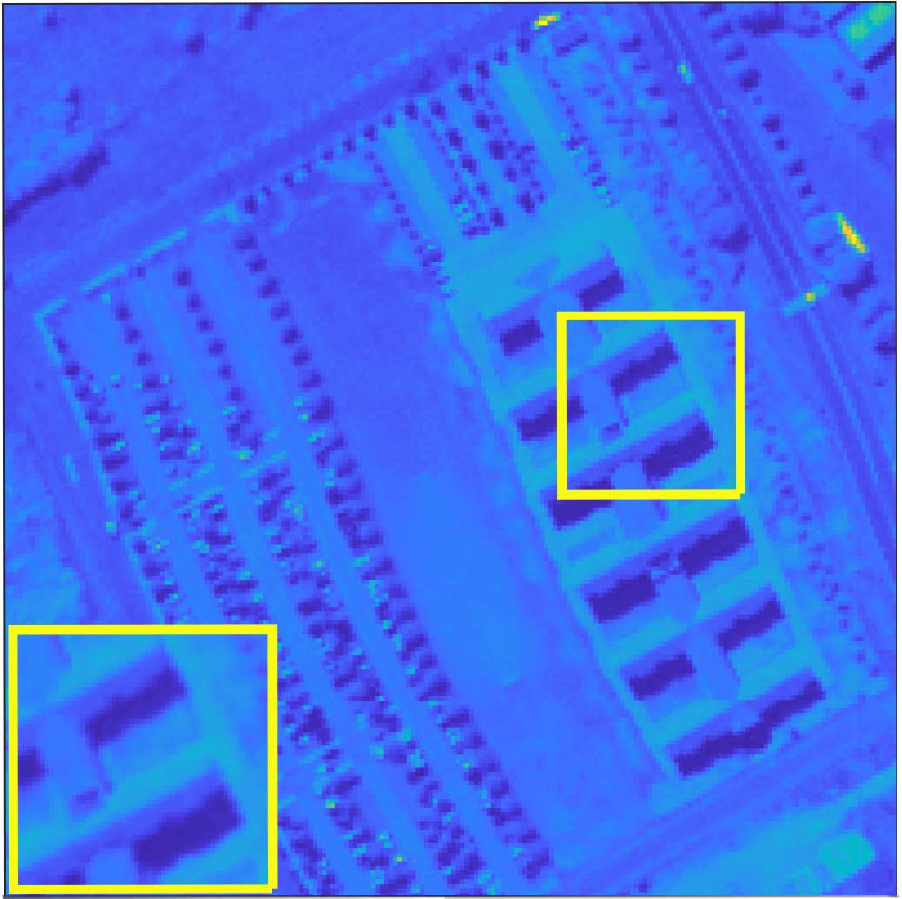}&
\includegraphics[width=0.135\textwidth]{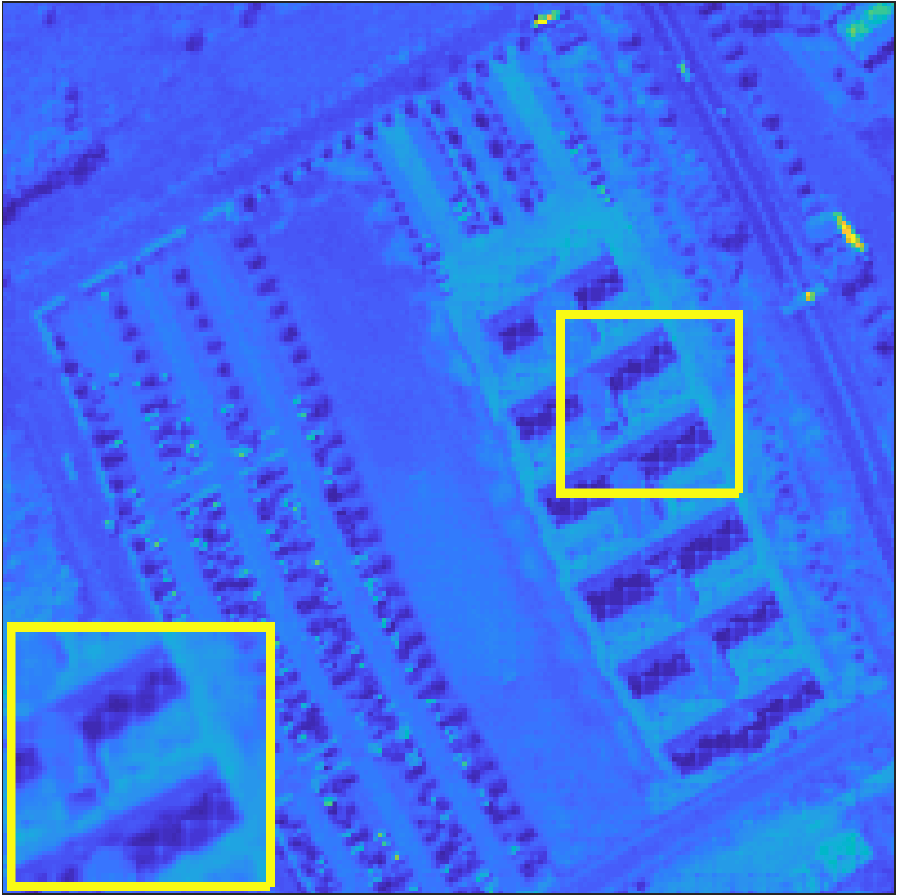}&
\includegraphics[width=0.135\textwidth]{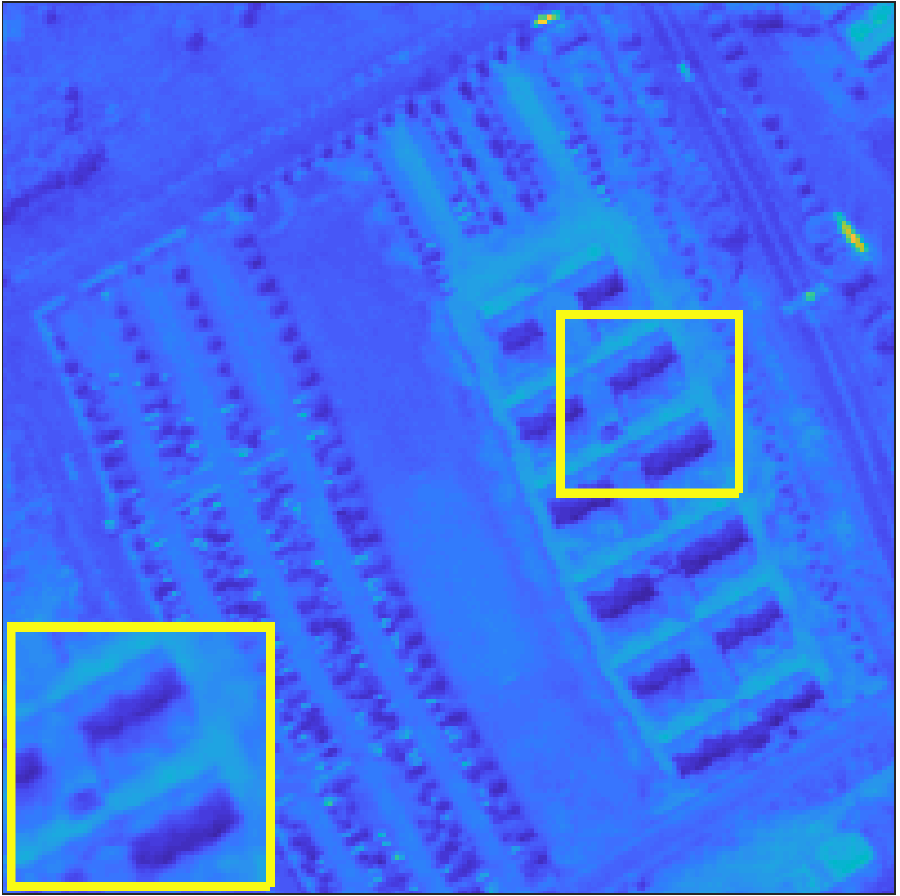}&
\includegraphics[width=0.135\textwidth]{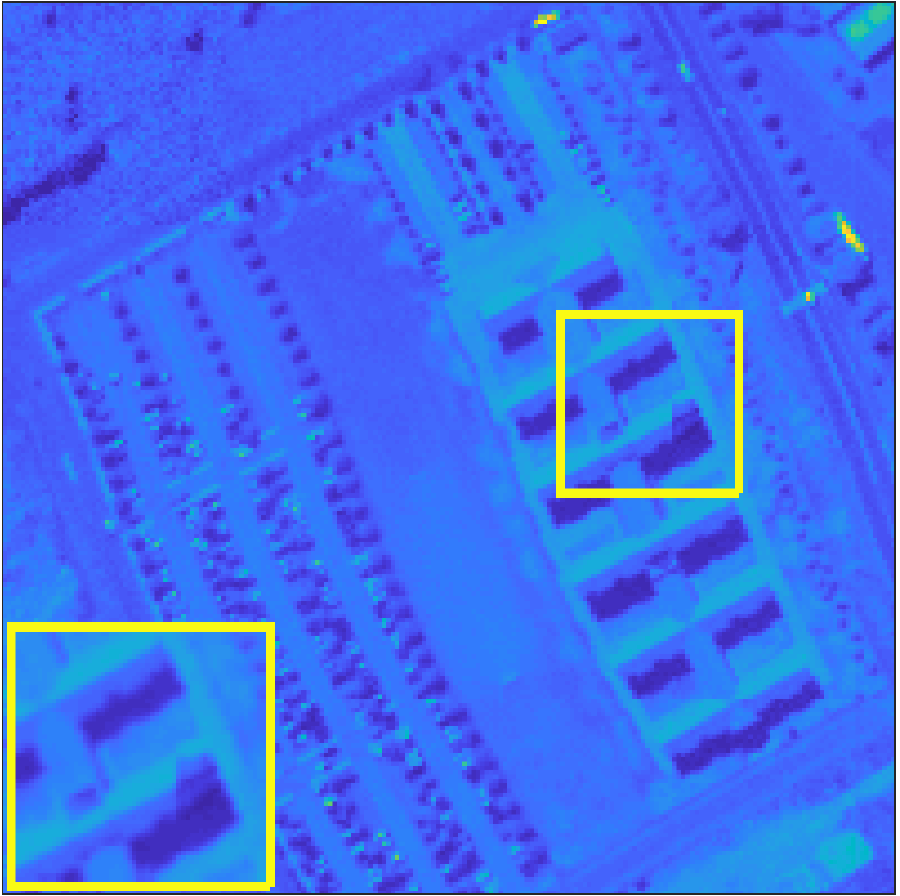}&
\includegraphics[width=0.135\textwidth]{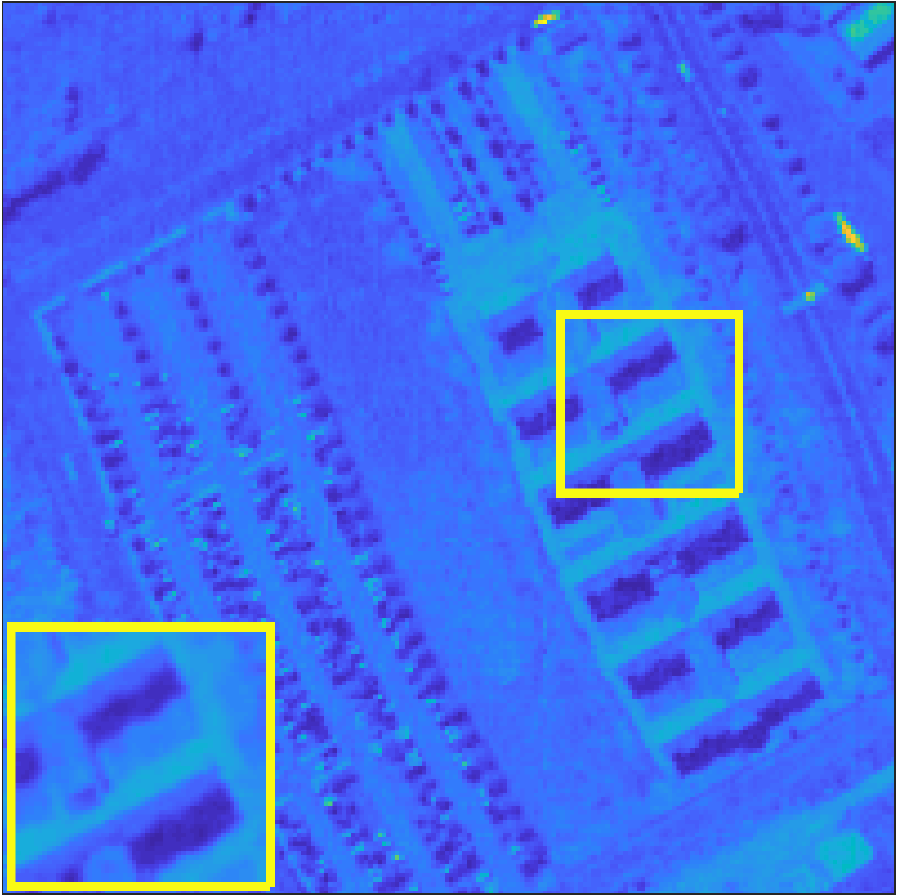}&
\includegraphics[width=0.135\textwidth]{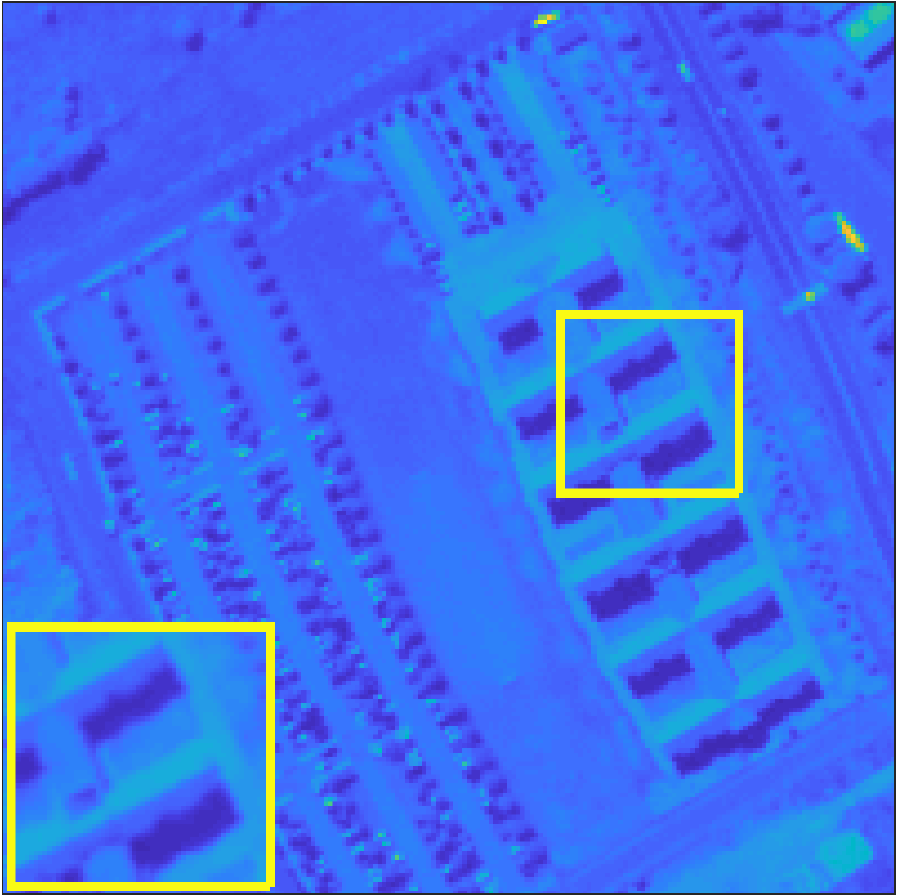}&
\includegraphics[width=0.135\textwidth]{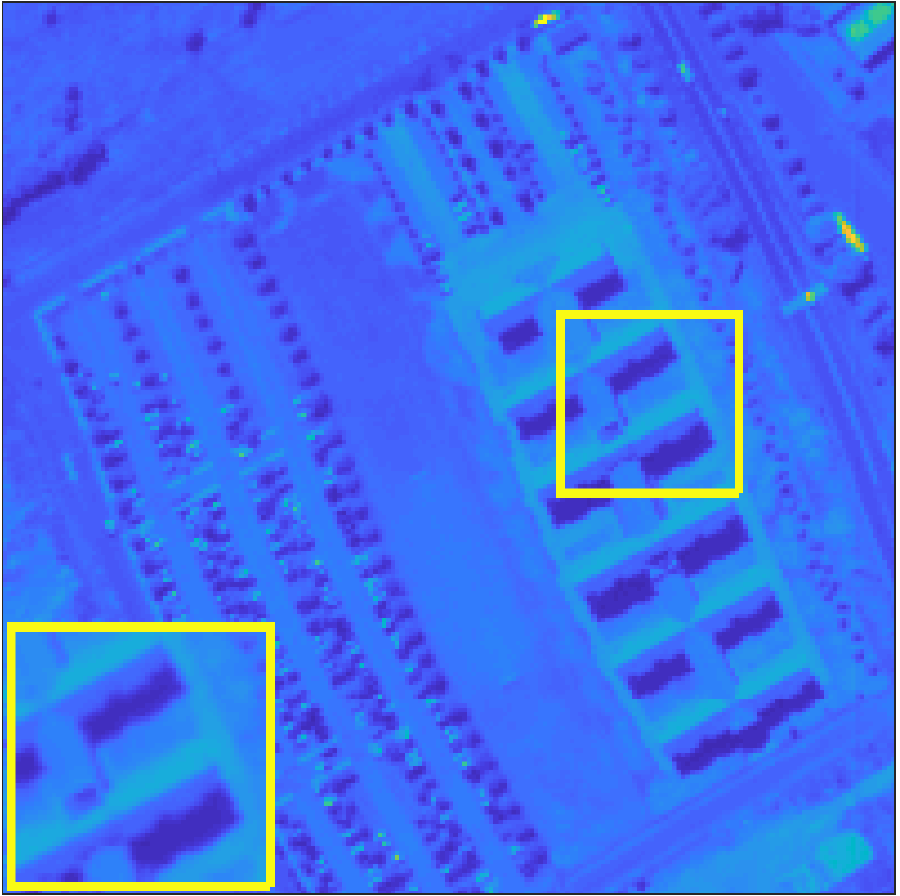}&
\hspace{0.01cm}
\includegraphics[width=0.018\textwidth]{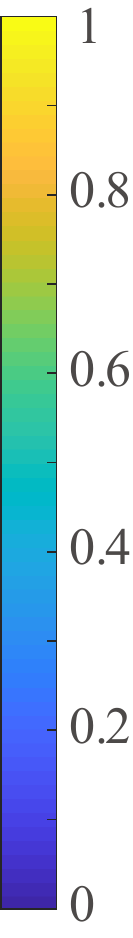}\\
\includegraphics[width=0.135\textwidth]{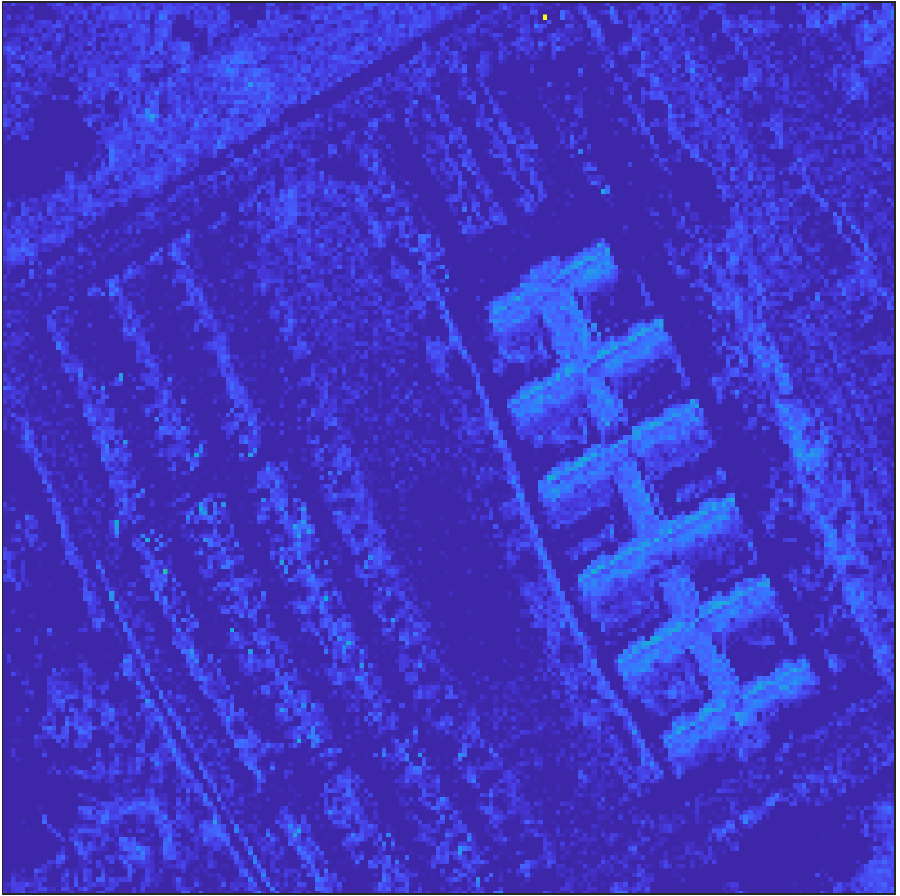}&
\includegraphics[width=0.135\textwidth]{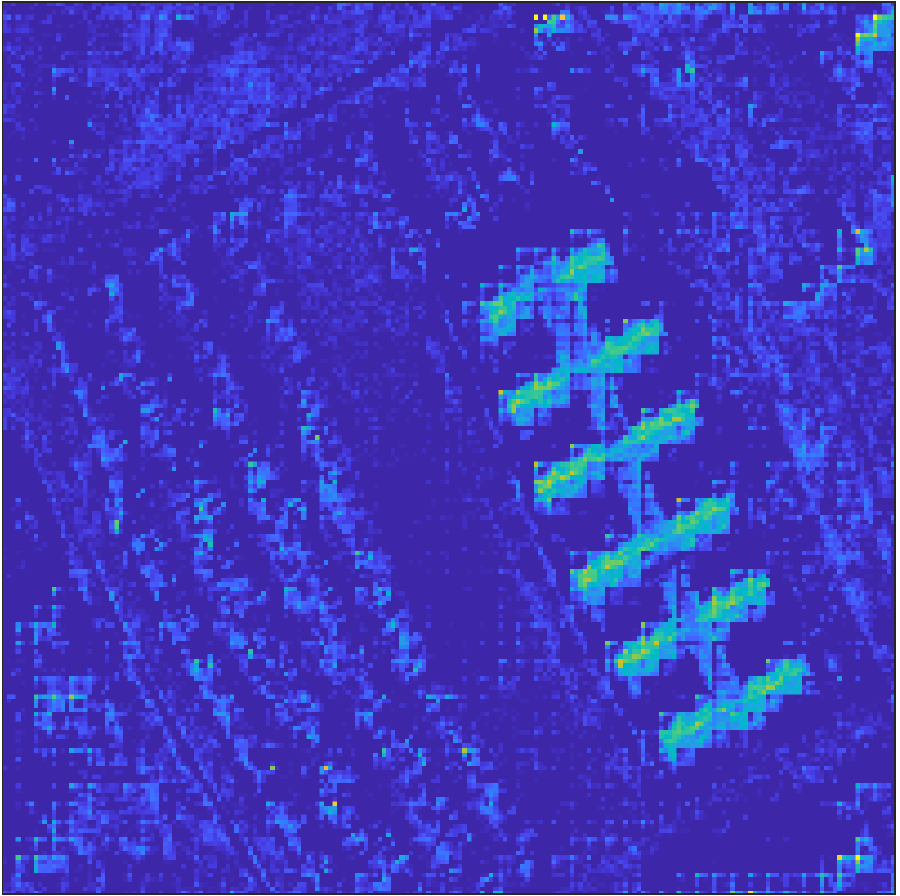}&
\includegraphics[width=0.135\textwidth]{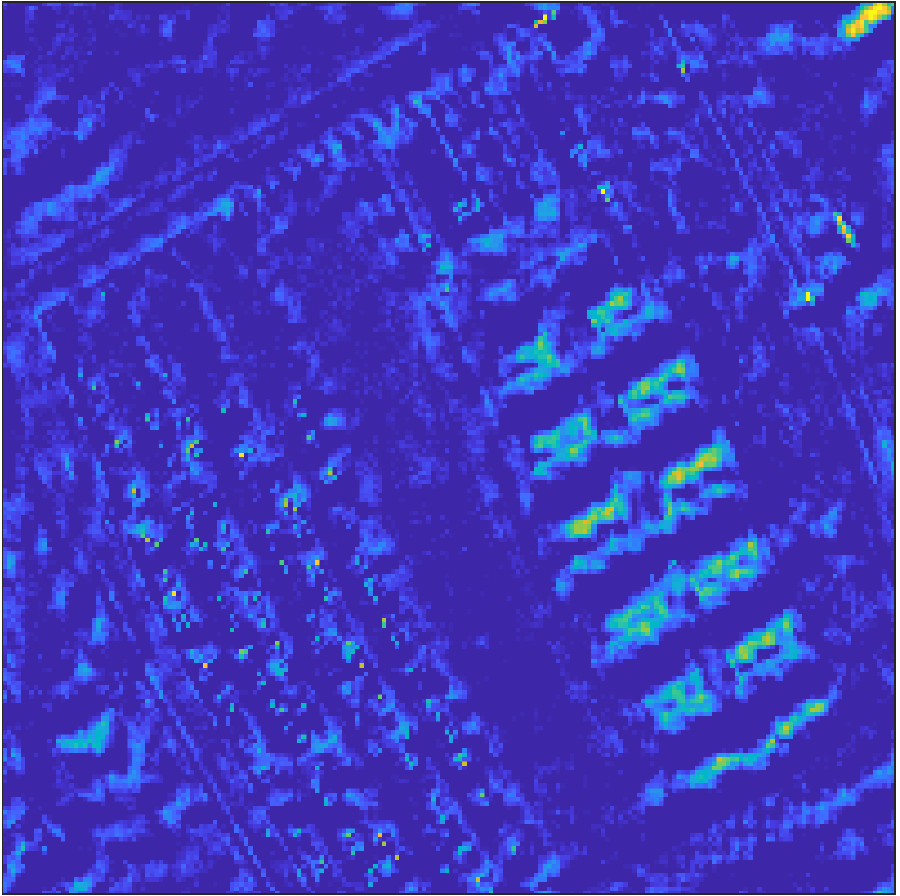}&
\includegraphics[width=0.135\textwidth]{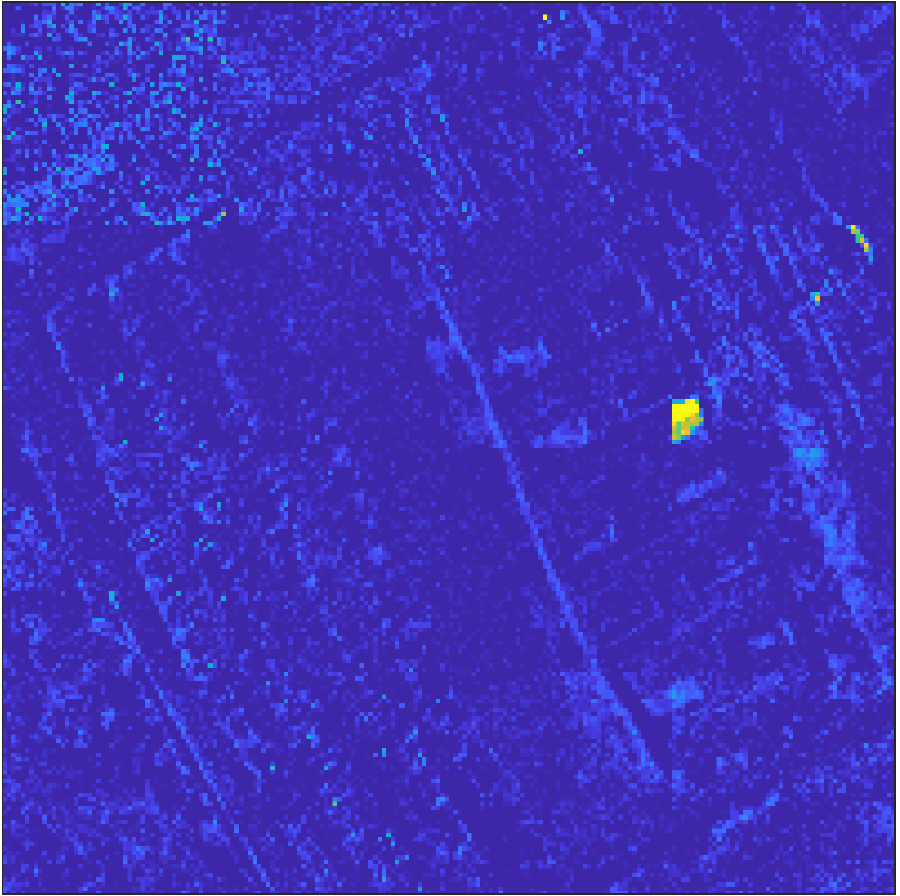}&
\includegraphics[width=0.135\textwidth]{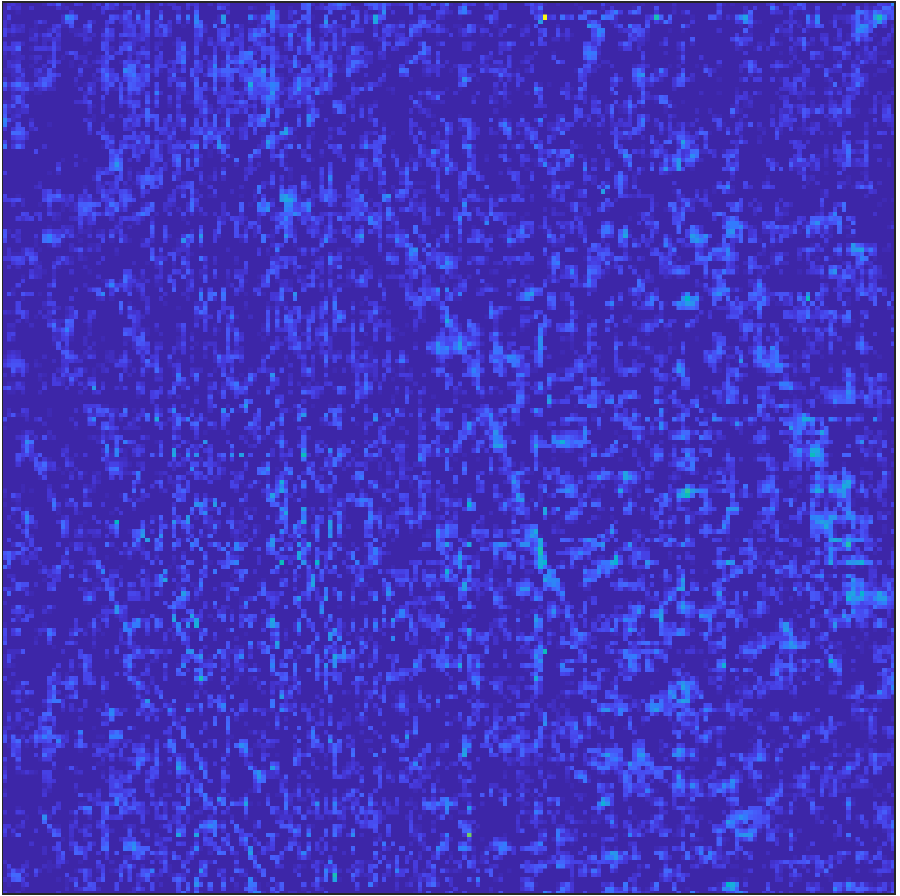}&
\includegraphics[width=0.135\textwidth]{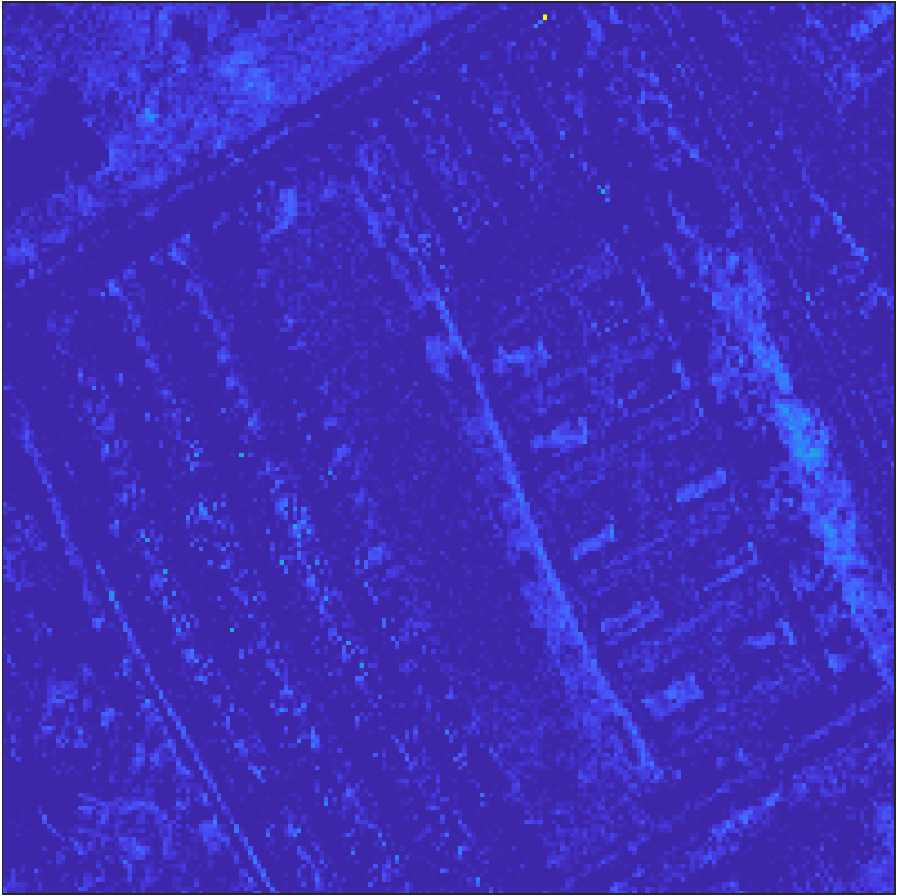}&
\includegraphics[width=0.135\textwidth]{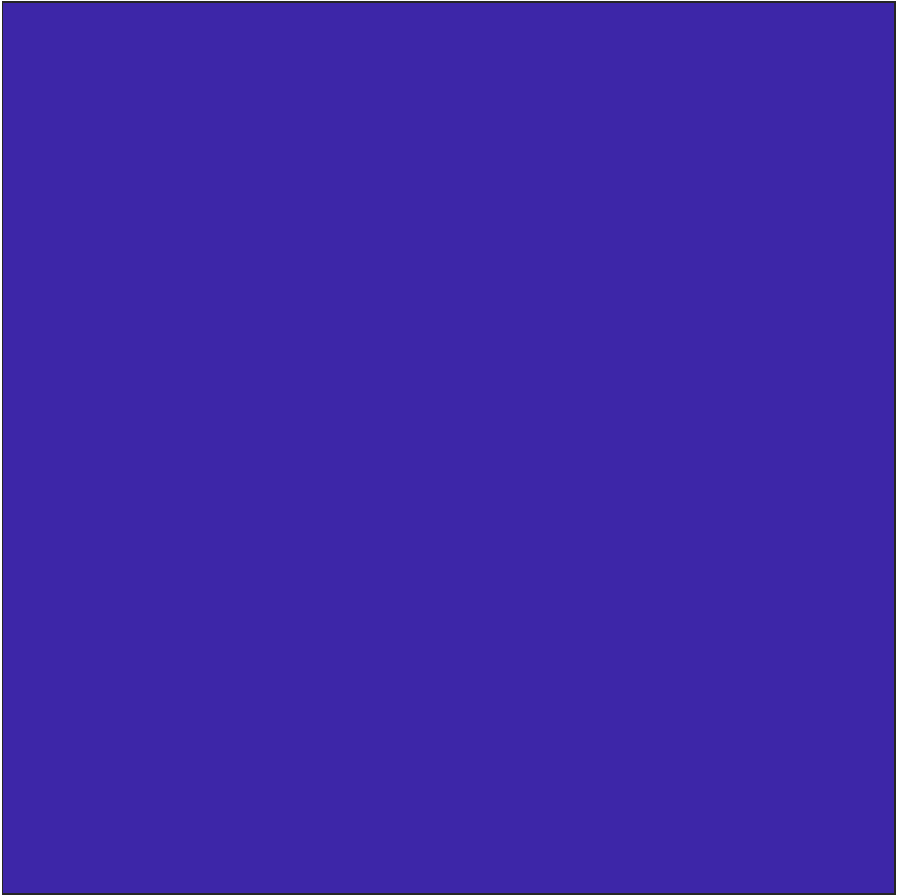}&
\hspace{0.05cm}
\includegraphics[width=0.021\textwidth]{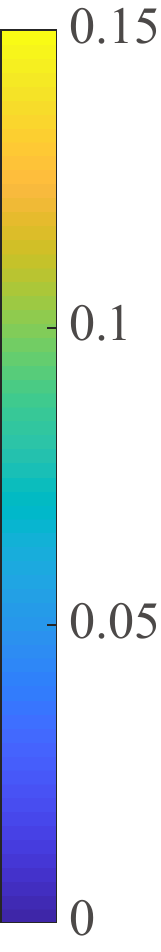}\\
\includegraphics[width=0.135\textwidth]{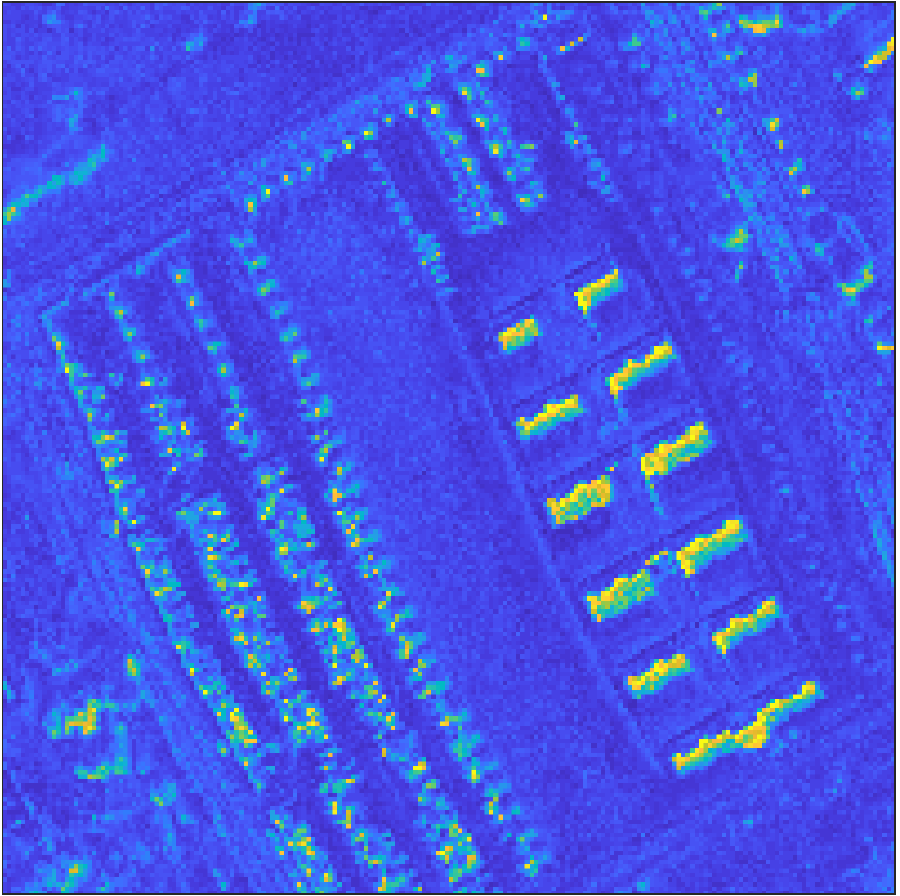}&
\includegraphics[width=0.135\textwidth]{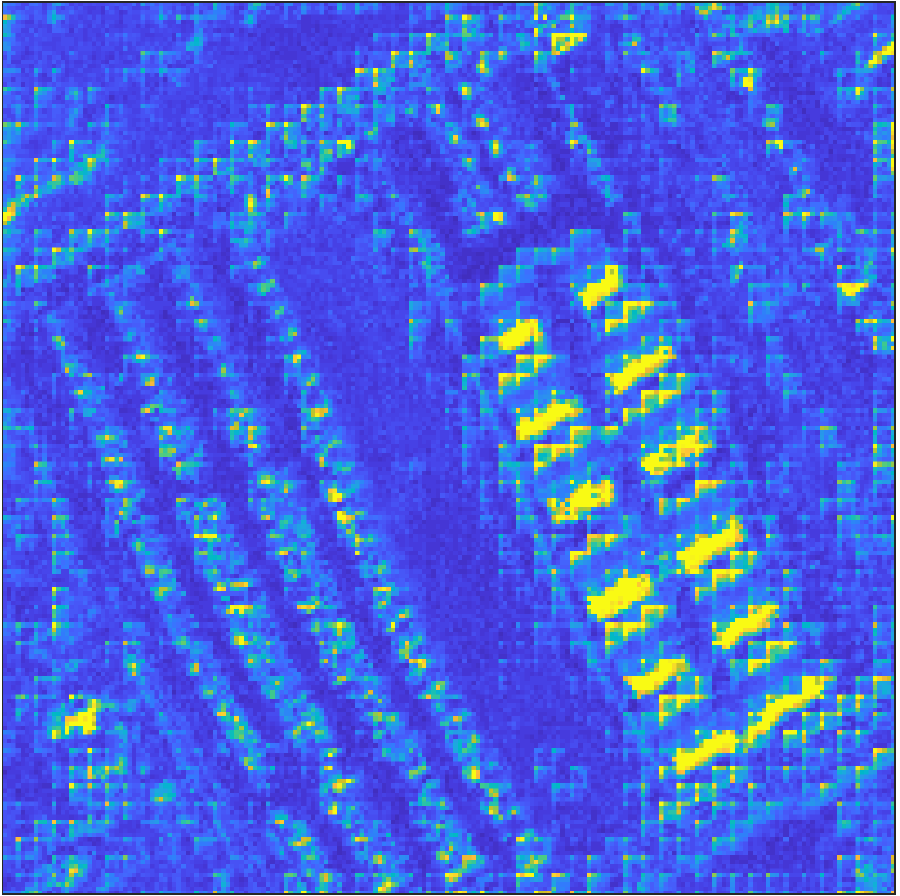}&
\includegraphics[width=0.135\textwidth]{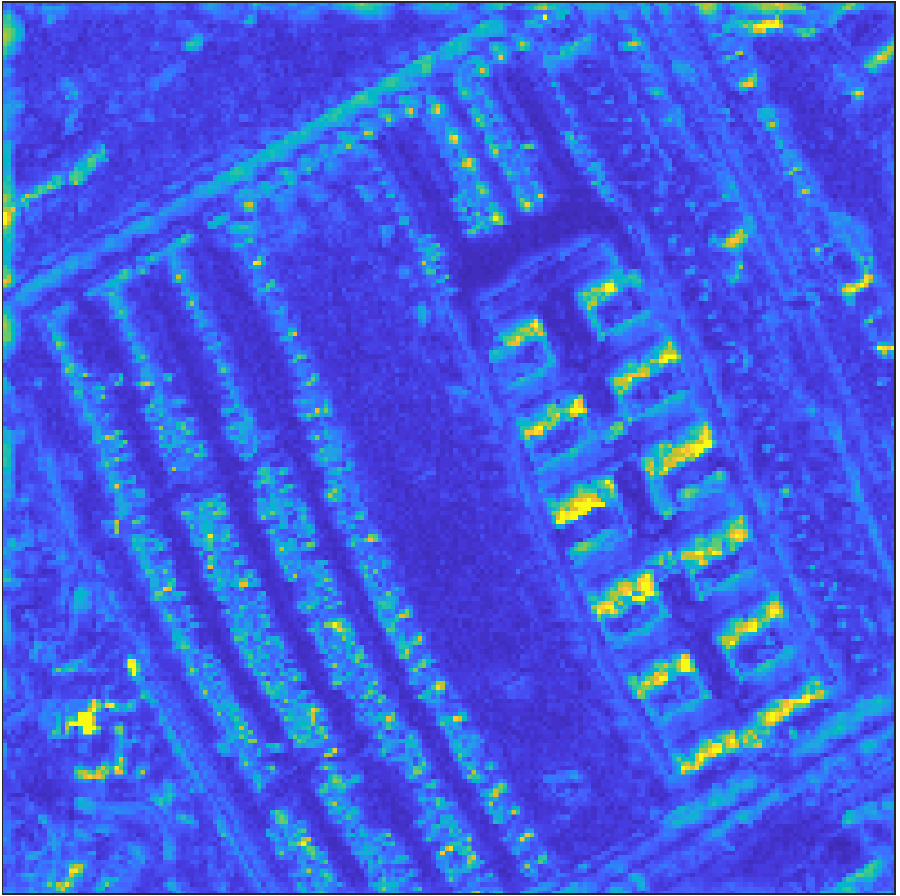}&
\includegraphics[width=0.135\textwidth]{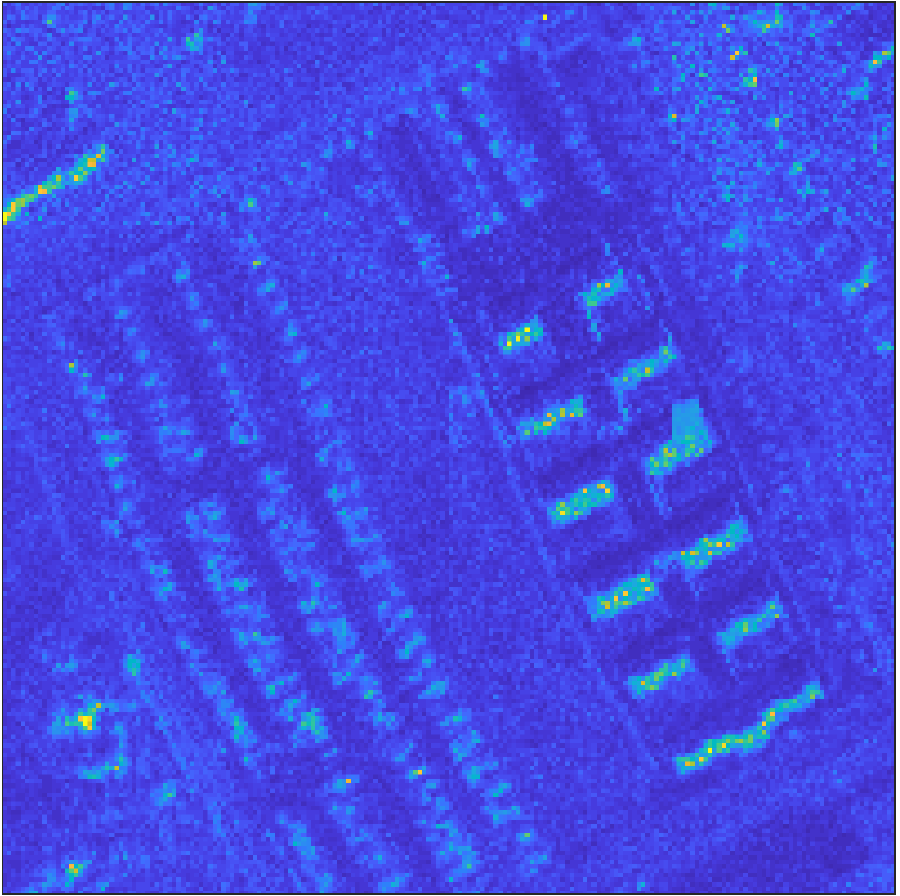}&
\includegraphics[width=0.135\textwidth]{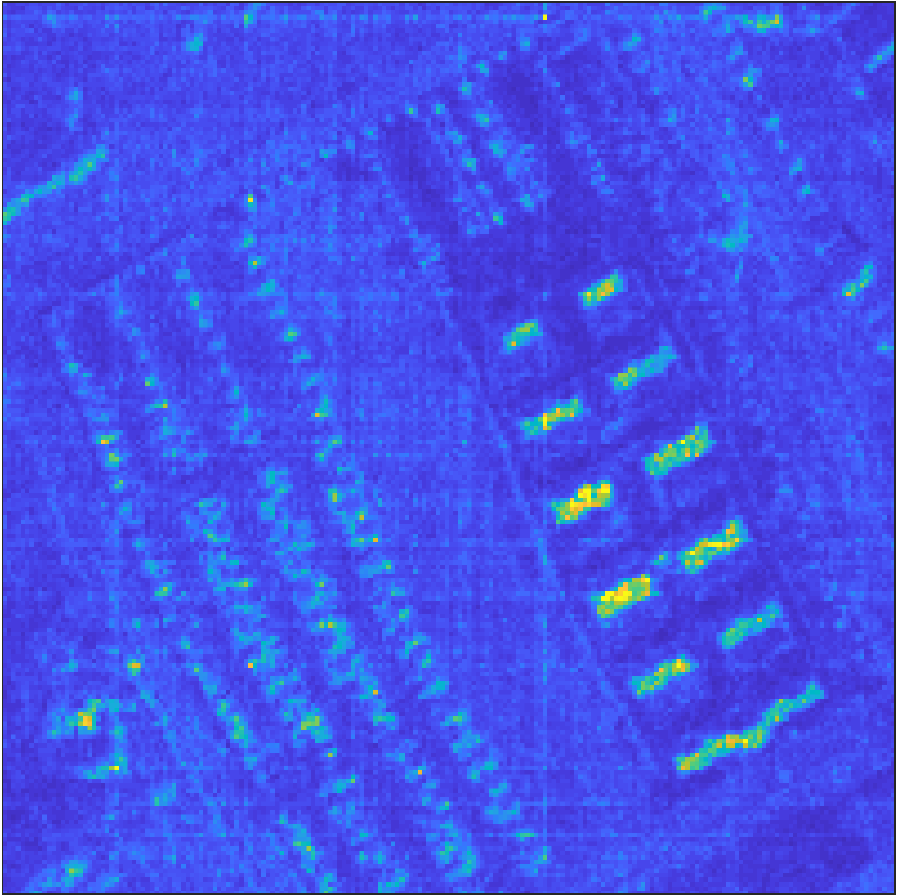}&
\includegraphics[width=0.135\textwidth]{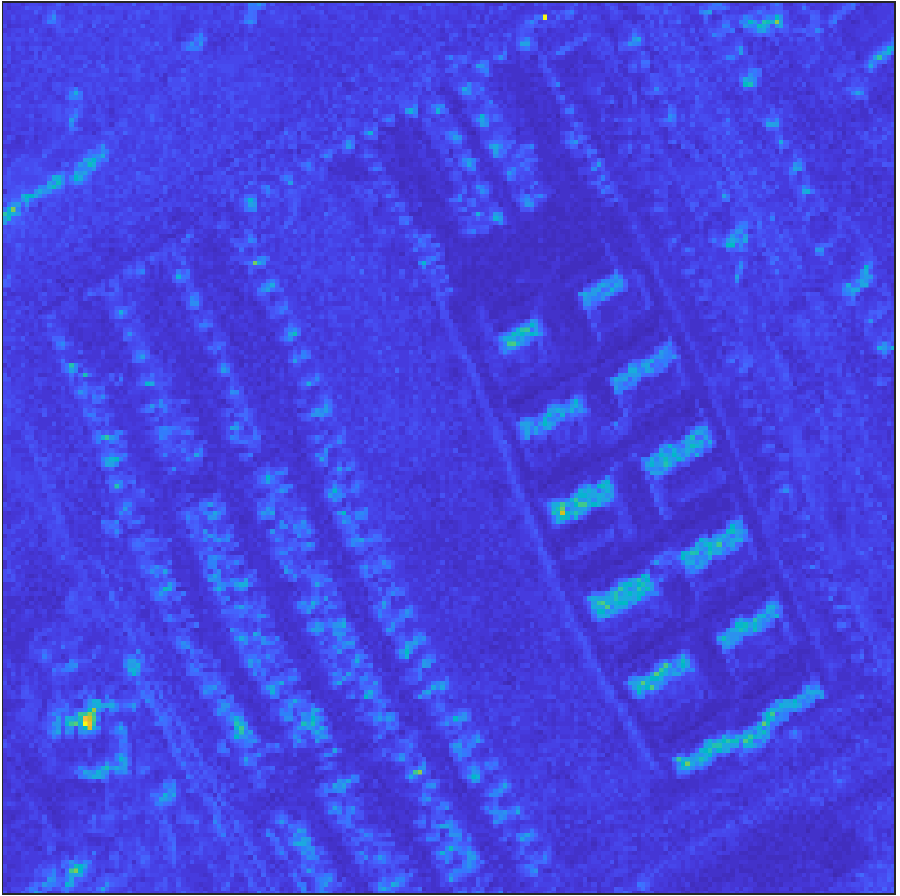}&
\includegraphics[width=0.135\textwidth]{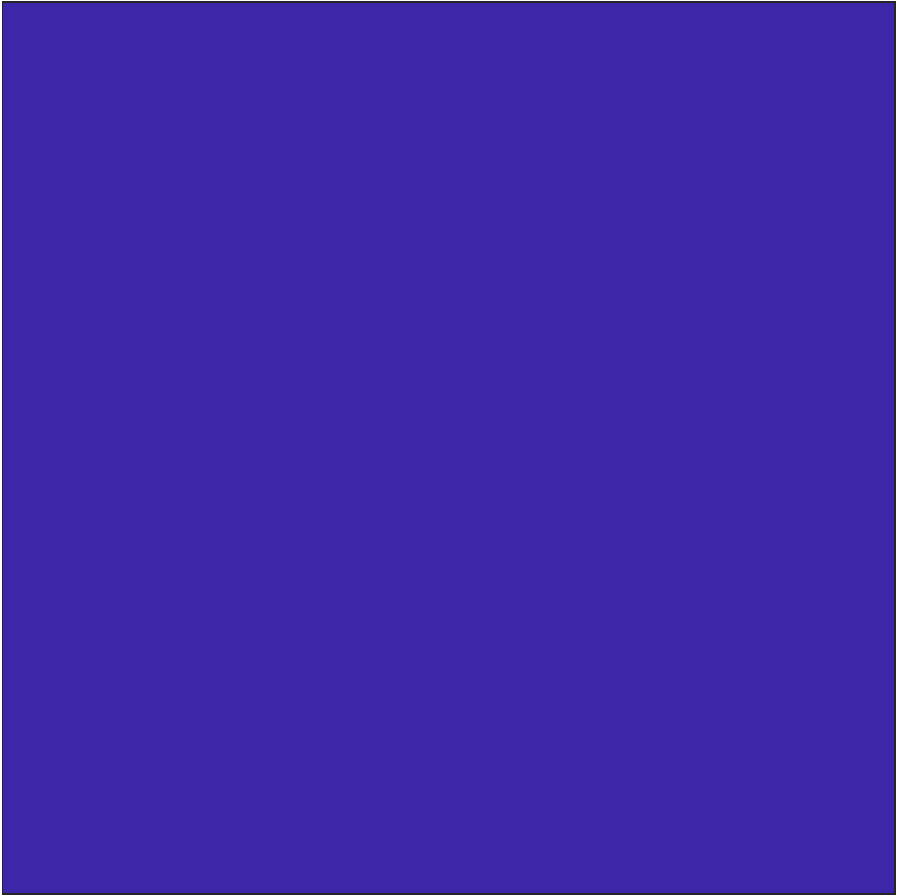}&
\hspace{0.01cm}
\includegraphics[width=0.018\textwidth]{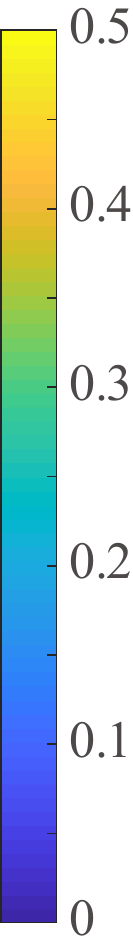}\\
(a) CNMF & (b) HySure & (c) FUSE & (d) BSCOTT & (e) BSTEREO & (f) \textsf{BSC-LL1} & (g) SRI\\
\end{tabular}
\caption{The results on the Pavia University dataset under semi-blind settings. First row: the recovered  SRIs of the 73-th band; Second row: the corresponding residual images of the 73-th band; Third row: the SAM maps.}
  \label{fig:Pavia_blind_image}
  \end{center}\vspace{-0.3cm}
\end{figure*}

In this subsection, we test the proposed \textsf{BSC-LL1} under the case where the spatial degradation matrices $\bm P_1$ and $\bm P_2$ are unknown. 
Note that the coupled CPD and coupled Tucker methods can also work under such a scenario.
Hence, we employ their ``semi-blind'' versions, namely, the BSTEREO \cite{Kanatsoulis2018HSR} and BSCOTT \cite{Prevost2020HSR} algorithms, as our baselines in the pertinent experiments. 
We use the same Pavia University and Jasper Ridge datasets as before.
All the settings remain the same, except that the spatial degradation operators are assumed to be unknown.

Fig. \ref{fig:Pavia_blind_image} shows the recovered SRIs, residual images, and the SAM maps obtained on the Pavia Unversity dataset.  One can see that the proposed \textsf{BSC-LL1} method keeps the edges of the SRI better compared to the baselines. 
This again shows our method's ability for striking a good balance between spatial smoothness and detail sharpness.

\begin{table}[!ht]
  \centering
  \caption{Performance for Pavia University with the spatial degradation unknown.}
  \resizebox{\linewidth}{!}{
    \begin{tabular}{c|c|c|c|c|c|c}\hline

    \hline
    Method (ideal)  & \multicolumn{1}{c|}{CNMF} & \multicolumn{1}{c|}{HySure} & \multicolumn{1}{c|}{FUSE}  & \multicolumn{1}{c|}{BSCOTT} & \multicolumn{1}{c|}{BSTEREO}  & \multicolumn{1}{c}{\textsf{BSC-LL1}} \\ \hline
    R-SNR $(\infty)$   & 20.23  & 15.98  & 17.18  & 24.83  & 23.91  & \textbf{26.05} \\
    SSIM $(1)$    & 0.9406 & 0.8927 & 0.9037 & 0.9554 & 0.9441 & \textbf{0.9708} \\
    CC $(1)$      & 0.9804 & 0.9447 & 0.9541 & 0.9908 & 0.9890 & \textbf{0.9932} \\
    UIQI $(1)$    & 0.9086 & 0.8537 & 0.8579 & 0.9216 & 0.9033 & \textbf{0.9420} \\
    RMSE  $(0)$   & 0.0220 & 0.0357 & 0.0311 & 0.0129 & 0.0143 & \textbf{0.0112} \\
    ERGAS $(0)$   & 0.5572 & 0.8603 & 0.7230 & 0.3180 & 0.3441 & \textbf{0.2699} \\
    SAM   $(0)$   & 0.0823 & 0.1117 & 0.0923 & 0.0626 & 0.0735 & \textbf{0.0547} \\
    \hline
    \end{tabular}}%
  \label{table:Pavia_unknown}%
\end{table}%

\begin{table}[!ht]
  \centering
  \caption{Performance for Jasper Ridge with the spatial degradation unknown.}
  \resizebox{\linewidth}{!}{
    \begin{tabular}{c|c|c|c|c|c|c}\hline

    \hline
    Method (ideal)  & \multicolumn{1}{c|}{CNMF} & \multicolumn{1}{c|}{HySure} & \multicolumn{1}{c|}{FUSE}  & \multicolumn{1}{c|}{BSCOTT} & \multicolumn{1}{c|}{BSTEREO}  & \multicolumn{1}{c}{\textsf{BSC-LL1}} \\ \hline
    R-SNR $(\infty)$   & 25.15  & 20.08  & 16.37  & 25.11  & 24.26  & \textbf{27.02} \\
    SSIM $(1)$    & 0.9531 & 0.9051 & 0.8298 & 0.9538 & 0.9303 & \textbf{0.9721} \\
    CC $(1)$      & 0.9883 & 0.9701 & 0.9523 & 0.9847 & 0.9805 & \textbf{0.9919} \\
    UIQI $(1)$    & 0.8663 & 0.7960 & 0.6899 & 0.8573 & 0.8239 & \textbf{0.8982} \\
    RMSE $(0)$    & 0.0161 & 0.0288 & 0.0441 & 0.0161 & 0.0178 & \textbf{0.0129} \\
    ERGAS  $(0)$  & 0.3819 & 0.6302 & 0.8286 & 0.4326 & 0.5844 & \textbf{0.3340} \\
    SAM   $(0)$   & 0.0808 & 0.1168 & 0.1386 & 0.0883 & 0.1039 & \textbf{0.0696} \\
    \hline
    \end{tabular}}%
  \label{table:Ridge_unknown}%
\end{table}%

Tables \ref{table:Pavia_unknown} and \ref{table:Ridge_unknown} show the corresponding evaluation results on the Pavia Unversity dataset and the Jasper Ridge dataset, respectively. Again, \textsf{BSC-LL1} performs better than other methods under all metrics.
For both datasets, the R-SNR performance of the proposed method largely exceeds that of the baselines.

\begin{figure*}[!ht]
\scriptsize\setlength{\tabcolsep}{0.3pt}
\begin{center}
\begin{tabular}{cccccccc}
\includegraphics[width=0.135\textwidth]{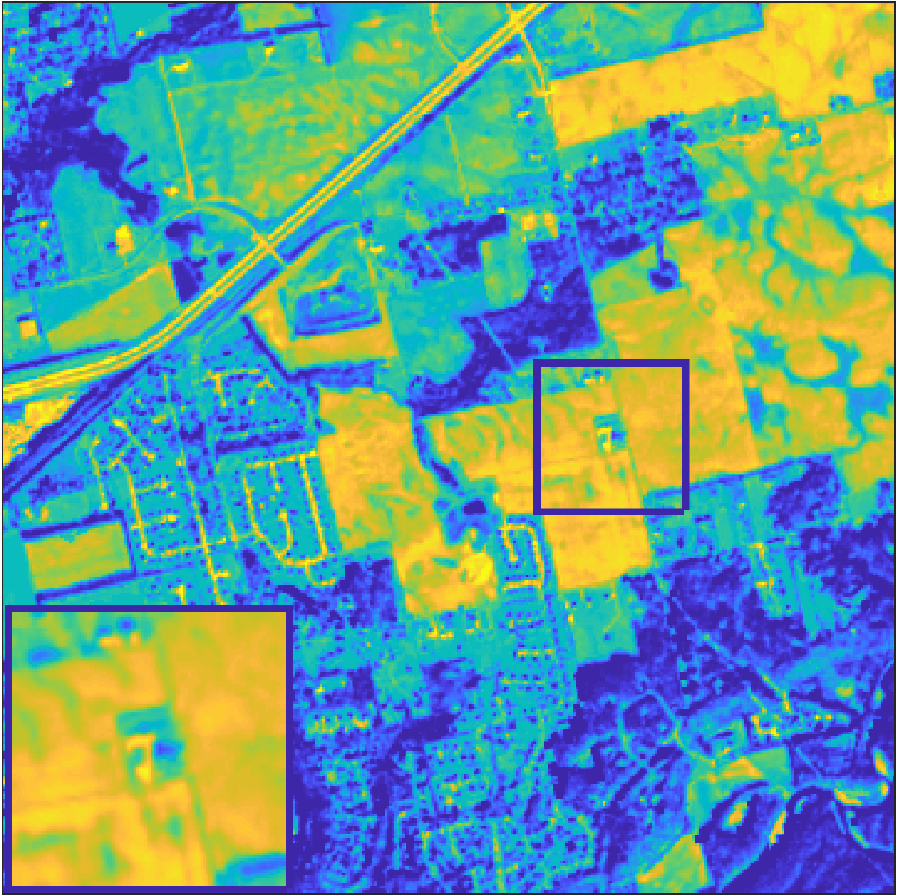}&
\includegraphics[width=0.135\textwidth]{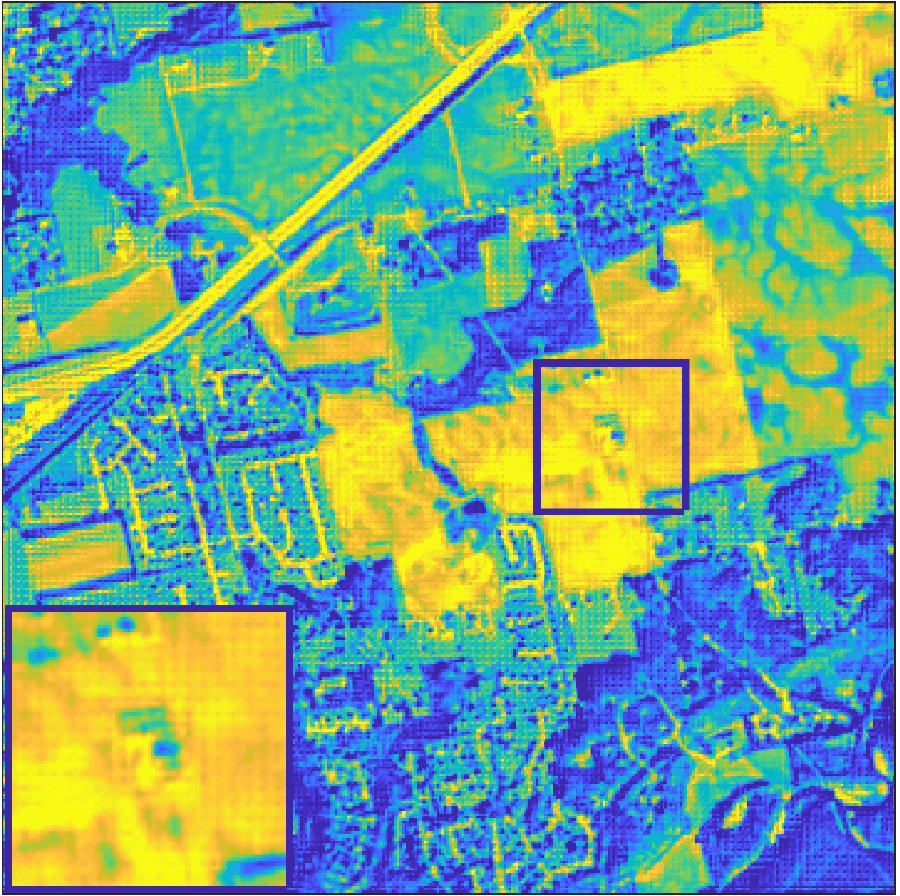}&
\includegraphics[width=0.135\textwidth]{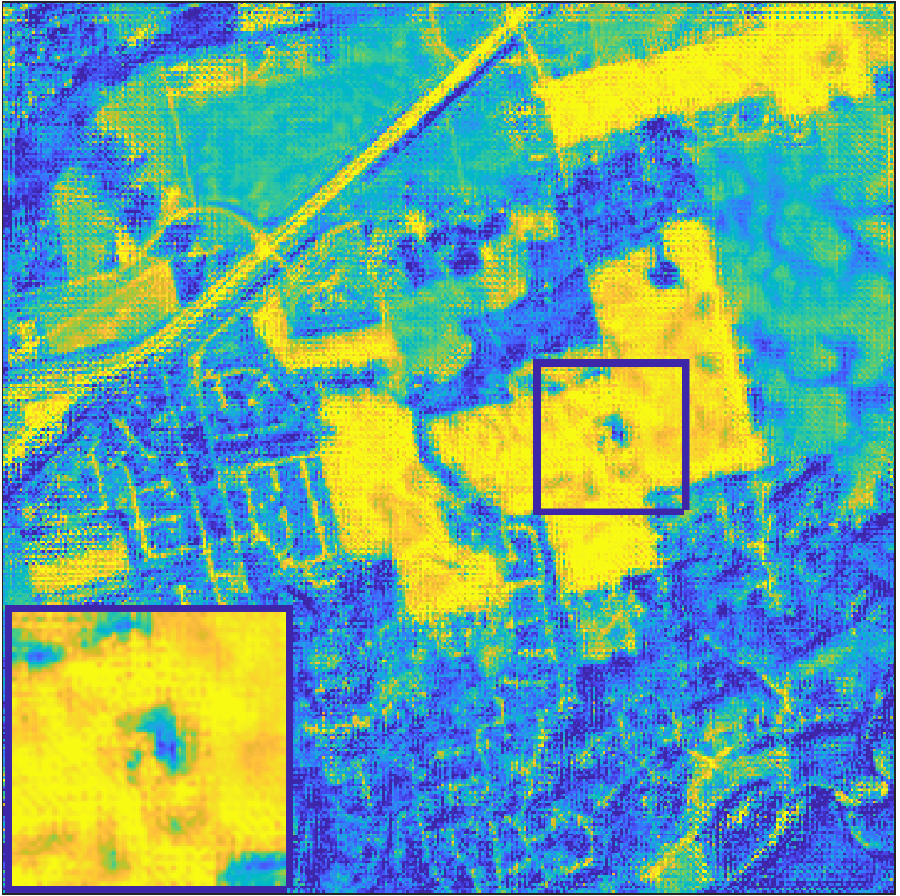}&
\includegraphics[width=0.135\textwidth]{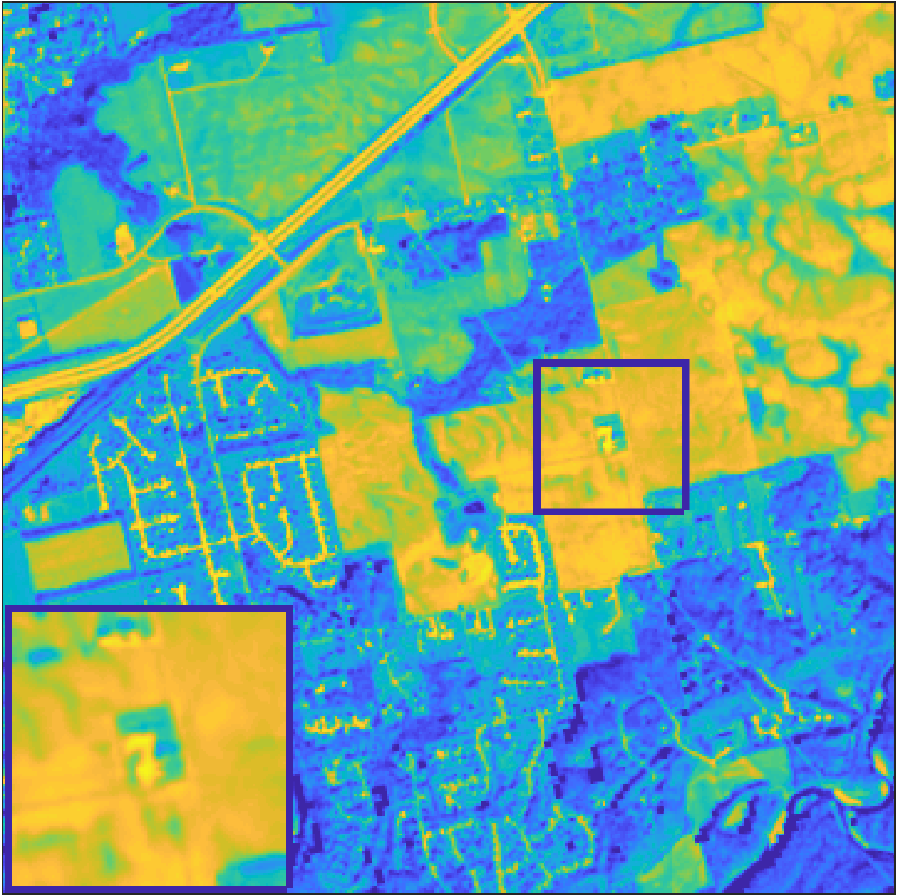}&
\includegraphics[width=0.135\textwidth]{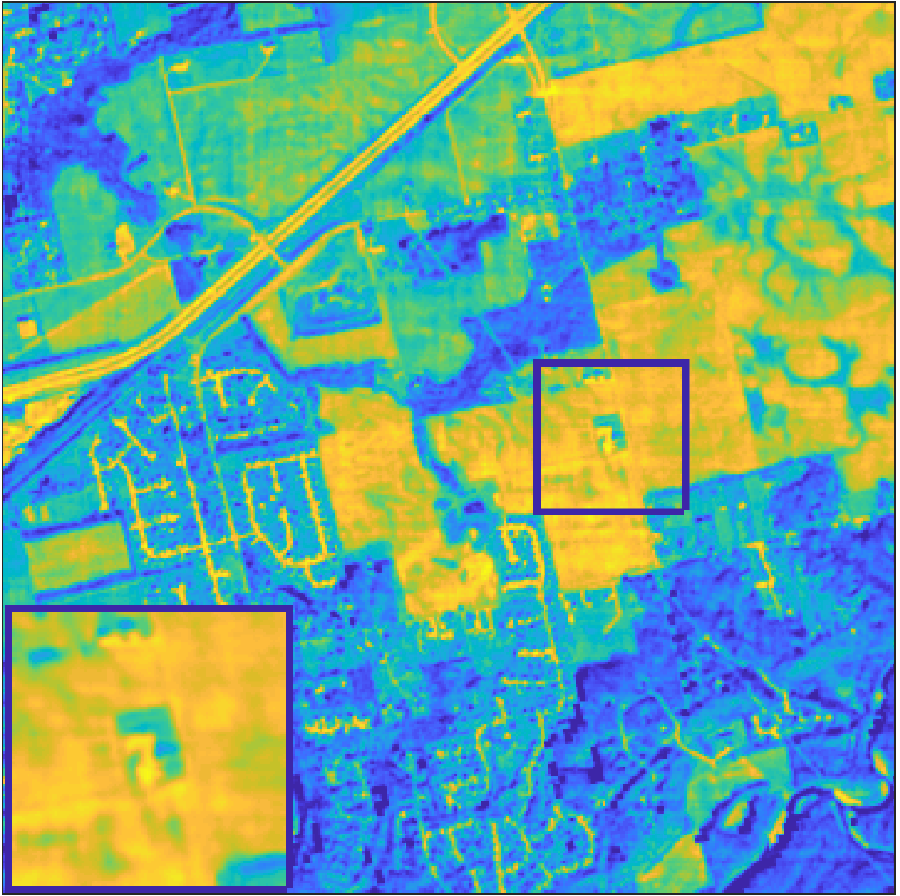}&
\includegraphics[width=0.135\textwidth]{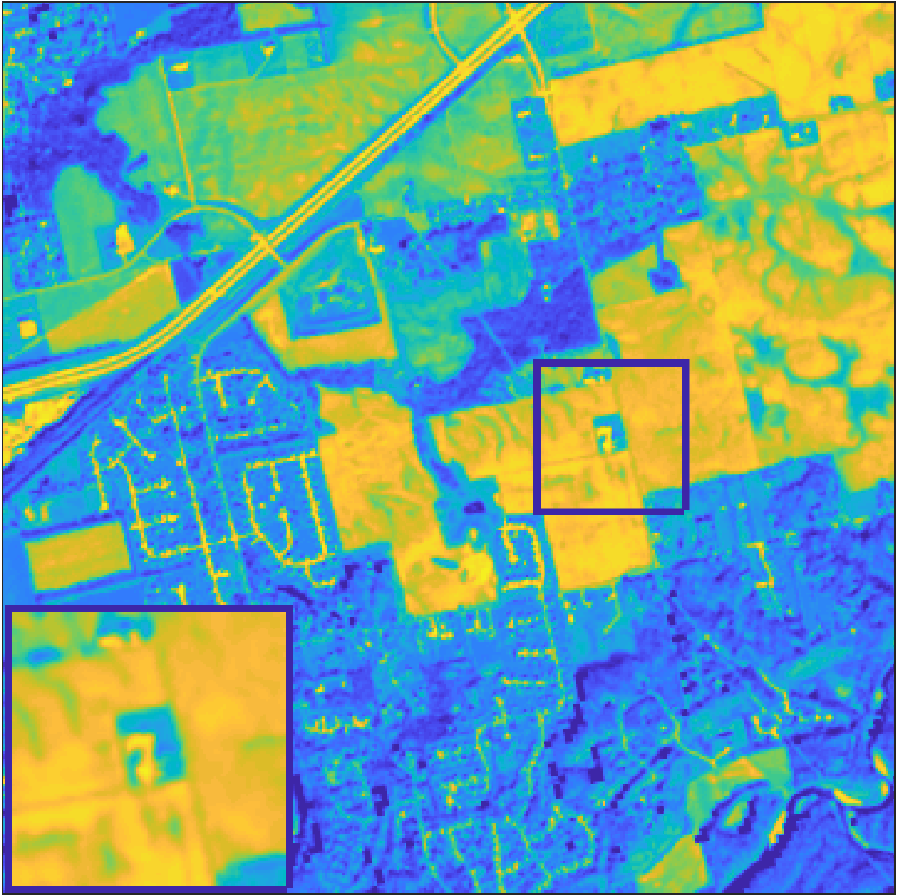}&
\includegraphics[width=0.135\textwidth]{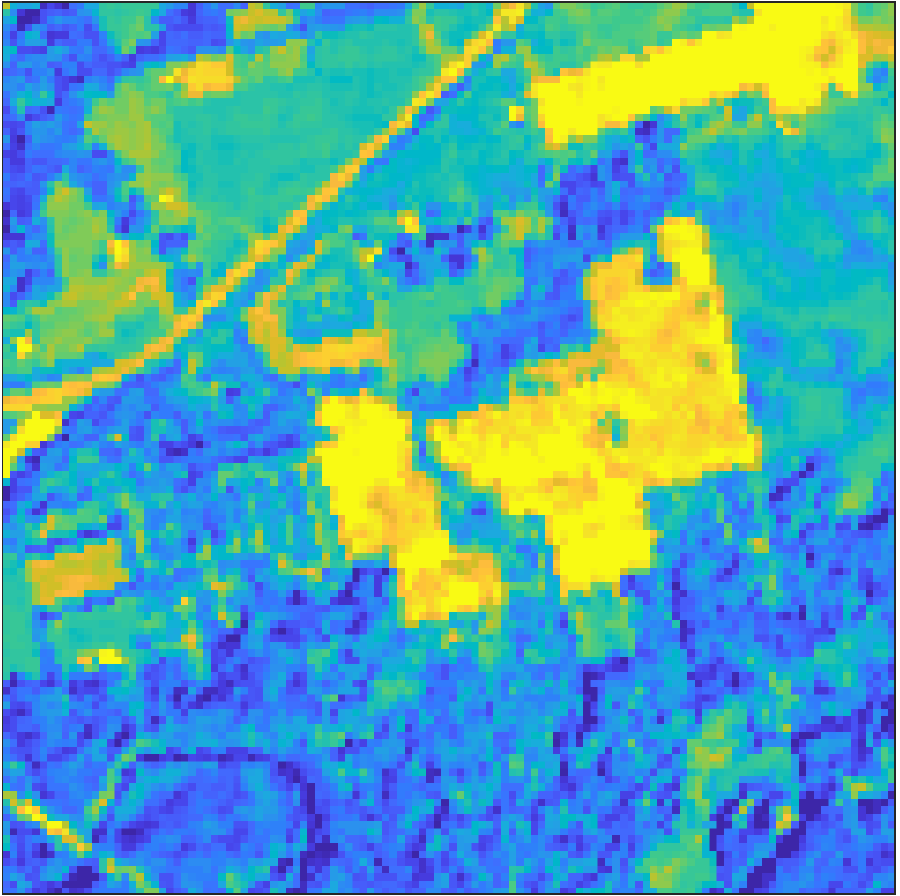}&
\hspace{0.05cm}
\includegraphics[width=0.018\textwidth]{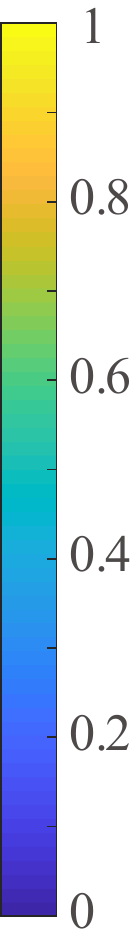}\\
\includegraphics[width=0.135\textwidth]{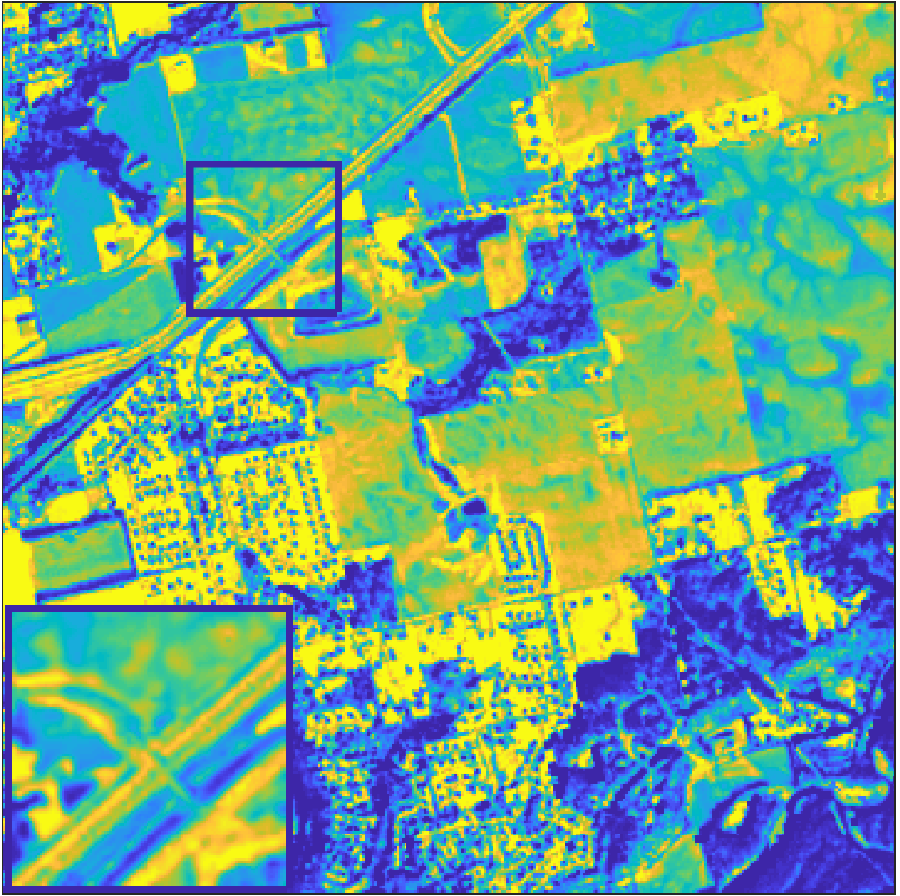}&
\includegraphics[width=0.135\textwidth]{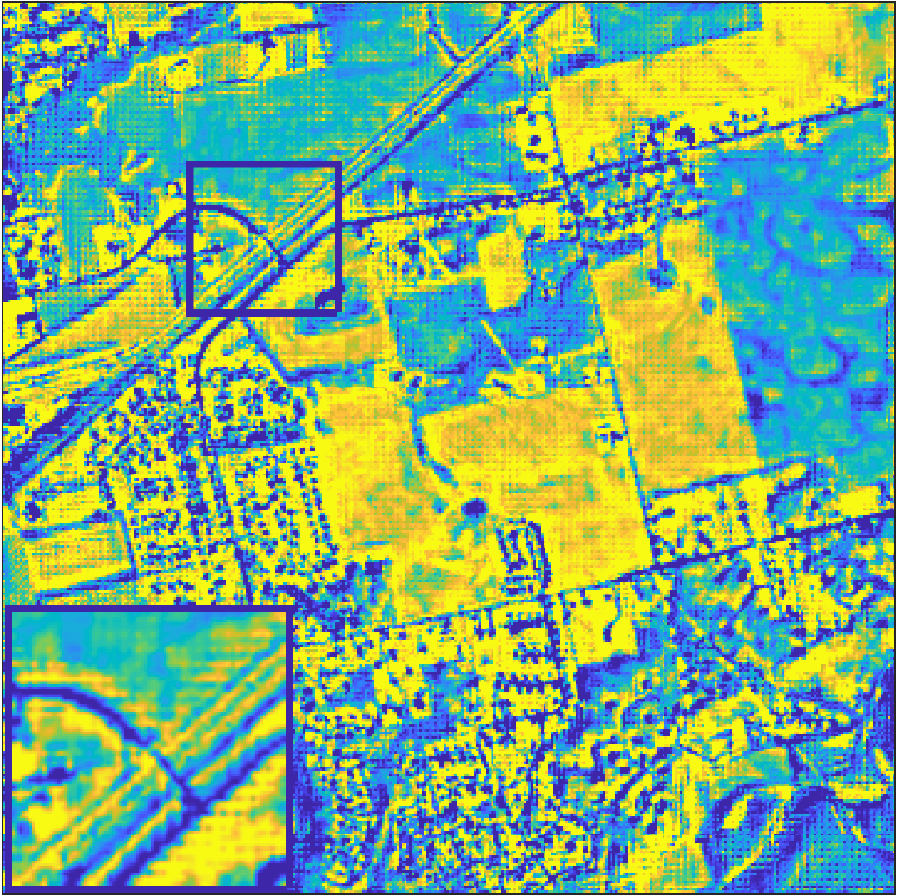}&
\includegraphics[width=0.135\textwidth]{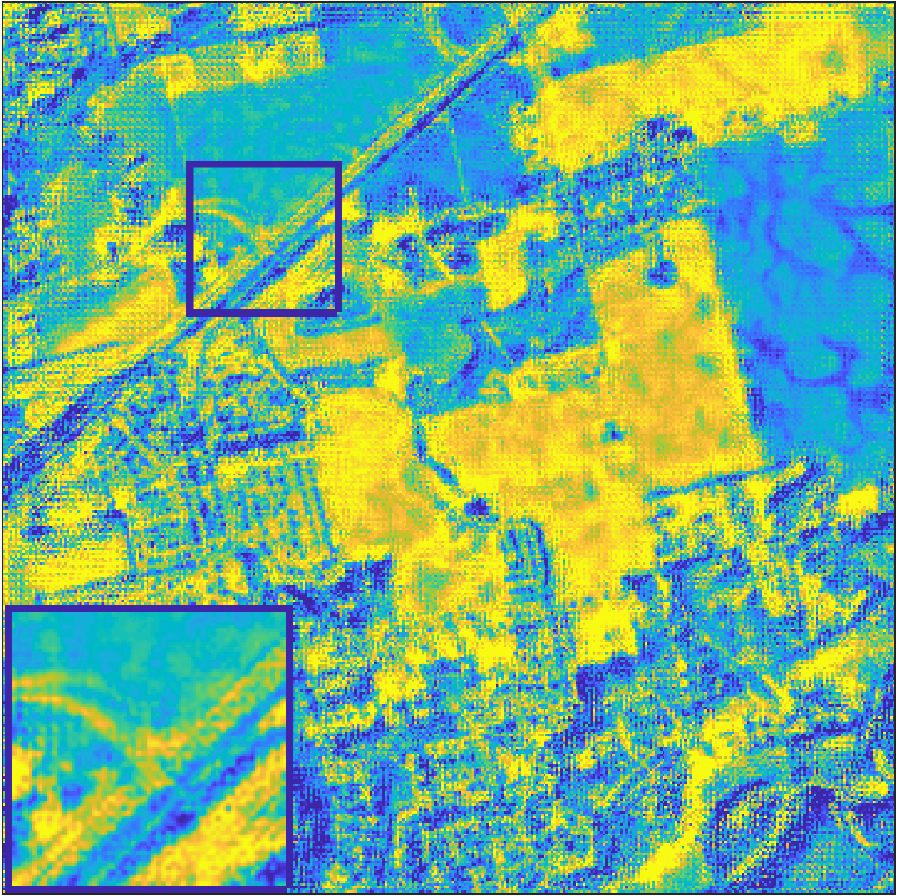}&
\includegraphics[width=0.135\textwidth]{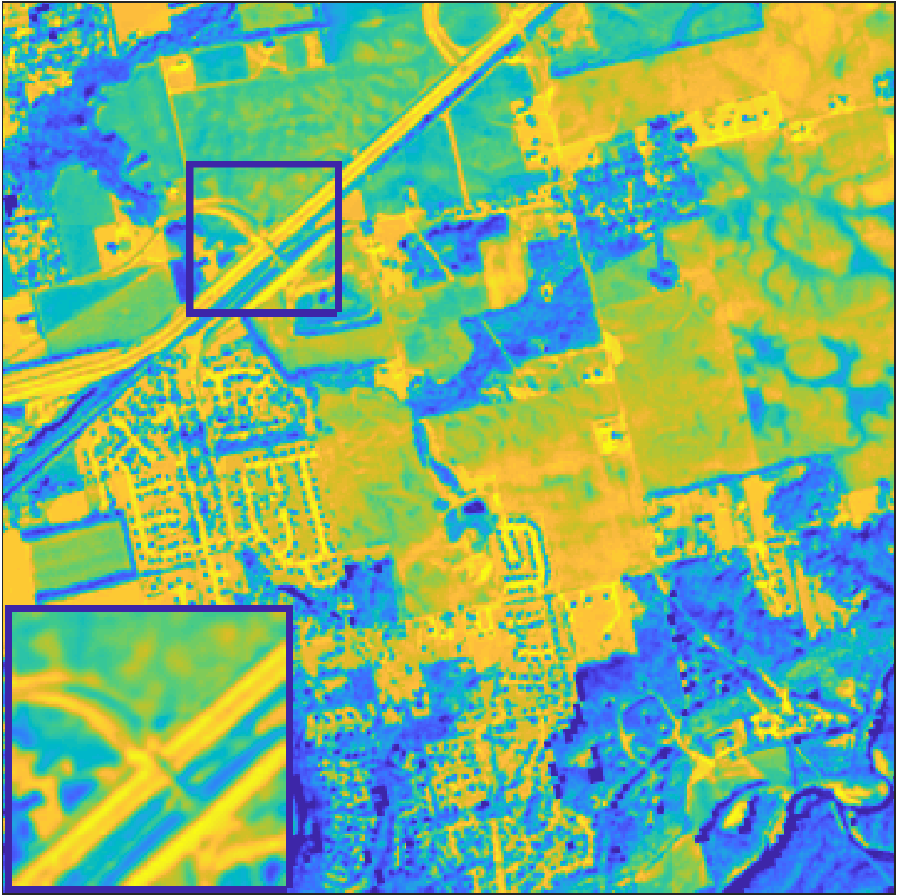}&
\includegraphics[width=0.135\textwidth]{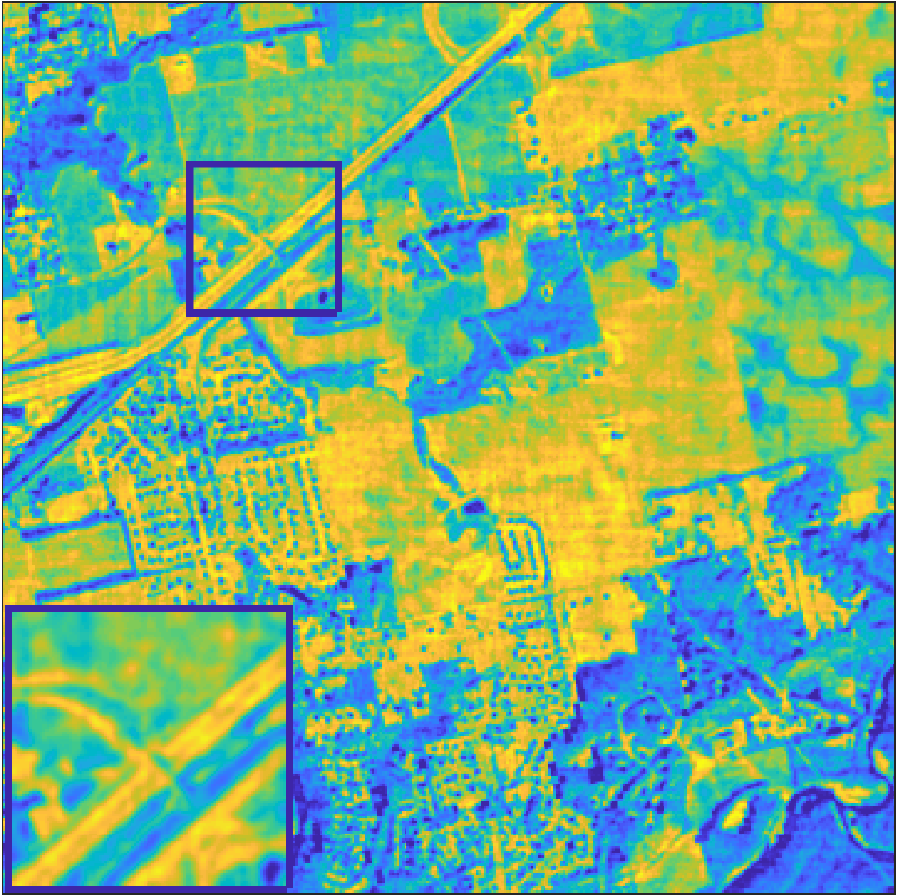}&
\includegraphics[width=0.135\textwidth]{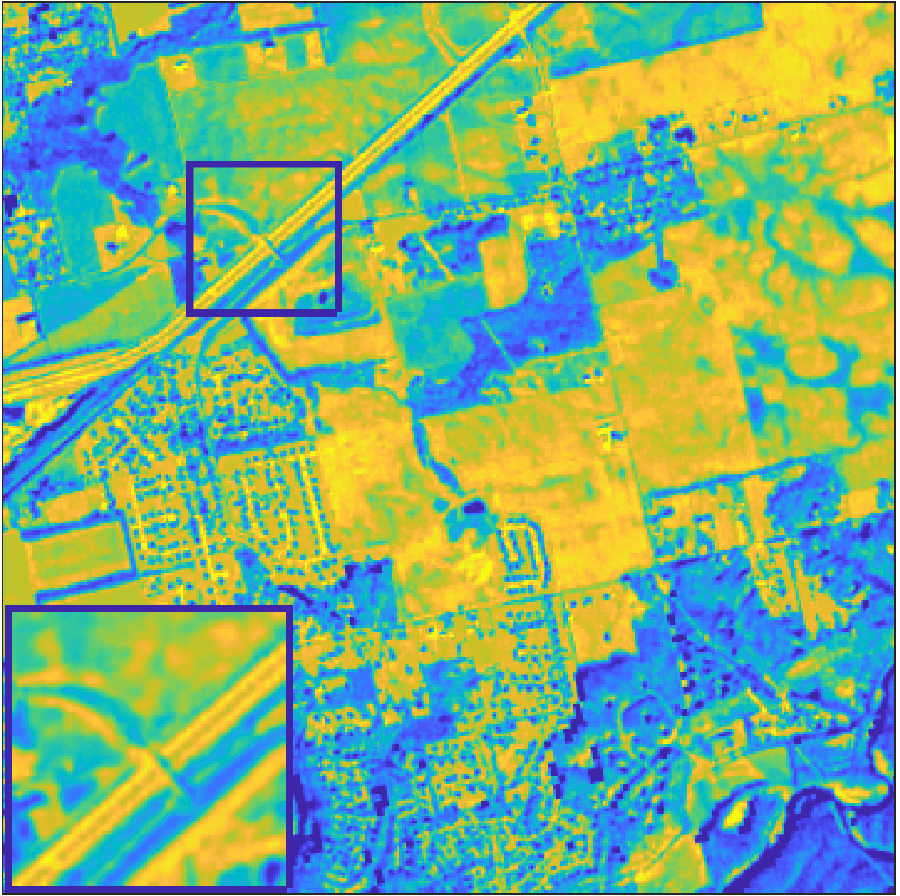}&
\includegraphics[width=0.135\textwidth]{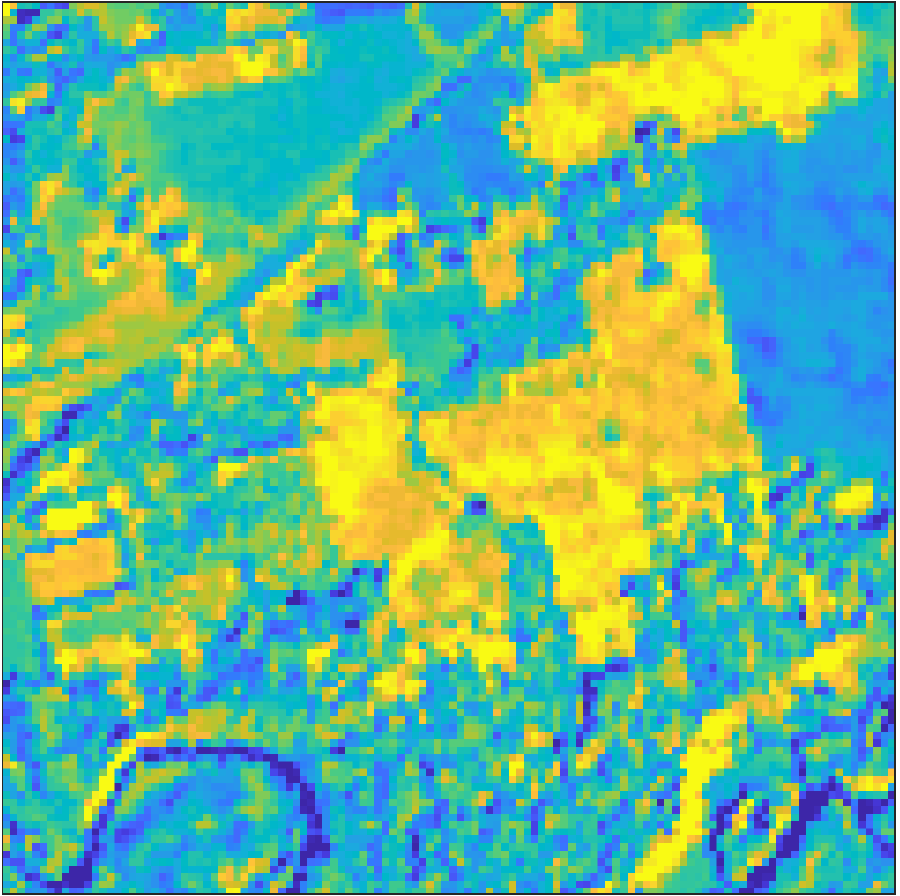}&
\hspace{0.05cm}
\includegraphics[width=0.018\textwidth]{figs/Real_colormap.pdf}\\
(a) CNMF & (b) HySure & (c) FUSE & (d) BSCOTT & (e) BSTEREO & (f) \textsf{BSC-LL1} & (g) HSI \\
\end{tabular}
\caption{The recovered  SRIs of real-data experiment. First row: the 23-th band of the recovered SRIs; Second row: the 50-th band of the recovered SRIs.}
  \label{fig:real_data}
  \end{center}\vspace{-0.3cm}
\end{figure*}

\subsection{Real-data Experiment}
We also test the algorithms on real co-registered hyperspectral and multispectral images, where are offered in a recent work \cite{Yang2018HSR}. The HSI data with a size of $120\times 120\times 89$ is acquired by the Hyperison sensor. The 89 spectral bands are obtained after removing heavily corrupted bands. The MSI data with a size of $360\times 360\times 4$ is acquired by the Sentinel-2A satellite, which provides MSIs with 13 bands. We 
follow the demosntration method in \cite{Wu2019Hyperspectral} and select 4 bands whose pixels are 10m$\times$10m areas on the ground. The central wavelengths of these 4 bands are 490 nm, 560 nm, 665 nm, and 842 nm, respectively.  Note that the Sentinel-2A image has more than 4 bands, but the other bands admit different spectral resolutions and thus are excluded in this experiment.  

Note that both the spatial and spectral degradation operators are unknown in the real experiment. Hence, we apply the proposed \textsf{BSC-LL1} for this dataset. As for the tensor-based baselines, we employ the semi-blind versions, i.e., BSCOTT \cite{Prevost2020HSR} and BSTEREO \cite{Kanatsoulis2018HSR}. For the matrix-based baselines, i.e., CNMF \cite{Yokoya2012HSR}, HySure \cite{Simoes2015HSR}, FUSE \cite{Wei2015HSRSylvester}, we use their own modules for estimating $\bm P_1$, $\bm P_2$ and $\bm P_M$ from the input HSI and MSI.
The spectral degradation matrix $\bm{P}_{M}$ for the tensor-based methods is estimated from the available HSI and MSI images using the algorithm in \cite{Simoes2015HSR}. For the number of materials, we set $R=4$, which is by visual inspection.

Fig. \ref{fig:real_data} shows the recovered performance of real data at two selected bands, namely, the 23-th and 50-th bands---which are not included in the MSI. 
One can see that all the algorithms work to a certain extent, outputting higher spatial resolution images compared to the HSI (see column (g)).
Nevertheless, one can observe that the FUSE and HySure algorithms both have undesired strip noise in the recovered SRI. The CNMF obtains sharp edges, which indicates satisfactory performance.
However, the recovered SRIs by CNMF loses some objects and details in the spatial domain. 
The tensor based baselines, i.e., BSCOTT and BSTEREO, both work reasonably well.
However, the proposed \textsf{BSC-LL1} seems to perform better in terms of keeping edges and smoothness of the recovered SRI; please see the zoom-in regions of recovered images.

\section{Conclusion}
\label{conclusion}
In this paper, we proposed a coupled {\sf LL1} block-term tensor decomposition framework for the HSR problem. 
We showed that the SRI recoveriability is guaranteed under mild conditions. Hence, using the advocated {\sf LL1} modeling instead of the CPD/Tucker modeling for spectral images does not lose theoretical guarantees.
Unlike the existing tensor-based approaches, a salient feature of the proposed framework is that the latent factors under the {\sf LL1} model admit physical meaning. Therefore, structural constraints and regularization terms that reflect prior information about the latent factors can be flexibly incorporated in our HSR formulations---thereby enhancing performance under noisy scenarios. We recast the proposed HSR criteria into optimization-friendly forms that take into consideration of the physical characterizations of the endmembers' spectral signatures and abundance maps (e.g., nonnegativity and spatial smoothness), and proposed a convergence-guaranteed accelerated alternating gradient projection algorithmic framework to tackle these problems.
We tested the proposed algorithms over extensive semi-real and real experiments, which both showed consistently promising performance on HSR tasks.


\bibliographystyle{IEEEtran}
\bibliography{SCBTD}

\clearpage
\appendices
{\bf Supplementary Materials for ``Hyperspectral Super-Resolution via InterpretableBlock-Term Tensor Modeling'' \\
M. Ding, X. Fu, T.-Z. Huang, J. Wang, and X.-L. Zhao
}

\section{Proof of Theorem \ref{the:non_blind_BTD}}
\label{app:identifiability_non_blind}
To proceed, let us consider the following important lemmas. 
\begin{lemma} \label{lemma:joint} \cite{Kanatsoulis2018HSR}
Let $\widetilde{\bm{A}}=\bm{P}\bm{A}$, where $\bm{A}$ is drawn from any absolutely continuous joint distribution in $\mathbb{R}^{I\times L}$ and $\bm{P}\in \mathbb{R}^{I'\times L}$ is full row rank. Then, $\widetilde{\bm{A}}$ follows a joint absolutely continuous joint distribution in $\mathbb{R}^{I'\times L}$.
\end{lemma}
\begin{lemma} \label{lemma:LL1} \cite{Lathauwer2008BTD2}
Let $(\{\bm{A}_{r}\in \mathbb{R}^{I \times L_{r}},\bm{B}_{r}\in \mathbb{R}^{J \times L_{r}}\}_{r=1}^{R},\bm{C}\in \mathbb{R}^{K \times R})$ be the latent factors of the {\sf LL1} tensor $\underline{\bm{Y}} = \sum_{r=1}^R (\A_r\B_r^\T)\circ \C(:,r)$ where $L_r=L$. Assume that $\bm{A}_{r}$, $\bm{B}_{r}$, and $\bm{C}$ are drawn from any absolutely continuous distributions. Then, the {\sf LL1} decomposition of $\underline{\bm Y}$ is essentially unique almost surely, if $IJ\geq L^{2}R$ and
\[
\min \left(\left\lfloor\frac{I}{L} \right\rfloor, R \right)+\min \left(\left\lfloor\frac{J}{L} \right\rfloor, R \right)+\min(K,R)\geq 2R+2.
\]
Here, essentially uniqueness means that if $\underline{\bm Y}=\sum_{r=1}^R (\A_r^\ast (\B_r^\ast)^\T)\circ \C^\ast(:,r)$, then, we must have
\[  \bm S^\ast = \bm S \bm \Pi\bm \Lambda,~\bm C^\ast = \bm C \bm \Pi\bm \Lambda^{-1}, \]
where $\bm \Pi$ and $\bm \Lambda$ denote a permutation matrix and a nonsingular scaling matrix, respectively, $\S^{\star}=[\textrm{vec}(\S_{1}^{\star}),\ldots,\textrm{vec}(\S_{R}^{\star})]$, $\S_{r}^{\star}=\A_{r}^{\star} \left(\B_{r}^{\star}\right)^{\top}$ and $\S=[\textrm{vec}(\S_{1}),\ldots,\textrm{vec}(\S_{R})]$, $\S_{r}=\A_{r}\B_{r}^{\top}$.
\end{lemma}

\bigskip

Let $(\{\A_{r},\B_{r}^{\top}\}_{r=1}^{R},\C)$ denote the ground-truth factors of the SRI tensor. 
Assume that $(\{\A_{r}^{\star},\left(\B_{r}^{\star}\right)^{\top}\}_{r=1}^{R},\C^{\star})$ represents an optimal solution of the formulation \eqref{Non_blind_BTD_model}. 
Then, we have
\begin{align}
    \underline{\bm{Y}}_M &= \sum_{r=1}^R \left( \A^\star_r (\B_r^\star)^\T  \right)\circ \bm P_M\C^\star(:,r) \nonumber \\
    &= \sum_{r=1}^R \left( \A_r \B_r^\T  \right)\circ \bm P_M\C(:,r),  \label{eq:optimal_Ym}\\
    \underline{\bm{Y}}_H & = \sum_{r=1}^R \left( \bm P_1\A^\star_r (\bm P_2\B_r^\star)^\T  \right)\circ \C^\star(:,r)  \nonumber \\
    &= \sum_{r=1}^R \left( \bm P_1\A_r (\bm P_2\B_r)^\T  \right)\circ \C(:,r). \label{eq:optimal_Yh}
\end{align}
To proceed, note that $\bm P_1$, $\bm P_2$ and $\bm P_M$ all have full row rank.
By Lemma \ref{lemma:joint}, $\bm{P}_{1}\A$, $\bm{P}_{2}\B$, and $\bm{P}_{M}\C$ all follow certain joint absolutely continuous distributions. 
Therefore, under the conditions $I_{M}J_{M}\geq L^{2}R$ and
\[\min \left(\left\lfloor\frac{I_{M}}{L} \right\rfloor, R \right)+\min \left(\left\lfloor\frac{J_{M}}{L} \right\rfloor, R \right)+\min(K_{M},R)\geq 2R+2,\]
and by Lemma \ref{lemma:LL1}, the {\sf LL1} decomposition of $\underline{\bm Y}_M$ is essentially unique---which means that from \eqref{eq:optimal_Ym} one can conclude the following:
\begin{align}\label{eq:sstar}
\S^{\star}=\S\mathbf{\Pi} \mathbf{\Lambda},\ \bm{P}_{M}\C^{\star}=\bm{P}_{M}\C\mathbf{\Pi} \mathbf{\Lambda}^{-1},    
\end{align}
where $\S^{\star}=[\textrm{vec}(\S_{1}^{\star}),\ldots,\textrm{vec}(\S_{R}^{\star})]$, $\S_{r}^{\star}=\A_{r}^{\star} \left(\B_{r}^{\star}\right)^{\top}$ and $\S=[\textrm{vec}(\S_{1}),\ldots,\textrm{vec}(\S_{R})]$, $\S_{r}=\A_{r}\B_{r}^{\top}$.
Define $\widetilde{\bm S}=[ (\bm P_2\B_1 \odot_c \bm P_1\A_1) \bm 1_L\ldots,(\bm P_2\B_R\odot_c \bm P_1\A_R)\bm 1_L ]$.
One can easily see that
\begin{equation}\label{eq:stilde}
    \widetilde{\S}^{\star}= \widetilde{\S}\mathbf{\Pi} \mathbf{\Lambda} ,   
\end{equation}  
where 
$ \widetilde{\bm S}^\star = [ (\bm P_2\B_1^\star \odot_c \bm P_1\A_1^\star) \bm 1_L\ldots,(\bm P_2\B_R^\star\odot_c \bm P_1\A_R^\star)\bm 1_L ].$

Next, we show that $\widetilde{\bm S}$ has full column rank almost surely. To see this, note that the matrix
\[  \bm P_2\bm B \odot \bm P_1\bm A = [\bm P_2\B_1 \otimes \bm P_1\A_1,\ldots,\bm P_2\B_R\otimes \bm P_1\A_R ]       \]
admits full column rank almost surely if all $\A_r$ and $\B_r$ are drawn from any joint absolutely continuously distribution and if $I_{H}J_{H}\geq LR$ \cite[Lemma 3.3]{Lathauwer2008BTD1}, where we have used the notation $\bm B=[\bm B_1,\ldots,\bm B_R]$ and $\A= [\A_1,\ldots,\A_R]$.
Since the matrix $$ \widetilde{\bm P}=[\bm P_2\B_1 \odot_c \bm P_1\A_1,\ldots,\bm P_2\B_R\odot_c \bm P_1\A_R ]    $$ is a submatrix of $\bm P_2\bm B \odot \bm P_1\bm A$, it also has full column rank.
Note that 
\[  \widetilde{\bm S} = [ (\bm P_2\B_1 \odot_c \bm P_1\A_1) \bm 1_L\ldots,(\bm P_2\B_R\odot_c \bm P_1\A_R)\bm 1_L ].      \]
Our claim is that $\widetilde{\bm S}$ has to be full column rank. This can be seen by contradiction. Suppose that $\widetilde{\bm S}$ is rank deficient.
Then, there exists $(\alpha_1,\ldots,\alpha_R)\neq \bm 0$ such that
\begin{equation}\label{eq:contra}
    (\bm P_2\B_1 \odot_c \bm P_1\A_1) \bm 1_L \alpha_1+ \ldots + (\bm P_2\B_R\odot_c \bm P_1\A_R)\bm 1_L\alpha_R =\bm 0. 
\end{equation}
However, \eqref{eq:contra} also means that the columns of $\widetilde{\bm P}$ (and thus those of $\bm P_2\bm B \odot \bm P_1\bm A$) are linearly dependent, which is a contradiction.

To continue, we rearrange the tensor using the following rule:
\begin{align}\label{eq:unfold}
\bm{Y}_{H}=\left[\underline{\Y}_{H}(1,1,:),\underline{\Y}_{H}(2,1,:),\ldots,\underline{\Y}_{H}(I_H,J_H,:)\right]^{\top}.    
\end{align}
The above matrix can also be represented as \cite{Lathauwer2008BTD2}:
\[    \bm Y_H = \widetilde{\bm S}\bm C^\T. \] 
By our identification criterion and \eqref{eq:optimal_Yh}, we also have
\[  \bm Y_H = \widetilde{\bm S}^\star (\bm C^\star) ^\T. \]
Combining the above two equalities and using \eqref{eq:stilde}, we have
\begin{align}
    &\widetilde{\bm S}\bm C^\T = \widetilde{\bm S}\bm \Pi\bm \Lambda (\bm C^\star)^\T \nonumber\\
    \Longrightarrow& \bm C^\star = \bm C\bm \Pi\bm \Lambda^{-1}, \label{eq:cstar}
\end{align}
where we have used the fact that $\widetilde{\bm S}$ has full column rank.

Combining \eqref{eq:sstar} and \eqref{eq:cstar}, one can see that the matrix unfolding of the SRI (following the same unfolding rule as in \eqref{eq:unfold}), i.e., $\bm Y_S$, can be recovered by
\[ \bm Y_S = \bm S^\star (\bm C^\star)^\T. \]
This completes the proof.

\section{Proof of Theorem \ref{the:blind_BTD}}
\label{app:identifiability_blind}
{The proof of Theorem \ref{the:blind_BTD} is by applying Lemma \ref{lemma:LL1} to the HSI and MSI individually. Then, the coupled tensor modeling is utilized for ``aligning'' the permutation and scaling ambiguities automatically.

Assume that $(\{\widetilde{\A}_{r}^{\star},\widetilde{\B}_{r}^{\star}\}_{r=1}^{R},\{\bm{A}_{r}^{\star},\bm{B}_{r}^{\star}\}_{r=1}^{R},\bm{C}^{\star})$ is an optimal solution of the formulation \eqref{blind_BTD_model}. Using Lemma \ref{lemma:LL1} and the assumptions, the following expressions hold 
\begin{equation}\label{eq:LL1_MSI}
\S^{\star}=\S\mathbf{\Pi}_{1} \mathbf{\Lambda}_{1},\ \bm{P}_{M}\C^{\star}=\bm{P}_{M}\C\mathbf{\Pi}_{1} \mathbf{\Lambda}_{1}^{-1},
\end{equation}
where $\bm \Pi_1$ and $\bm \Lambda_1$ are a permutation matrix and a nonsingular scaling matrix associated with the MSI decomposition, i.e., the decomposition model in \eqref{eq:msi_constraint}.

Moreover, applying Lemma \ref{lemma:LL1} to the HSI and the equality constraint in \eqref{eq:hsi_constraint}, the following expressions also hold
\begin{equation}\label{eq:LL1_HSI}
\widetilde{\S}^{\star}=\widetilde{\S}\mathbf{\Pi}_{2} \mathbf{\Lambda}_{2},\ \C^{\star}=\C\mathbf{\Pi}_{2} \mathbf{\Lambda}_{2}^{-1},
\end{equation}
where $\widetilde{\S}^{\star}$ and $\widetilde{\S}$ are denoted as in Appendix \ref{app:identifiability_non_blind}, and $\bm \Pi_2$ and $\bm \Lambda_2$ are a permutation matrix and a nonsingular scaling matrix associated with the decomposition model in \eqref{eq:hsi_constraint}.

Plugging $\C^{\star}$ into \eqref{eq:LL1_MSI}, we have
\begin{align}\label{eq:thm2key}
\bm{P}_{M}\C\mathbf{\Pi}_{2} \mathbf{\Lambda}_{2}^{-1}=\bm{P}_{M}\C\mathbf{\Pi}_{1} \mathbf{\Lambda}_{1}^{-1}.    
\end{align}
By Lemma \ref{lemma:joint}, $\bm{P}_{M}\bm{C}$ is drawn from an absolutely continuous joint distribution. 
Hence, the Kruskal rank of $\bm{P}_{M}\bm{C}$ is $\min(K_M,R)$ almost surely. 

First notice that if $R=1$, then $\bm \Pi_1=\bm \Pi_2$ and $\bm \Lambda_1 =\bm \Lambda_2$ hold trivially.

Second, consider the case where $R\geq 2$. Under such cases and by assuming that $K_M\geq 2$, we have 
\[   \min(K_M,R)\geq 2. \]
Hence, any two columns of $\bm{P}_{M}\bm{C}$ are linearly independent by the fact that $\bm{P}_{M}\bm{C}$ is drawn from a joint absolutely continuous distribution.
Our claim is that $\bm \Pi_1=\bm \Pi_2$ and $\bm \Lambda_1 =\bm \Lambda_2$ still hold under such circumstances. To see this, let us re-write \eqref{eq:thm2key} as follows:
\begin{align}
\bm{P}_{M}\C\bm Z_1(:,r) =\bm{P}_{M}\C\bm Z_2(:,r),  ~r=1,\ldots,R,     
\end{align}
where $\bm Z_i = \bm \Pi_i \bm \Lambda_i^{-1}$ for $i=1,2$.
Consequently, we have
\begin{align}\label{eq:thm2key_rewrite}
\bm{P}_{M}\C\left(\bm Z_1(:,r) -\bm Z_2(:,r)\right)=\bm 0,  ~r=1,\ldots,R.     
\end{align}
Note that the vector $\left(\bm Z_1(:,r) -\bm Z_2(:,r)\right)$ has at most two nonzero elements. However, any two columns of $\bm{P}_{M}\C$ are linearly independent.
This means that \eqref{eq:thm2key_rewrite} holds if and only if $\bm Z_1(:,r) = \bm Z_2(:,r)$ for $r=1,\ldots,R$, which leads to
\[ \bm \Lambda_1 =\bm \Lambda_2,\quad \bm \Pi_1 = \bm \Pi_2.     \]
As a result, we have the following equality:
\[\C^{\star}=\C\mathbf{\Pi}_{1} \mathbf{\Lambda}_{1}^{-1}.\]
Then, the matrix unfolding of the SRI, i.e., $\bm{Y}_S$, can be recovered by
\[ \bm Y_S = \bm S^\star (\bm C^\star)^\T. \]
This completes the proof.
}

\section{Gradients in \eqref{eq:alternating_non_blind}}
\label{app:gradient_non_blind}
The $\bm{C}$-subproblem is a quadratic program. The gradient can be readily derived:
\begin{align}\label{eq:gradient_C_non_blind}
\nabla_{\C} {\cal J}_{1}(\S^{(t)},\C^{(t)})=&\bm{C}^{(t)}(\bm{S}^{(t)})^{\top}(\bm{P}_{2}\otimes \bm{P}_{1})^{\top}(\bm{P}_{2}\otimes \bm{P}_{1})\bm{S}^{(t)} \nonumber\\
+&\bm{P}_{M}^{\top}\bm{P}_{M}\bm{C}^{(t)}(\bm{S}^{(t)})^{\top}\bm{S}^{(t)}+\lambda\bm{C}^{(t)}\\
-&\bm{Y}_{H}^{\top}(\bm{P}_{2}\otimes \bm{P}_{1})\bm{S}^{(t)}+\bm{P}_{M}^{\top}\bm{Y}_{M}^{\top}\bm{S}^{(t)}, \nonumber
\end{align}
where $\bm{Y}_{M}=\left[\underline{\Y}_{M}(1,1,:),\underline{\Y}_{M}(2,1,:),\ldots,\underline{\Y}_{M}(I_M,J_M,:)\right]^{\top}$.

\begin{table}[!ht]
  \centering
  \caption{Complexity of computing $\nabla_{\C} {\cal J}_{1}(\S^{(t)},\C^{(t)})$.}
  \resizebox{\linewidth}{!}{
    \begin{tabular}{c|c}\hline

    \hline
    Term  & \multicolumn{1}{c}{Complexity}  \\ \hline
    $\bm{C}^{(t)}(\bm{S}^{(t)})^{\top}\bm{P}_{H}^{\top}\bm{P}_{H}\bm{S}^{(t)}$   & $\mathcal{O}(R((d^2+R)I_H J_H +RK_H))$   \\ \hline
    $\bm{P}_{M}^{\top}\bm{P}_{M}\bm{C}^{(t)}(\bm{S}^{(t)})^{\top}\bm{S}^{(t)}$    & $\mathcal{O}(RI_M J_MK_H)$ \\ \hline 
    $\bm{Y}_{H}^{\top}\bm{P}_{H}\bm{S}^{(t)}$  &
    $\mathcal{O}(K_H(RI_M J_M+d^2I_HJ_H))$  \\ \hline
    $\bm{P}_{M}^{\top}\bm{Y}_{M}^{\top}\bm{S}^{(t)}$   &
    $\mathcal{O}(RI_M J_MK_H)$\\

    \hline
    \end{tabular}}%
  \label{table:complexity_gradient_C}%
\end{table}%

The detailed computational complexity analysis for instantiating $\nabla_{\C} {\cal J}_{1}(\S^{(t)},\C^{(t)})$ can be found in Table~\ref{table:complexity_gradient_C}. {The computational complexity may be reduced by exploiting the structure of $\bm{P}_H=\bm{P}_{2}\otimes \bm{P}_{1}$.  Note that $\bm{P}_{1}$ and $\bm{P}_{2}$ are often sparse \cite{Wei2015HSRSylvester,Kanatsoulis2018HSR}. For example, $d\times d$ Gaussian blurring kernels and downsampling operators are widely used to model $\bm P_1$ and $\bm P_2$, and such modeling makes these matrices sparse---the number of nonzero element $\textrm{nnz}(\bm{P}_1)=d I_H$ and $\textrm{nnz}(\bm{P}_2)=d J_H$; see, e.g., \cite{Kanatsoulis2018HSR}, for details.}

To see the analytical form of $\nabla_{\S} {\cal J}_{1}(\S,\C^{(t+1)})$, we construct a tight upper bounded function ${\cal F}(\S,\C^{(t+1)};\S^{(t)})$ at $t$-th iteration such that
\begin{align*}
   {\cal F}(\S^{(t)},\C^{(t+1)};\S^{(t)}) & \geq {\cal J}_{1}(\S^{(t)},\C^{(t+1)}),\\
    \nabla_{\S} {\cal F}(\S^{(t)},\C^{(t+1)};\S^{(t)}) &= \nabla_{\S} {\cal J}_{1}(\S^{(t)},\C^{(t+1)}).
\end{align*}
Then, we will compute $\nabla_{\S} {\cal J}_{1}(\S,\C^{(t+1)})$ through computing $\nabla_{\S} {\cal F}(\S^{(t)},\C^{(t+1)};\S^{(t)})$.

It was shown in \cite{Mohan2012Iterative,Wu2019Hyperspectral} that, for $0<p\leq 1$, $\phi_{p,\tau}(\bm{Z})$ admits a quadratic tight upper bound (i.e., a {\it majorizer}) at $\bm{Z}^{(t)}$:
\begin{equation}\label{majprizer_Sp}
\begin{split}
& \widetilde{\phi}(\bm{Z},\bm{W}^{(t)}) \\
&=\frac{p}{2}\textrm{tr}\big(\bm{W}^{(t)}(\bm{Z}\bm{Z}^{\top}+\tau\bm{I})\big)+\frac{2-p}{p}\textrm{tr}\big((\bm{W}^{(t)})^{\frac{p}{p-2}}\big),
\end{split}
\end{equation}
where $\bm{W}^{(t)}=\big(\bm{Z}^{(t)}(\bm{Z}^{(t)})^{\top}+\tau\bm{I}\big)^{\frac{p-2}{2}}$. In addition, it is shown in \cite{Fu2015Joint} that a quadratic majorizer $\widetilde{\varphi}(\bm{z},\bm{z}^{(t)})$ of $\varphi_{q,\varepsilon}(\bm{z})$ ($0<q\leq 1$) is as follows
\begin{align}\label{majprizer_Lp}
\widetilde{\varphi}(\bm{z},\bm{z}^{(t)}) &= \sum_{i} [\bm{w}^{(t)}]_{i}[\bm{z}]_{i}^{2}+\frac{2-q}{2}\Big(\frac{2}{q}[\bm{w}^{(t)}]_{i}\Big)^{\frac{q}{q-2}}+\varepsilon [\bm{w}^{(t)}]_{i}\nonumber \\
&=\frac{q}{2}\bm{z}^{\top}\bm{U}^{(t)}\bm{z}+{\rm const},
\end{align}
where $[\bm{w}^{(t)}]_{i}=\frac{q}{2}\big(([\bm{z}^{(t)}]_{i})^{2}+\varepsilon\big)^{\frac{q-2}{2}}$,  $\bm{U}^{(t)}$ is a diagonal matrix with $[\bm{U}^{(t)}]_{i,i}=[\bm{w}^{(t)}]_{i}$ and ${\rm const}$ is a constant. Combining \eqref{majprizer_Sp} and \eqref{majprizer_Lp}, we obtain the quadratic majorizer function ${\cal F}(\S,\C^{(t+1)};\S^{(t)})$ as follows:
\begin{align}\label{eq:quadratic_majorizer_S}
&{\cal F}(\S,\C^{(t+1)};\S^{(t)})=\frac{1}{2}\|\bm{Y}_{H}-(\bm{P}_{2}\otimes \bm{P}_{1})\bm{S} (\bm{C}^{(t+1)})^{\top}\|_{F}^{2} \nonumber \\
+&\frac{1}{2}\|\bm{Y}_{M}-\bm{S}(\bm{C}^{(t+1)})^{\top}\bm{P}_{M}^{\top}\|_{F}^{2} 
+\sum_{r=1}^{R}\widetilde{\phi}(\bm{S}_{r},\bm{W}_{r}^{(t)}) \nonumber \\
+&\sum_{r=1}^{R} \big(\widetilde{\varphi}(\bm{H}_{x}{\bm q}_r,\bm{H}_{x}{\bm q}_r^{(t)})+ \widetilde{\varphi}(\bm{H}_{y}{\bm q}_r,\bm{H}_{y}{\bm q}_r^{(t)}) \big).
\end{align}
The gradient of ${\cal F}(\S,\C^{(t+1)};\S^{(t)})$ w.r.t. $\S$ can be expressed as follows:
\[\label{eq:gradient_S_non_blind}
\begin{split}
\nabla_{\S}& {\cal J}_{1}(\S^{(t)},\C^{(t+1)})\\
=&(\bm{P}_{2}\otimes \bm{P}_{1})^{\top}\big((\bm{P}_{2}\otimes \bm{P}_{1})\bm{S}^{(t)} (\bm{C}^{(t+1)})^{\top}-\bm{Y}_{H}\big)\bm{C}^{(t+1)}\\
+&\big(\bm{S}^{(t)}(\bm{C}^{(t+1)})^{\top}\bm{P}_{M}^{\top}-\bm{Y}_{M}\big)\bm{P}_{M}\bm{C}^{(t+1)}\\
+&p[\eta_{1}\textrm{vec}(\bm{W}^{(t)}_{1}\bm{S}_{1}^{(t)}),\ldots,\eta_{R}\textrm{vec}(\bm{W}^{(t)}_{R}\bm{S}_{R}^{(t)})]\\
+&q[\theta_{1}\bm{H}_{x}^{\top}\bm{U}_{1}^{(t)}\bm{H}_{x}\bm{q}_{1}^{(t)},\ldots,\theta_{R}\bm{H}_{x}^{\top}\bm{U}_{R}^{(t)}\bm{H}_{x}\bm{q}_{R}^{(t)}]\\
+&q[\theta_{1}\bm{H}_{y}^{\top}\bm{V}_{1}^{(t)}\bm{H}_{y}\bm{q}_{1}^{(t)},\ldots,\theta_{R}\bm{H}_{y}^{\top}\bm{V}_{R}^{(t)}\bm{H}_{y}\bm{q}_{R}^{(t)}],
\end{split}
\]
where
$\bm{W}^{(t)}_{r}=(\bm{S}^{(t)}_{r}(\bm{S}^{(t)}_{r})^{\top}+\tau\bm{I})^{\frac{p-2}{2}}$,
$[\bm{U}_{r}^{(t)}]_{i,i}=\big([\bm{H}_{x}{\bm q}_r^{(t)}]_{i}^{2}+\varepsilon\big)^{\frac{q-2}{2}}$,
and $[\bm{V}_{r}^{(t)}]_{i,i}=\big([\bm{H}_{y}{\bm q}_r^{(t)}]_{i}^{2}+\varepsilon\big)^{\frac{q-2}{2}}$, $r=1,\ldots, R$. 

\begin{table}[!ht]
  \centering
  \caption{Complexity of computing $\nabla_{\S} {\cal J}_{1}(\S^{(t)},\C^{(t+1)})$.}
  \resizebox{\linewidth}{!}{
    \begin{tabular}{c|c}\hline

    \hline
    Term  & \multicolumn{1}{c}{Complexity}  \\ \hline
    $\bm{P}_{H}^{\top}\big(\bm{P}_{H}\bm{S}^{(t)} (\bm{C}^{(t+1)})^{\top}-\bm{Y}_{H}\big)\bm{C}^{(t+1)}$   & $\mathcal{O}(I_HJ_H(d^2 R+R K_H+d^2 K_H))$   \\ \hline
    $\big(\bm{S}^{(t)}(\bm{C}^{(t+1)})^{\top}\bm{P}_{M}^{\top}-\bm{Y}_{M}\big)\bm{P}_{M}\bm{C}^{(t+1)}$    & $\mathcal{O}(RI_M J_MK_H)$ \\ \hline 
    $\textrm{vec}(\bm{W}^{(t)}_{r}\bm{S}_{r}^{(t)})$  &
    $\mathcal{O}(RI_M^2(I_M+J_M))$  \\ \hline
    $\bm{H}_{x}^{\top}\bm{U}_{r}^{(t)}\bm{H}_{x}{\bm q}_r^{(t)}$, $\bm{H}_{y}^{\top}\bm{V}_{r}^{(t)}\bm{H}_{y}{\bm q}_r^{(t)}$   &
    $\mathcal{O}(RI_MJ_M)$\\

    \hline
    \end{tabular}}%
  \label{table:complexity_gradient_S}%
\end{table}%

The complexity of constructing $\nabla_{\C} {\cal J}_{1}(\S^{(t)},\C^{(t)})$ can be seen in Table~\ref{table:complexity_gradient_S}.

\section{Proof of Proposition \ref{pro:non_blind_convergence}}
\label{proof:non_blind_convergence}
With the derivations in Appendix~\ref{app:gradient_non_blind}, it is readily seen that the largest eigenvalue of the Hessian of ${\cal J}_1(\bm S^{(t)},\bm C^{(t)})$ w.r.t. $\bm C^{(t)}$ is upper boounded by the following:
\begin{align*}
&\big(\sigma_{\textrm{max}}\big((\bm{S}^{(t)})^{\top}(\bm{P}_{2}\otimes \bm{P}_{1})^{\top}(\bm{P}_{2}\otimes \bm{P}_{1})\bm{S}^{(t)}\big)\\
&+\sigma_{\textrm{max}}(\bm{P}_{M}^{\top}\bm{P}_{M})\times\sigma_{\textrm{max}}\big((\bm{S}^{(t)})^{\top}\bm{S}^{(t)}\big)+\lambda\big),
\end{align*}
since the subproblem is a quadratic program. Hence, the above is a legitimate Lipschitz constant $L_{\bm C}^{(t)}$ of the corresponding gradient.

Similarly, a Lipschitz constant $L_{\S}^{(t)}$ of $\nabla_{\S} {\cal J}_{1}(\S^{(t)},\C^{(t+1)})$ can be derived---which is as specified in Proposition~\ref{pro:non_blind_convergence}.

With these two constants being bounded $L_{\bm C}^{(t)}<\infty$ and $L_{\bm S}^{(t)}<\infty$, it can be seen that the alternating gradient projection algorithm in Algorithm~\ref{our_non_blind_algorithm} falls into the category of inexact block coordinate descent with {\it sufficient decrease} guarantees in each iteration \cite{Xu2013BCD}---if the step sizes are chosen following the rules stated in the statement of Proposition~\ref{pro:non_blind_convergence}. Then, invoking the convergence theory (i.e., Theorem~2.8 \cite{Xu2013BCD}) there, one can easily show that every limit point of solution sequence produced by the proposed algorithm is a stationary point of the optimization problem of interest.

\section{Gradients and Lipschitz constants in Algorithm \ref{our_blind_algorithm}}
\label{pro:our_blind_algorithm}
When the spatial degradation operators are unknown, similar gradient calculations as in Appendices \ref{app:gradient_non_blind}-\ref{proof:non_blind_convergence} can be applied. 
One can show the following:
\[
\begin{split} 
\nabla_{\C} &{\cal J}_{2}(\S^{(t)},\widetilde{\S}^{(t)},\C^{(t)})\\
&=\bm{C}^{(t)}(\widetilde{\bm{S}}^{(t)})^{\top}\widetilde{\bm{S}}^{(t)}\\
&+\bm{P}_{M}^{\top}\bm{P}_{M}\bm{C}^{(t)}(\bm{S}^{(t)})^{\top}\bm{S}^{(t)}+\lambda\bm{C}^{(t)}\\
&-(\bm{Y}_{H})^{\top}\widetilde{\bm{S}}^{(t)}+\bm{P}_{M}^{\top}(\bm{Y}_{M})^{\top}\bm{S}^{(t)},\\
\nabla_{\S}& {\cal J}_{2}(\S^{(t)},\widetilde{\S}^{(t)},\C^{(t+1)})\\
&=\big(\bm{S}^{(t)}(\bm{C}^{(t+1)})^{\top}\bm{P}_{M}^{\top}-\bm{Y}_{M}\big)\bm{P}_{M}\bm{C}^{(t+1)}\\
&+p[\eta_{1}\textrm{vec}(\bm{W}^{(t)}_{1}\bm{S}_{1}^{(t)}),\ldots,\eta_{R}\textrm{vec}(\bm{W}^{(t)}_{R}\bm{S}_{R}^{(t)})]\\
&+q[\theta_{1}\bm{H}_{x}^{\top}\bm{U}_{1}^{(t)}\bm{H}_{x}\bm{q}_{1}^{(t)},\ldots,\theta_{R}\bm{H}_{x}^{\top}\bm{U}_{R}^{(t)}\bm{H}_{x}\bm{q}_{R}^{(t)}]\\
&+q[\theta_{1}\bm{H}_{y}^{\top}\bm{V}_{1}^{(t)}\bm{H}_{y}\bm{q}_{1}^{(t)},\ldots,\theta_{R}\bm{H}_{y}^{\top}\bm{V}_{R}^{(t)}\bm{H}_{y}\bm{q}_{R}^{(t)}],\\
\nabla_{\widetilde{\S}_{r}}& {\cal J}_{2}(\S^{(t+1)},\widetilde{\S}^{(t)},\C^{(t+1)})\\
&=\big(\widetilde{\S}^{(t)}(\bm{C}^{(t+1)})^{\top}-\bm{Y}_{H}\big)\bm{C}^{(t+1)}\\
&+p[\eta_{1}\textrm{vec}(\widetilde{\W}^{(t)}_{1}\widetilde{\S}_{1}^{(t)}),\ldots,\eta_{R}\textrm{vec}(\widetilde{\W}^{(t)}_{R}\widetilde{\S}_{R}^{(t)})].\\
\end{split}
\]
In addition, we have
\[
\begin{split} 
L_{\bm C}^{(t)}&=\sigma_{\textrm{max}}(\bm{P}_{M}^{\top}\bm{P}_{M})\times\sigma_{\textrm{max}}\big((\bm{S}^{(t)})^{\top}\bm{S}^{(t)}\big)\\
&+\sigma_{\textrm{max}}\big((\widetilde{\bm{S}}^{(t)})^{\top}\widetilde{\bm{S}}^{(t)}\big)
+\lambda,\\
L_{\bm S}^{(t)}&=\sigma_{\textrm{max}}((\bm{C}^{(t+1)})^{\top}\bm{P}_{M}^{\top}\bm{P}_{M}\bm{C}^{(t+1)})\\
&+p\max_{r=1,\ldots,R}\eta_{r}\sigma_{\textrm{max}}(\bm{W}^{(t)}_{r})\\
&+q\max_{r=1,\ldots,R}\theta_{r}\sigma_{\textrm{max}}(\bm{H}_{x}^{\top}\bm{U}_{r}^{(t)}\bm{H}_{x})\\
&+q\max_{r=1,\ldots,R}\theta_{r}\sigma_{\textrm{max}}(\bm{H}_{y}^{\top}\bm{V}_{r}^{(t)}\bm{H}_{y}),\\
L_{\widetilde{\S}}^{(t)}&=\sigma_{\textrm{max}}((\bm{C}^{(t+1)})^{\top}\bm{C}^{(t+1)})\\
&+p\max_{r=1,\ldots,R}\eta_{r}\sigma_{\textrm{max}}(\widetilde{\W}^{(t)}_{r}),\\
\end{split}
\]
where
$\widetilde{\W}^{(t)}_{r}=(\widetilde{\S}^{(t)}_{r}(\widetilde{\S}^{(t)}_{r})^{\top}+\tau\bm{I})^{\frac{p-2}{2}}$, $r=1,\ldots, R$.

\end{document}